\documentclass[twoside,12pt]{article}
\usepackage{epsfig}
\usepackage{amsmath,amssymb}
   
\def\Journal#1#2#3#4{{#1} {#2} (#4) #3 }

\def\NPA{{\em Nucl.\ Phys.} A}
\def\NPB{{\em Nucl.\ Phys.} B}

\def\PRO{{\em Prog.\ Theor.\ Phys.}}
\def\NPB{{\em Nucl.\ Phys.} B}
\def\PLA{{\em Phys.\ Lett.} A}
\def\PLB{{\em Phys.\ Lett.} B}

\def\PRL{\em Phys.\ Rev.\ Lett.}

\def\PREP{\em Phys.\ Rep.}

\def\PRC{{\em Phys.\ Rev.} C}
\def\PRD{{\em Phys.\ Rev.} D}
\def\PRE{{\em Phys.\ Rev.} E}
\def\ZPC{{\em Z.\ Phys.} C}

\def\ANNP{\em Ann.\ Phys.\ (N.Y.)}
\def\RMP{{\em Rev.\ Mod.\ Phys.}}

\def\JHEP{{\em JHEP}}
\def\JPG{{\em J.\ Phys.} G}
\def\NCS{{\em Nuovo Cim.\ Suppl.}}
\def\UFN{{\em Usp.\ Fiz.\ Nauk}}
\def\ZP{{\em Z.\ Phys.}}
\def\BJP{\em Braz.\ J.\ Phys.}
\def\JCP{{\em J.\ Comput. Phys.}}
\def\EPC{{\em Eur.\ Phys.\ J.} C} 

\def\Nat{{\em Nature}}

\newcommand{\be}{\begin{equation}}
\newcommand{\ee}{\end{equation}}
\newcommand{\bea}{\begin{eqnarray}}
\newcommand{\eea}{\end{eqnarray}}

\newcommand{\vba}{\mathbf{b}_A}
\newcommand{\vbb}{\mathbf{b}_B}
\newcommand{\vs}{\mathbf{s}}
\newcommand{\vecr}{\mbox {\boldmath$r$}}
\newcommand{\vj}{\mbox {\boldmath$j$}}
\newcommand{\vn}{\mbox {\boldmath$n$}}
\newcommand{\vsigma}{\mbox {\boldmath$\sigma$}}

\begin{document}

\title{ \vspace{1cm} Evaluating Results from the Relativistic Heavy Ion
  Collider with Perturbative QCD and Hydrodynamics}
\author{R.\ J.\ Fries,$^{1,2}$ C.\ Nonaka$^3$
\\
\\
$^1$Cyclotron Institute, Texas A\&M University, College Station TX, USA\\
$^2$RIKEN/BNL Research Center, Brookhaven National Laboratory, Upton NY, USA\\
$^3$Department of Physics, Nagoya University, Nagoya, Japan}
\maketitle
\begin{abstract} 
We review the basic concepts of perturbative quantum chromodynamics (QCD) 
and relativistic hydrodynamics, and their applications to hadron production in 
high energy nuclear collisions. We discuss results from
the Relativistic Heavy Ion Collider (RHIC) in light of these theoretical
approaches. Perturbative QCD and hydrodynamics together explain a large 
amount of experimental data gathered during the first decade of RHIC running,
although some questions remain open.
We focus primarily on practical aspects of the calculations,
covering basic topics like perturbation theory, initial state 
nuclear effects, jet quenching models, ideal hydrodynamics, dissipative
corrections, freeze-out and initial conditions. We conclude by comparing
key results from RHIC to calculations.
\end{abstract}
\eject
\tableofcontents
\eject

\section{Introduction\label{sec:introduction}}

The Relativistic Heavy Ion Collider (RHIC) at Brookhaven National Laboratory
started operations about a decade ago. The amount of data collected and the quality
of the data have been outstanding. Besides a successful proton-proton and
proton-nucleus program, RHIC has mostly provided data on nuclear collisions,
from a few GeV center of mass energy up to 200 GeV per nucleon-nucleon pair.
We have strong evidence that the central goal of the RHIC program, the
discovery  of quark gluon plasma (QGP), a deconfined state of nuclear matter, 
has been achieved. In order to draw this conclusion a wide variety of
observables have been weighed against theoretical expectations and we will
discuss a few of those in this article.
Some key experimental discoveries at RHIC over the past decade were
(i) the extremely strong jet quenching, many times
that of ordinary nuclear matter \cite{Adler:2003qi,Adams:2003kv};
(ii) the very large elliptic flow of the fireball that confirms
collective behavior at energy densities larger
than expected at the phase transition \cite{white_papers};
(iii) the surprising quark number scaling of elliptic flow that
seems to indicate that the collective flow is carried by quarks
\cite{Abelev:2008ed,Afanasiev:2009wq} (see \cite{Dusling:2009df} for an
attempt of an alternative explanation); and (iv) direct photon measurements
that suggest large initial temperatures  \cite{:2008fqa}.
Even before the partonic nature of the fireball could be established there was
mounting evidence that the hot matter at RHIC was not behaving like 
a weakly interacting gas, but rather like a strongly interacting
liquid. This has led to the conjecture that quark gluon plasma is a nearly perfect liquid
\cite{Gyulassy:2004zy}, at least close to the phase transition temperature. 
Conservative estimates for the initial energy density in the center of head-on 
collisions at top RHIC energy find a lower bound $\sim 3$ GeV/fm$^3$, 
which is above the estimated critical energy density \cite{white_papers}.

Perturbative quantum chromodynamics (pQCD) and relativistic hydrodynamics
have been two important tools to understand and interpret RHIC data. 
It was found that the bulk of the produced particles at RHIC
(for transverse momenta $P_T$ smaller than $\approx 2$ GeV/$c$) show
signatures of collective behavior. The mean-free path of particles seems to 
be small enough for the dynamics to be described by relativistic fluid dynamics.
This was a non-trivial finding since hydrodynamic descriptions for lower
energy nuclear collisions routinely overestimated the amount of collectivity.
While hydrodynamic modeling 10 years ago was still rough, based on 
(2+1)-dimensional ideal fluid dynamics with simple initial conditions
and freeze-out, there has been an amazing amount of progress since
then by going to full (3+1)-dimensional modeling, taking into account
dissipative corrections, fine-tuning of initial conditions all the way
to event-by-event calculations, and a deeper understanding of the hadronic
phase with separate chemical and thermal freeze-outs, and through the advent
of hybrid hydro+cascade models. We will highlight many of these
improvements in this article.
The progress has enabled hydrodynamic models --- and the entire RHIC program
--- to enter a phase in which quantitative measurements are finally close.
Prime candidates for such quantitative measurements are the equation of 
state of hot QCD, including the order of the phase transition between hadronic
matter and QGP and the existence and location of a critical point,  and the
shear viscosities of these phases. The measurement of other bulk transport 
coefficients, like the bulk viscosity and relaxation times, are in principle possible but
remain elusive for now. We will discuss the status and potential problems of
such measurements.

Hydrodynamics describes the bulk of the particles in a collision
(more than 98\% of them).
The tail of the particle $P_T$-spectra in nuclear collisions,
which clearly contain particles that have not thermalized, should not simply
be disregarded. In fact it was proposed a long time ago that they can 
serve as ``hard probes'' of the bulk matter created. In elementary
$p+p$ or $p+\bar p$ collisions hadrons with transverse momenta of 5 GeV/$c$ 
or more are created through a single hard scattering of two partons within 
the wave functions of the colliding hadrons, which then fragment in the
vacuum away from the collision into collimated bunches of hadrons, called jets.
This entire process can be calculated
in perturbative QCD due to the large momentum transfer involved, while the
unavoidable non-perturbative contributions can be treated in a controlled
way through a formalism called collinear factorization. Perturbative
QCD based on collinear factorization has been a great success story
in elementary collisions \cite{Brock:1993sz}. Hard initial scatterings
of partons from the initial nuclear wave functions should proceed in
a way very similar to elementary collisions, with the understanding that
the wave functions of free nucleons and those in nuclei might differ 
somewhat. However, the big difference arises in the final state, when
an outgoing parton or jet finds itself embedded in a fireball of
hot and dense quark gluon plasma. Clearly we expect those partons to
rescatter and lose energy through elastic collisions or bremsstrahlung.
The final state effects on high-$P_T$ hadrons and jets should encode 
valuable information about the QGP phase. The most prominent example 
is the transport coefficient $\hat q$ that parameterizes the average momentum
transfer per unit path length to a fast parton in the medium. The so-called
LPM effect, coming from the finite formation time of induced radiation, 
leads to a signature quadratic dependence on the thickness of the medium.
We will focus our attention here on the leading particle description which has
received the most attention by theoreticians and is the most relevant effect
for observables measured so far. Despite this restriction to the apparently
simplest problem we will see that we do not yet have a consistent description
of this problem. We will not deal in detail with more comprehensive approaches 
that follow full jet showers in the medium.

In this review we want to lay out the basic concepts of both perturbative QCD 
and relativistic hydrodynamics and their applications to hadron observables in
nuclear collisions at RHIC. This will enable us to discuss some important
results from RHIC
and to draw conclusions. The article is organized as follows. In Section
\ref{sec:pqcd} we review the fundamentals of collinear factorization,
parton distributions and fragmentation functions and simple pQCD 
cross section computations. We will then see how these processes change
in a nuclear environment leading us to nuclear shadowing and the Cronin
effect. We then proceed to discuss final state energy loss and the LPM
effect. We focus on four common models of leading parton energy loss. Finally we
give a quick overview of photon production in heavy ion collisions. In Section
\ref{sec:hydro} we present the basic concepts of both ideal and viscous
hydrodynamics, and quickly comment on possible numerical implementations.
Then we connect hydrodynamics to the bigger picture of heavy ion collisions
and discuss necessary details like the equation of state, initial conditions
and freeze-out procedures. We also briefly touch upon quark recombination.
In Section \ref{sec:results} we discuss data from RHIC in light of the
previous two sections. We present single particle spectra for light
hadrons, azimuthal asymmetries, hadron correlations, and photons and
their correlations. We have omitted heavy quarks and dileptons in this
review which are very interesting topics in their own right but would
have significantly increased the size of this article.
Section \ref{sec:sum} contains our conclusions and summary.
Along the way we try to emphasize practical applications of theory over
technical derivations. We hope that this article serves as a useful guide 
for the practitioner.

\section{Particle Production in Perturbative Quantum Chromodynamics
\label{sec:pqcd}}

Perturbation theory is a well established tool to deal with interacting
quantum field theories. In quantum electrodynamics (QED) it
has produced some of the most accurate predictions confirmed by experimental 
data. The basic concept is an expansion of observables in powers of the 
coupling constant $g$ of the theory if $g \ll 1$.
Naturally, this method becomes unreliable if $g$ is too large. 
Unfortunately, in quantum chromodynamics the strong coupling
$\alpha_s = g^2/(4\pi)$ grows logarithmically as the momentum transfer
squared $Q^2$ decreases. This behavior immediately raises serious questions 
about the usefulness of perturbation theory in any realistic situation.
Weak coupling methods should work in the asymptotic limit $Q^2 \to \infty$. 
But QCD bound states, hadronization, and the transport properties around the 
QCD phase transition temperature are completely outside the perturbative
region.
Nevertheless perturbation theory in QCD, truncated after the few 
lowest orders, together with a rigorous factorization program, to separate
off infrared divergences representing the long distance behavior of QCD, 
has been shown to work down to $Q^2 \approx 1$ GeV$^2$ in 
some applications. Not all processes feature a rigorous and unambiguous 
factorization, but we have hope that hadron and photon production at
large transverse momentum $P_T$ in nuclear collisions can be described 
by perturbative methods. The goal of this program is it to use high-$P_T$
hadrons and jets as hard probes, whose final state interactions with the
bulk of the event reveals important information about quark gluon plasma.

In the first subsection we review some of the basic principles of computing
the cross sections or yield of high-$P_T$ hadrons and jets at 
hadron colliders. In the second part we discuss modifications expected 
in collisions involving nuclei, focusing in particular on initial state, or
cold nuclear matter effects. In the third part we address
final state effects and parton energy loss for hadrons and jets, which we
critique and compare further in the fourth part. In
the last subsection we briefly discuss the production of photons.

\subsection{\it Factorization in pQCD}

\begin{figure}[tb]
\begin{center}
\begin{minipage}[t]{8 cm}
\epsfig{file=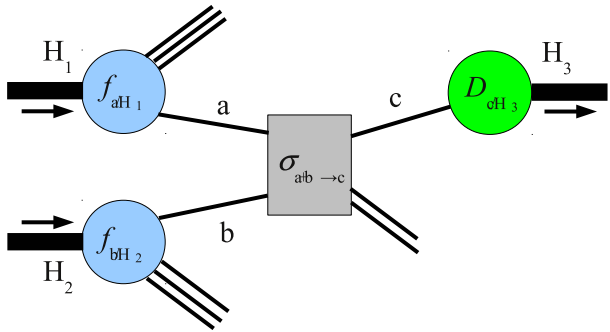,scale=0.35}
\end{minipage}
\begin{minipage}[t]{16.5 cm}
\caption{Schematic sketch for the amplitude of the leading factorized 
processes for production of a single hadron $H_3$ in the collision of
two hadrons $H_1$ and $H_2$. The thin lines represent single partons
connecting the soft (i.e.\ long distance) processes within the
hadrons with the hard (i.e.\ short-distance) scattering. Soft and
hard processes are depicted by round blobs and boxes respectively. The
probability distributions referring to the soft and hard processes
are defined as the parton distributions $f$ or fragmentation functions $D$,
and the hard partonic cross section $\sigma_{a+b\to c}$ resp.
\label{fig:factor}}
\end{minipage}
\end{center}
\end{figure}

Even though the creation of hadrons at large transverse momentum $P_T \gg
1$ GeV involves a large momentum transfer $Q\sim P_T$, one has to deal 
with the fact that the initially colliding hadrons $H_1$ and $H_2$, and the 
final hadrons $H_3$, $H_4$, $\ldots$ are multi-parton states in QCD bound by 
non-perturbative dynamics. Fortunately, for several key processes it has been 
possible to prove factorization theorems
\cite{Mueller:1981sg,Ellis:1982wd,Collins:1985ue,Collins:1989gx,Qiu:1990xxa,Qiu:1990xy},
see also \cite{Brock:1993sz} for a more didactic introduction. They allow us
to separate perturbative and non-perturbative processes in a well-defined, 
systematic way by factorizing all infrared and long-range dynamics into 
universal, well-defined and observable matrix elements.
They establish an expansion (in powers of $1/Q$) of the underlying ``hard''
partonic process. The leading process in $1/Q$, often called leading twist,
is usually the one with the \emph{fewest possible partons} connecting
to the long-range part. 
E.g.\ for single (or di-hadron) production from two hadrons,
$H_1 + H_2 \to H_3 +X$ (or $H_1 + H_2 \to H_3 + H_4 + X$),
the leading underlying parton process is that of $2 \to 2$ scattering of 
partons $a + b \to c + d$ with parton $a$, $b$, $c$, ($d$) being associated 
with bound states $H_1$, $H_2$, $H_3$, ($H_4$), resp., see Fig.\
\ref{fig:factor}.
The blobs in Fig.\ \ref{fig:factor} represent the association of one
parton with its parent hadron. In the initial state these are called
parton distributions $f_{a/H}$, in the final state they are
fragmentation functions $D_{c/H}$.

The factorization of the cross section can schematically be written as
\be
  d \sigma_{H_1+H_2\to H_3+X} = 
  \sum_{a,b,c} f_{a/H_1} \otimes f_{b/H_2} \otimes d\sigma_{a+b\to c+x}
  \otimes D_{c/H_3} + \mathrm{power\> suppr.\> terms}
 \label{eq:fact}
\ee
where $\sigma_{a+b\to c+x}$ is the hard partonic cross section\footnote{More
precisely this is the cross section modulo some collinear and infrared
divergences which have been factorized into the parton distributions and
fragmentation functions} involving a large momentum transfer.
Processes involving other ``associations'', most notably those with more 
partons taken from one hadron are of higher twist and suppressed by 
powers of $1/Q \sim 1/P_T$. The convolution signs mean that the parton
momenta connecting blobs and hard cross sections have to be
integrated, if not fixed by kinematics. This will become more clear 
when concrete examples are discussed further below. 

A few additional remarks are in order. 
\begin{itemize}
\item  Here we will only deal with collinear 
  factorization. This is sufficient for processes with a single hard scale 
  $Q$, or even for processes with different scales $Q_1$, $Q_2$ 
  (then resummations are needed) as long as all scales are large.
  So called $k_T$-factorization is needed for processes with one hard and one
  soft scale and will not be discussed here \cite{Belitsky:2002sm,Ji:2004wu}.
  Practically this means partons can kinematically always be treated as 
  collinear with their parent hadrons, which simplifies the momentum 
  integrals in the factorization formulas tremendously. 
\item At first we will only discuss leading twist processes. With scales
  of the order of a few GeV this is sufficient for single hadrons. However,
  in the case of nuclei we will see that some higher twist processes are
  enhanced and become important. 
\item We will also refrain from discussing particle production in the limit
  of very large center of mass energy, when the gluon distribution of hadrons
  saturates. For large nuclei this limit might be reached at RHIC energies and
  particle production from this Color Glass state could be dominant for 
  particles at lower $P_T$ (from scattering of partons with low Bjorken-$\xi$ in
  the initial wave function). With the saturation scale $Q_s$ for gold
  nuclei at RHIC energies estimated to be smaller than 2 GeV 
  \cite{Kharzeev:2000ph}, 
  the high-$P_T$ domain should be safely in the region of collinear 
  factorization. We will revisit this topic in the section about initial
  conditions for hydrodynamics.
  For the most recent reviews of the Color Glass Condensate see e.g.\
  \cite{Gelis:2010nm,Lappi:2010ek}.
\item Collinear factorization has been rigorously proven in very few
  cases, and even for simple processes there are examples where factorization
  breaks at a certain order in $1/Q^n$ \cite{Doria:1980ak}. This is particularly
  worrisome for collisions of nuclei where multiple scattering and higher
  twist corrections are enhanced. We will always assume factorization 
  of initial states and hard processes here. On the other hand the study 
  of final state effects in nuclear collisions is by definition an 
  investigation of how factorization and universality are broken for 
  long-distance final states.
\end{itemize}

\subsubsection{Cross Sections of Partons}

The factorization theorems mentioned above make sure that the underlying
hard parton cross sections are infrared-safe. They can be calculated
in a perturbative expansion. Singularities from radiative corrections
can be factored off into the long-distance part that is described by
parton distributions and fragmentation functions. E.g., for our example of
single hadron production at large momentum the underlying parton
cross section can be written as
\be
  d\sigma_{a+b\to c+x} = \alpha_S^2 d\sigma_{a+b\to c+x}^\mathrm{(LO)}
  +  \alpha_S^3 d\sigma_{a+b\to c+x}^\mathrm{(NLO)} + \ldots.
  \label{eq:pertexp}
\ee

Leading order (parton model) cross sections are easily calculated. For further
processing parton cross sections $a+b\to c+d$ are most easily parameterized in
terms of the Lorentz-invariant Mandelstam variables $s=(p_a+p_b)^2$,
$t=(p_a-p_c)^2$, $u=(p_a-p_d)^2$ where $p_a$, $p_b$, etc.\ are the
four-momenta of partons $a$, $b$, etc. For example for the scattering
of two different quark (or antiquark) species $q$ and $q'$
\be
  \frac{d\sigma^\mathrm{(LO)}_{q+q'\to q+q'}}{dt} = \frac{\pi\alpha_s^2}{s^2}
  \frac{N_c^2-1}{2N_c^2}\frac{s^2+u^2}{t^2}
\ee
where $N_c=3$ is the number of colors, and by definition we have averaged over
ingoing spins and colors and summed over outgoing spins and colors
(we have kept the coupling constant as part of the cross section unlike
indicated in (\ref{eq:pertexp})). A
comprehensive table for production of light partons and photons can be 
found in the review article by Owens \cite{Owens:1986mp}.

Next-to-leading (NLO) calculations of parton production is much more
challenging. The basic matrix elements can be found in the work by
Ellis and Sexton \cite{Ellis:1985er}. The one- and two-jet cross sections 
were e.g.\ worked out in \cite{Ellis:1990ek,Ellis:1992en}.
Several numerical codes are available for jet, hadron or photon production at
NLO accuracy, performing the required phase space integrals and cancellation
of singularities. An excellent starting point for the interested reader is
the PHOX collection by Aurenche and collaborators \cite{PHOX}.

We have to discuss an important point here. From the NLO-level on cross sections
with parton final states are no longer well defined, i.e. infrared-safe.
In fact we can only define cross sections either into hadrons or jets, i.e.\
sprays of hadrons defined by energy in a restricted region of 
phase space. At leading order one can make the convenient identification
\emph{jet} = \emph{parton}. At NLO two partons can be so close together
in phase space that they have to be replaced with one jet.

In nuclear collisions with its emphasis on final state effects the convenient
identification becomes a necessary simplification that allows for the
treatment of energy loss and other effects on the basis of single partons.
This is also one reason why a large fraction of literature on heavy ion
collisions uses leading order calculations. Recently, more and more NLO-based
calculations have been presented. In that case caution is in order if 
they are combined with final state effects based on a single parton picture.

The basis for the use of LO cross sections is the fact that for single 
and double hadron and photon $P_T$-spectra LO accuracy yields
reasonably good results. It turns out that for collisions of single
hadrons
\be
  \frac{d\sigma^\mathrm{(NLO)}}{d^2 P_T} = K 
  \frac{d\sigma^\mathrm{(LO)}}{d^2 P_T}
\ee
with a $K$-factor that is close to one and only weakly dependent on the 
momentum $P_T$ of the produced hadron \cite{Barnafoldi:2000dy}. For 
convenience $K$ is hence often approximated by a constant.

\subsubsection{Parton Distributions and Fragmentation Functions}

In the schematic factorization formula (\ref{eq:fact}) $f_{a/H_1}$ and
$f_{b/H_2}$ are parton distribution functions (PDFs) which describe the 
probabilities that partons $a$, $b$ can be found in hadrons
$H_1$, $H_2$, resp., with given momenta. Note that factorization at
leading twist provides a very satisfying probabilistic picture
(there is no interference between amplitude and complex conjugated amplitude
of the parton line connecting the hard cross section with bound states).
Parton distributions are well-defined and gauge invariant matrix elements 
in QCD. They are also universal, i.e.\ their definition is independent of 
the particular process in which they occur. 

\begin{figure}[tb]
\begin{center}
\begin{minipage}[t]{16 cm}
\epsfig{file=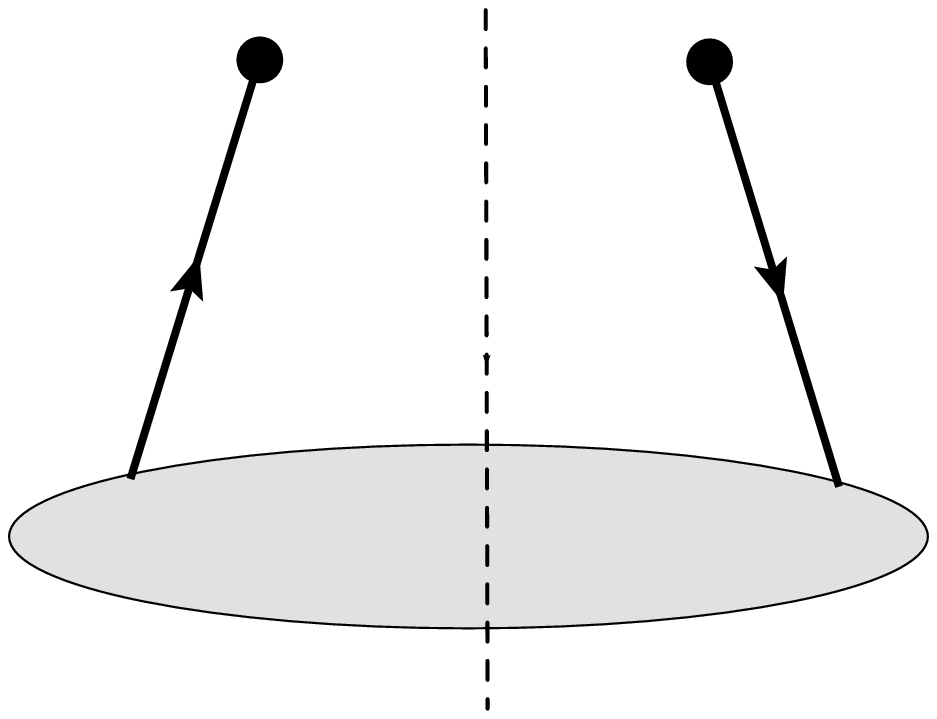,scale=0.35} \qquad\qquad
\epsfig{file=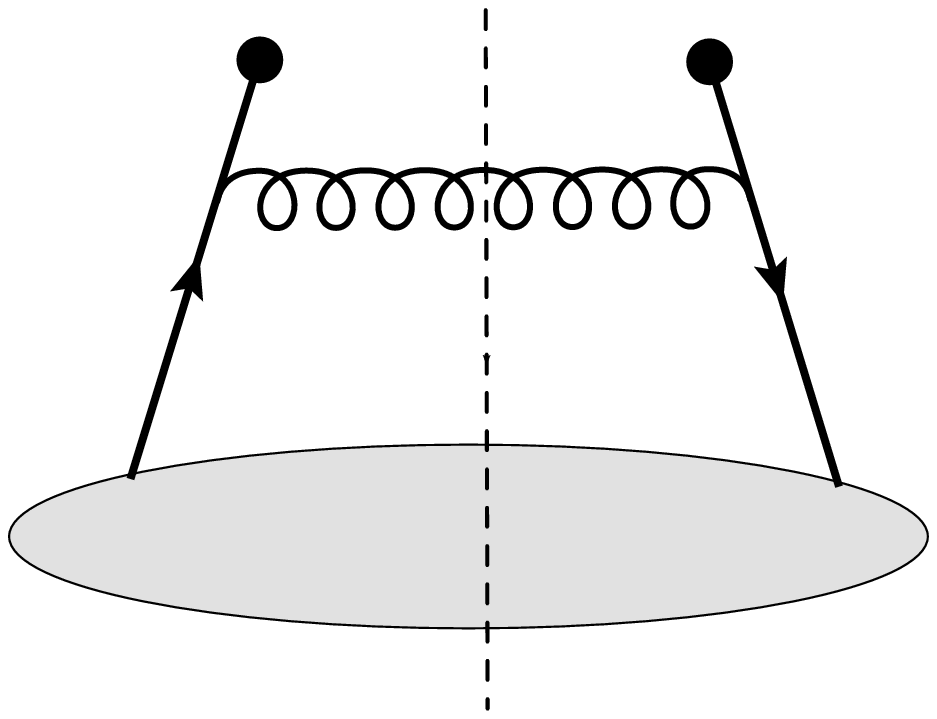,scale=0.35}
\epsfig{file=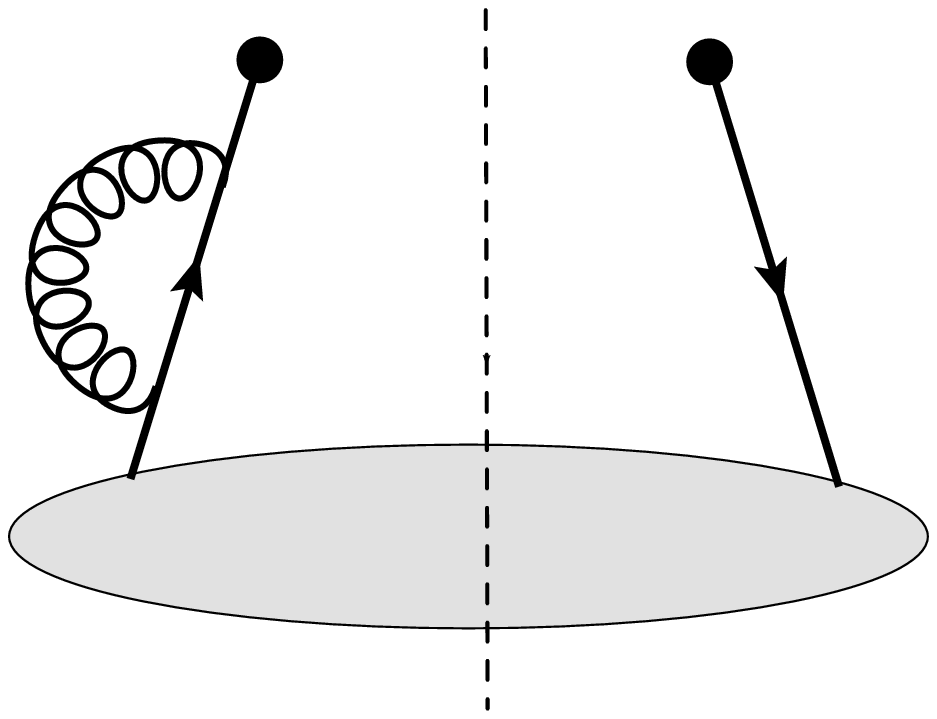,scale=0.35}
\epsfig{file=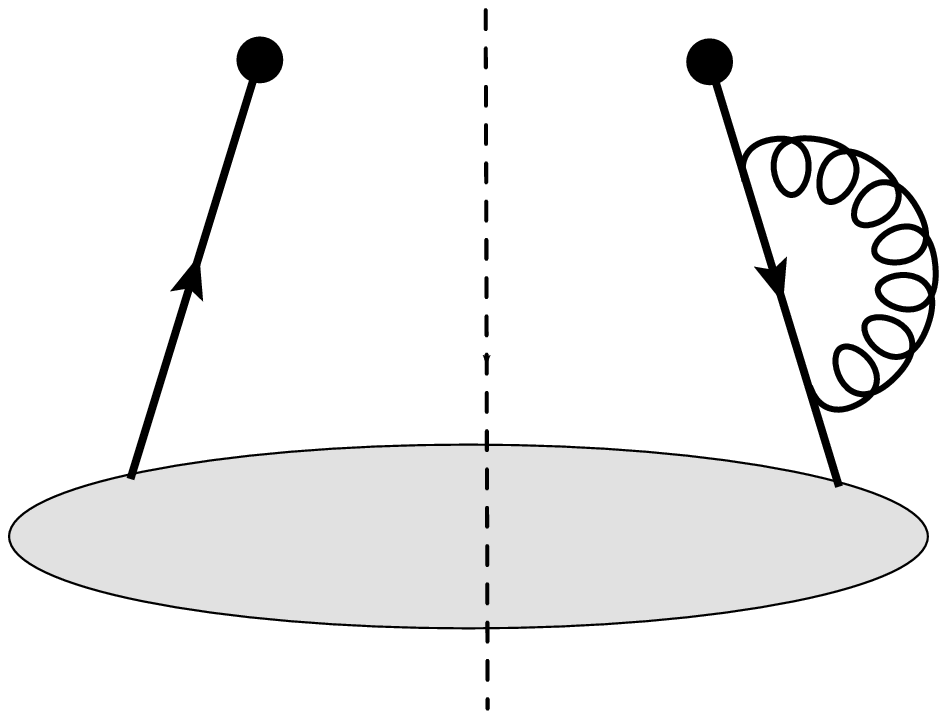,scale=0.35}
\end{minipage}
\begin{minipage}[t]{16.5 cm}
\caption{Left panel: diagrammatic representation of a quark parton 
distribution. The two ``end'' points represent the positions of the quark
fields in Eq.\ (\ref{eq:quarkpdf}). Right panel: diagrams with lowest order 
radiative corrections for the quark parton distribution in light cone gauge.
Note that there are generally more diagrams in different gauges due to the
presence of a gauge link.
\label{fig:dglap}}
\end{minipage}
\end{center}
\end{figure}

Suppose hadron $H$ is moving with large momentum $P$ along the positive
$z$ axis such that $P^+ \to \infty$. We introduce the light cone
components of a four-vector $p^\mu$ as $p^\pm = (p^0 \pm p^3)/\sqrt{2}$.
The parton distribution for quarks and gluons in a light cone gauge ($A^+ =0$) 
are defined as
\begin{align}
  f_{q/H}(\xi,\mu) &= \int \frac{dy^-}{4\pi} e^{- i \xi P^+ y^-}
  \left\langle H(P) \right| \bar q(y^-) \gamma^+ q(0)  \left|H(P) \right\rangle
  \label{eq:quarkpdf}
  \\
  f_{g/H}(\xi,\mu) &= \frac{1}{\xi P^+} \int \frac{dy^-}{2\pi} e^{- i \xi P^+ y^-}
  \left\langle H(P) \right| F_a^{+\nu}(y^-) F_{a\,\nu}^{\>\>\>\>+}(0) \left|H(P) 
  \right\rangle
\end{align}
Here $\left|H(P) \right\rangle$ is the suitably normalized single hadron state,
(note that an averaging over hadron spins is usually silently implied in the 
notation $ \left\langle H(P) \right| \ldots   \left|H(P) \right\rangle$),
and $q$ and $F$ are the operators for quark and gluon fields.
$\xi$ with $0<\xi<1$ is the momentum fraction of the parton in the parent
hadron. In light cone gauge it is straight forward to interpret these 
matrix elements as quark and gluon counting operators. 
Note again that these matrix elements can not be evaluated perturbatively 
for hadrons or nuclei. The left panel of Fig.\ \ref{fig:dglap} shows
a diagrammatic representation of a parton distribution function in
light cone gauge.

Radiative corrections introduce a weak, logarithmic scale dependence.
The first radiative corrections are shown in the right panel of 
Fig.\  \ref{fig:dglap} for a light cone gauge. Resummation of these
diagrams lead to the DGLAP evolution equations which determine the running of 
parton distributions $f_{a/H}(\xi,\mu)$ with the scale $\mu$
\cite{Gribov:1972ri,Altarelli:1977zs,Dokshitzer:1977sg}. 
For a parton $a$ they are
\begin{equation}
  \label{eq:dglap}
  \frac{\partial f_a(\xi,\mu)}{\partial \ln \mu} = \frac{\alpha_s}{\pi}
  \int_x^1 \frac{dy}{y} \sum_b P_{b\to a}(y) f_b\left(\frac{\xi}{y},\mu\right)
\end{equation}
where the set of $P_{b\to a}(y)$ are called the splitting functions. This notion
is easily explained with a look at the right panel of Fig.\ \ref{fig:dglap} 
whose diagrams represent the $q\to q$ splitting function which is
\begin{equation}
  \label{eq:qqsplitting}
  P_{q\to q} (y) = C_F \left( \frac{1+y^2}{1-y} \right)_+ + C_F \frac{3}{2}\delta(y-1) 
\end{equation}
where $C_F = 4/3$ is the color factor. Virtual corrections make a contribution
at $y=1$ which introduces the $\delta$-function term and regularizes
the singularity in the first term via the $+$ description:
$(f(y)/[1-y])_+ \to [f(y)-f(1)]/[1-y]$ for the integral over any function $f(y)/[1-y]$.
More details and the full set of splitting functions are discussed in \cite{Brock:1993sz}.

While the $\mu$-dependence is hence perturbatively calculable, the 
$\xi$-dependence can only be extracted from fits to data. This relies heavily
on data from the ``clean'' deep-inelastic scattering (DIS) process. 
Very accurate parameterizations including estimates 
of uncertainties are available for protons and, via isospin symmetry, for
neutrons in a wide range of about $10^{-5} < x < 0.5$. The most
used parameterizations are from the CTEQ \cite{Lai:1999wy,Pumplin:2002vw}
and MRST collaborations \cite{Martin:2002dr,Martin:2006qz}. The Durham
data base has comprehensive information about PDFs \cite{DurhamDB}.
Parton distributions of nuclei are discussed further below.

Fragmentation functions $D_{c/H}(z,\mu)$ give the reverse probability that 
hadron $H$ hadronizes from parton $c$ in the vacuum with a certain momentum
fraction $z$ of the parent parton \cite{Collins:1981uw}. Unlike the case of 
parton distributions
a complete sum over states can not be removed and hence fragmentation
functions can not be written as a single forward matrix element. Instead
we have
\begin{equation}
  D_{q/H}(z,\mu) = z \int\frac{dy^-}{4\pi} e^{-i P^+ y^-/z}
  \left\langle H(P) \right| \bar q(y^-) \left| 0 \right\rangle \gamma^+
  \left\langle 0 \right| q(0) \left|H(P) \right\rangle
\end{equation}
for a quark $q$ with large light cone momentum $P^+$. 
As for parton distributions, vacuum fragmentation functions have been
parameterized, mostly from hadron production data in $e^++e^-$ collisions,
but uncertainties in the fits are appreciable, even for quite common hadrons
like protons and kaons. In addition, some sets are not isospin-separated,
i.e. they only parameterize processes like $u+\bar u \to \pi^+ + \pi^-$.
This uncertainty in the theoretical baseline makes the search for nuclear 
effects more challenging.
The most widely used parameterizations in heavy ion physics are the sets 
by Kniehl et al.\ (KKP) \cite{Kniehl:2000fe} and Albino et al.\ (AKK)
\cite{Albino:2005me}, Hirai et al.\ (HKNS) \cite{Hirai:2007cx} and deFlorian
et al.\ (DSS) \cite{deFlorian:2007aj,deFlorian:2007hc}. The latter ones
include iso-spin separation, partially also including data from $p+p$ collisions.

\subsubsection{Factorized Cross Sections}

In this subsection we will summarize some often used factorization formulas
for hadron or jet production at leading order. They can be used
together with the list of leading order parton cross sections in \cite{Owens:1986mp} 
and the parton distributions and fragmentation functions referenced above
to make estimates for rates of hadron and jet production.
Our starting point is the differential production cross section of two partons
$c$ and $d$ from two hadrons $A$ and $B$
\begin{equation}
  \label{eq:fact2}
  d\sigma_{AB\to cd} = \sum_{a,b} f_{a/A}(\xi_a) f_{b/B}(\xi_b) 
  d\sigma_{ab\to cd}
  \, .
\end{equation}
We need to introduce some notation for the kinematics. Let the momenta
of the parent hadrons be $P_A$ and $P_B$ in positive and negative direction, 
resp., along the $z$-axis in the center of mass frame of the hadrons. 
We assume that
$P \equiv P_A^+ = P_B^-$ is larger than any relevant masses.
Note that the kinematics in (\ref{eq:fact2}) is fixed at leading order with 
$\xi_a = p_a^+/P$ and $\xi_b = p_b^-/P$, resp. where $p_a$, $p_b$, $p_c$ and
$p_d$ are the momenta of the four partons.

One can easily deduce the cross section for a di-jet event with final
rapidities $y_c$ and $y_d$ and transverse momentum $p_T$ (the transverse
momenta of $c$ and $d$ are equal and opposite),
\begin{equation}
  \label{eq:dijet}
  \frac{d\sigma_{AB\to cd}}{2\pi p_T dp_T dy_cdy_d } = \sum_{a,b} \xi_a
  f_{a/A}(\xi_a)  \xi_b f_{b/B}(\xi_b) \frac{1}{\pi} \frac{d\sigma_{ab\to cd}}{dt}
\end{equation}
where the momentum fractions are fixed to be
\begin{equation}
  \label{eq:xiaxib}
  \xi_a = \frac{2p_T}{s} \left( e^{y_c} + e^{y_d} \right) \, , \qquad
  \xi_b = \frac{2p_T}{s} \left( e^{-y_c} + e^{-y_d} \right) \, .
\end{equation}
Note that all Mandelstam variables $s$, $t$, $u$ are defined on the level of 
partons $a, b, c, d$. In particular
\begin{equation}
  t = - \xi_a p_T\sqrt{S}e^{-y_c}
\end{equation}
where $S = (P_A + P_B)^2= 2P^2$ is the total center of mass energy squared 
of the two hadrons.

For single jet events we have to integrate one of the final parton momenta.
This introduces effectively one non-trivial integral which is usually
rewritten as an integral over one of the initial parton momentum fractions.
For a jet with transverse momentum $p_T$ and rapidity $y$ we find
\begin{equation}
  \label{eq:jet}
  \frac{d\sigma_{AB\to c+X}}{2\pi p_T dp_T dy} = \sum_{a,b,d}
  \int_{\xi_{\mathrm{min}}}^1 d\xi_a f_{a/A}(\xi_a)  f_{b/B}(\xi_b)
  \frac{2}{\pi} \frac{\xi_a\xi_b}{2\xi_a-\xi_T e^y}
  \frac{d\sigma_{ab\to cd}}{dt}
\end{equation}
where $\xi_b$ is fixed to
\begin{equation}
  \label{eq:xib}
  \xi_b =  \frac{\xi_a\xi_T e^{-y}}{2\xi_a-\xi_T e^y} \, ,
\end{equation}
and the integration boundary (to keep $\xi_b <1$ for fixed $y$, $p_T$) is 
\begin{equation}
  \label{eq:xiamin}
  \xi_{\mathrm{min}} = \frac{\xi_T e^{y}}{2-\xi_T e^{-y}}.
\end{equation}
We have introduced the useful scaling variable $\xi_T = 2p_T/\sqrt{S}$.

In order to arrive at cross sections for hadrons we have to multiply the
differential cross section for partons with the corresponding fragmentation
functions. The resulting phase space integrals are often shifted to be
integrals over the initial parton momentum fractions $\xi_a$ and $\xi_b$.
For applications in heavy ion physics we rather adopt 
a different way that keeps factorizability between the fragmentation functions
on one hand and the parton cross section plus parton distributions on the
other hand explicit. For two hadrons $C$ and $D$ with momenta $P_{TC}$ 
and $P_{TC}$ we can write
\begin{multline}
  \label{eq:indepfrag}
  \frac{dN_{CD}}{2\pi P_{TC} dP_{TC} P_{TD} dP_{TD}dy_Cdy_D } = 
  \sum_{c,d} \int_{z_{c,\mathrm{min}}}^1 \frac{dz_c}{z_c^2}
  \int_{z_{d,\mathrm{min}}}^1 \frac{dz_d}{z_d^2} \\  \times
  \frac{dN_{cd}}{2\pi p_{Tc} dp_{Tc} p_{Td} dp_{Td}dy_Cdy_D } 
  D_{c/C}(z_c) D_{d/D}(z_d)  \, .
\end{multline}
This is a very general formula that connects a distribution function of
partons $c$, $d$ with momenta $p_c=P_C/z_c$, $p_d=P_D/z_d$, resp.\ in the 
final state to hadrons $C$, $D$. Of course the applicability of 
this formula still requires the collinear fragmentation picture to hold 
in this much generalized setting. Nevertheless,
derivatives from Eq.\ (\ref{eq:indepfrag}) are often used to model final state
interactions for hadron production in nuclear collisions.
There might be kinematic constraints that lead to lower bounds on the 
integrals over $z_c$ and $z_d$ whose exact specification will 
depend on the distribution $N_{cd}$ of partons.

For completeness and further clarification let us discuss the more
familiar special case of hadron production in a regime where final state 
interactions can be neglected, e.g.\ in $p+p$ collisions. The formula 
above holds also for cross sections, $N_{CD} \to \sigma_{AB\to CD}$, and the 
partonic cross section is given by (\ref{eq:dijet}) times an obvious
phase space factor $1/p_{Td}\, \delta(p_{Tc} - p_{Td})$. This factor can be used 
to cancel the integral over $z_d$ to lead to
\begin{multline}
  \label{eq:dihadron}
  \frac{d\sigma_{AB\to CD}}{2\pi P_{TC} dP_{TC} P_{TD} dP_{TD} dy_Cdy_D } \\ = 
  \frac{1}{P_{TC}P_{TD}}
  \sum_{a,b,c,d} \int_{z_{\mathrm{min}}}^{z_{\mathrm{max}}} \frac{dz_c}{z_c}
  \xi_a f_{a/A}(\xi_a)  \xi_b f_{b/B}(\xi_b) \frac{1}{\pi} D_{c/C}(z_c)
  D_{d/D}(z_d) \frac{d\sigma_{ab\to cd}}{dt}
\end{multline}
where
\begin{equation}
  z_d = z_c \frac{P_{TD}}{P_{TC}} \, , \qquad
  z_{\mathrm{max}} = \mathrm{min}\left\{ 1, \frac{P_{TC}}{P_{TD}} \right\}
  \, , \qquad
  z_{\mathrm{min}} = \frac{2P_{TC}}{\sqrt{S}}\mathrm{max}\left\{ \cosh y_C,
  \cosh y_D \right\} \, .
\end{equation}
Note that $\xi_a$ and $\xi_b$ are given by (\ref{eq:xiaxib}) with $p_{Tc} = 
P_{TC}/z_c$ and $p_{Td} = P_{TD}/z_d$.

For single hadron production we can provide a similar general formula
for fragmentation from a distribution of partons $N_c$,
\begin{equation}
  \label{eq:indepfrag2}
  \frac{dN_{C}}{2\pi P_{T} dP_{T} dy} = 
  \sum_{c} \int_{z_{\mathrm{min}}}^1 \frac{dz}{z^2} D_{c/C}(z)
  \frac{dN_{c}}{2\pi p_{T} dp_{T} dy} \, .
\end{equation}
It can be applied if collinear fragmentation is the correct description
of hadronization of an ensemble of partons and obviously $p = P/z$.
The special case of single hadron production in collisions with negligible
final state interactions gives the formula
\begin{equation}
  \label{eq:singlehadron}
  \frac{d\sigma_{AB\to C+X}}{2\pi P_T dP_T dy} = \sum_{a,b,c,d}
  \int_{z_\mathrm{min}}^1 \frac{dz}{z^2} D_{c/C}(z)
  \int_{\xi_{\mathrm{min}}}^1 d\xi_a f_{a/A}(\xi_a)  f_{b/B}(\xi_b)
  \frac{2}{\pi} \frac{\xi_a\xi_b}{2\xi_a-\xi_T e^y}
  \frac{d\sigma_{ab\to cd}}{dt}
\end{equation}
where
\begin{equation}
  z_{\mathrm{min}} = \frac{2P_{T}}{\sqrt{S}} \cosh y \, ,
\end{equation} 
and the other kinematic variables can be inferred from (\ref{eq:xib}) and
(\ref{eq:xiamin}).

For an alternative way of handling the phase space integrals in terms of
hadron production see \cite{Owens:1986mp}.
Let us point out once more that the convenient identification 
of single partons and jets is only valid in the context of leading order
calculations.

\subsubsection{Photons}

In principle, photons with high transverse momentum $P_T$ can be treated
in a fashion very similar to hadrons or jets. We usually do not consider
photons from decays of hadrons (predominantly $\pi^0$) long after the
collision. After subtracting those decay photons we are left with the 
``direct'' photons produced in the collision. Photon yields produced directly
in the hard process can be calculated via Eq.\ (\ref{eq:jet}) together
with the corresponding parton level processes. At leading order, the
cross sections of the
annihilation and Compton diagrams, $q+\bar q \to \gamma + g$ and $g+q \to
\gamma + q$, resp.\ can be found in the work by Owens \cite{Owens:1986mp}.

Another way to produce direct photons is as bremsstrahlung in
hard process like $g+g \to q+\bar q$. One of the outgoing quarks can 
radiate a collinear photon while fragmenting. This process can be described
by photon fragmentation functions \cite{Owens:1986mp,Bourhis:1997yu}. 
Eq.\ (\ref{eq:indepfrag2}) together with the usual set of hard parton 
processes and photon fragmentation functions are used to compute this 
contribution. At next-to-leading order, bremsstrahlung and hard photon 
radiation in the final state have to be calculated in a consistent scheme 
to separate large angle and collinear photon radiation
\cite{Aurenche:1983ws,Berger:1983yi}. NLO direct photon calculations have
had some difficulties in the past to describe all aspects of photon
production in hadronic collisions \cite{Aurenche:1998gv}. The theoretical
understanding has been improved in recent years by the use of various 
resummation techniques \cite{Laenen:2000de,Sterman:2004yk}. The PHOX codes 
can also be used for NLO calculations of direct photon production.

Photon-hadron and photon-jet pair production is a particularly hot topic
in heavy ion collisions as we will discuss in more detail later.
Their yields with both initial hard and bremsstrahlung photons can also be
calculated in a straight forward way from the factorization formulas in
the last subsection. Fragmented photons can in principle be distinguished
from prompt hard photons since the latter are not accompanied by hadrons
close by in phase space while the former are usually part of a jet cone.
Experimentally, isolation cuts for photons can help to suppress the
bremsstrahlung contribution and give access to more detailed information.

In nuclear collisions there are additional sources of direct photons.
We will discuss jet conversion into photons in the subsection about
final state interactions. There is also thermal radiation from the
hot hadronic matter, and, if energy densities are large enough, from 
the partonic QGP phase. In fact the latter is one of the key observables
that we would like to study at RHIC, since the thermal photon spectrum
can work as a direct (though time- and space-averaged) thermometer of the 
quark gluon plasma. To compute the photon spectrum the time-evolution of
the QGP fireball has to be folded with rates as a function of the local 
temperature and chemical potential. The time evolution is naturally done
through a hydrodynamic model as discussed in Sec.\ \ref{sec:hydro} 
of this review.
The rates can be calculated perturbatively if the temperature $T$ is
large enough to render the strong coupling constant small, $g \ll 1$.
Although there is some doubt whether the maximum temperatures at RHIC 
($T < 450$ MeV) are sufficient to warrant a perturbative description the 
perturbative computation of thermal photon rates has been a sustained 
effort over many years \cite{Kapusta:1991qp,Baier:1991em,Aurenche:1998nw}. 
The complete leading order results have been given by Arnold, Moore and 
Yaffe in \cite{Arnold:2001ba,Arnold:2001ms}.
Additional photon radiation could be emitted in the pre-equilibrium phase. In 
particular, the fact that quarks and gluons are not in chemical equilibrium
early on could affect photon rates and would not be mimicked well by hydro
codes initialized at very early times. Estimates for pre-equilibrium photon
yields in a transport approach can be found in \cite{Bass:2002pm}.

\subsection{\it Nuclear Collisions: Initial State Effects}

Perturbative techniques and factorization have first been developed for
scattering involving individual hadrons. The basic principles should be
valid if one or both of the scattering partners are bound in a nucleus.
In fact, the small binding energies and slow relative motion of nucleons 
should not have large impact on scattering at large momentum transfer. 
However, the larger volumes filled with nuclear matter surrounding
the point-like hard interaction should potentially lead to rescattering 
both in the initial and final state. In this subsection we will discuss
the most relevant nuclear effects in the initial state.

\subsubsection{Shadowing and Nuclear Parton Distributions}

Individual nucleons are clearly distinguishable building blocks of nuclei.
Hence we expect parton distributions in a nucleus with $Z$ protons and $A-Z$
neutrons to be well represented by a linear superposition of parton 
distributions of the individual protons $p$ and neutrons $n$,
\begin{equation}
  f_{a/A}(\xi,\mu) = \left( \frac{Z}{A} f_{a/p} (\xi,\mu) + \frac{A-Z}{A} f_{a/n}
  (\xi,\mu) \right) R_A(\xi,\mu)
\end{equation}
The remaining non-trivial nuclear modification $R_A$ was expected to be
small until it was found by the EMC collaboration that deep-inelastic
scattering off nuclei leads to sizable differences between free and
bound nucleons \cite{Aubert:1983xm}. Note that nuclear parton distributions 
are usually normalized to one nucleon.

\begin{figure}[tb]
\epsfysize=9.0cm
\begin{center}
\begin{minipage}[t]{10 cm}
\epsfig{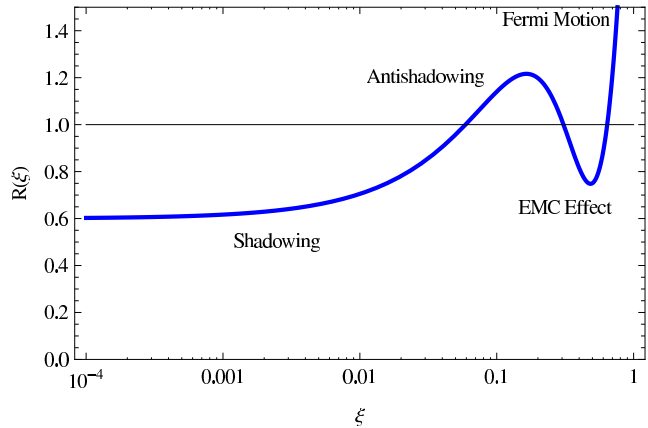}
\end{minipage}
\begin{minipage}[t]{16.5 cm}
\caption{Schematic sketch of the expected behavior of the ratio $R_A$ of 
nuclear parton distributions compared with free nucleon parton distributions. 
The distinct regions, the Fermi motion region, the EMC region and the shadowing
and antishadowing regions are visible. The ratio $R_A$ is directly 
reflected in the ratio of single particle spectra in $p+$A collisions to
$p+p$ collisions as a function of $P_T$.
 \label{fig:npdf}}
\end{minipage}
\end{center}
\end{figure}

Despite considerably larger uncertainties compared to free nucleon parton
distributions we can now identify four distinct regions of behavior in the 
momentum fraction $\xi$ which are indicted in Fig.\ \ref{fig:npdf},
see \cite{Akulinichev:1985xq,Frankfurt:1988nt,Armesto:2006ph} and references
therein.
\begin{itemize}
\item Fermi motion enhancement, $\xi > 0.8$: when the parton carries most 
  of the momentum of the nucleon the Fermi motion of the nucleon itself in the 
  nucleus becomes important.
\item EMC effect (proper), $0.3 < \xi < 0.8$: the kinematic region of the 
  original discovery, named after the experiment, exhibits a suppression 
  $R_A(\xi) < 1$ which is usually explained with nuclear binding effects.
\item Antishadowing, $0.1 < \xi < 0.3$: a region of enhancement of nuclear
  parton distributions required by momentum sum rules.
\item Shadowing, $\xi < 0.1$: a region of possibly large suppression of
  parton distributions. It can be understood through multiple scattering
  in the nuclear rest frame, or parton fusion in an infinite momentum frame.
  In the deep shadowing (small-$\xi$) region this might lead to a color
  glass condensate picture. We refer to \cite{Armesto:2006ph} for a modern 
  review of models for the shadowing effect.  
\end{itemize}

\begin{figure}[tb]
\epsfysize=9.0cm
\begin{center}
\begin{minipage}[t]{16 cm}
\epsfig{file=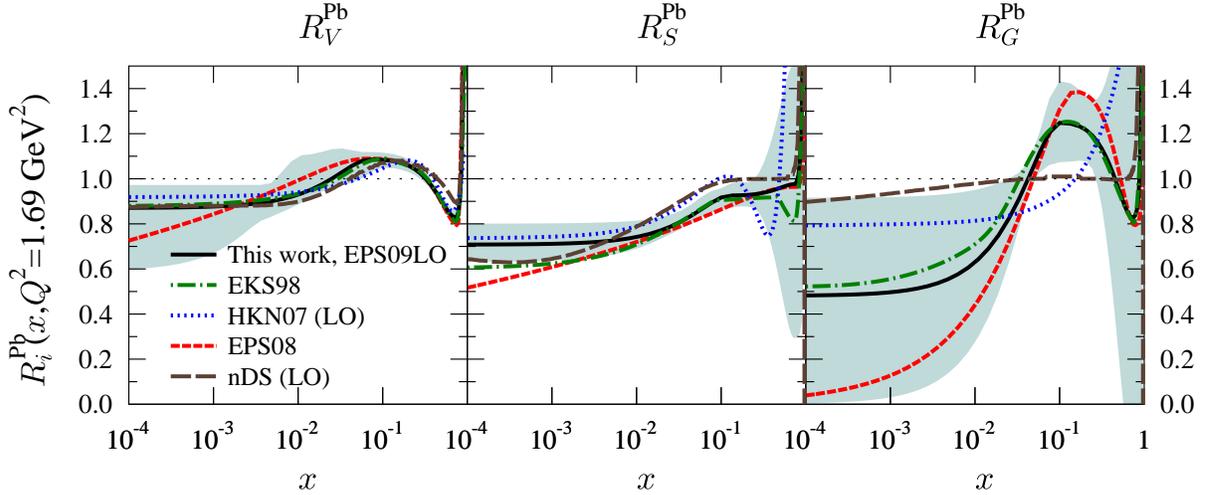,scale=0.75}
\end{minipage}
\begin{minipage}[t]{16.5 cm}
\caption{\label{fig:eps09}
Comparison of different leading order DGLAP parameterizations of the
nuclear modification $R_{\mathrm{Pb}}(x,\mu)$ for lead nuclei at 
$\mu = 1.3$ GeV. The parameterization correspond to EKS98
\cite{Eskola:1998df}, EPS08 \cite{Eskola:2008ca}, EPS09 \cite{Eskola:2009uj}, 
HKN07 \cite{Hirai:2007sx}, and nDS \cite{deFlorian:2003qf}. The large
theoretical uncertainties at low momentum fraction $x$, in particular for the
gluons, is clearly demonstrated. Figure reprinted from \cite{Eskola:2009uj}
with permission from JHEP.}
\end{minipage}
\end{center}
\end{figure}

A parton with 10 GeV/$c$ transverse momentum produced at midrapidities
($y=0$) in collisions at RHIC energies ($\sqrt{s_{NN}} = 200$ GeV) comes
from initial parton momentum fractions around $\xi_T =2p_T/\sqrt{S_{NN}} =0.1$.
Hence it is easy to see that for perturbative calculations at RHIC mostly the
shadowing and anti-shadowing regions are of importance. For not too
small momentum fractions $\xi$ nuclear parton distributions are still
in the universal DGLAP regime. They can be measured in deep inelastic
scattering on nuclei, while the perturbative evolution in the scale
$\mu$ can be used as a consistency check. The parameterizations (which
are often parameterizations of the modification $R_A$ for specific sets
of free nucleon parton distributions) can then be used  
for hadron-nucleus and nucleus-nucleus collisions.
DGLAP parameterizations are available from several groups 
\cite{Eskola:1998df,Eskola:2007my,Eskola:2008ca,Eskola:2009uj,Hirai:2001np,
Hirai:2004wq,Hirai:2007sx,deFlorian:2003qf}.
Some, like the EPS08 and EPS09 parameterizations 
\cite{Eskola:2008ca,Eskola:2009uj}, already include 
some RHIC data in the DGLAP fit. This has been done to improve the lack of
suitable deep-inelastic scattering data on nuclei. Previous deep-inelastic
scattering experiments off nuclei cover only large $\xi$ and have very 
little power to constrain the nuclear gluon distribution. This situation
leaves us with huge theoretical uncertainties on the nuclear gluon 
distribution below $\xi \approx 0.05$. Fig.\ \ref{fig:eps09} shows
several fits for modification factors for valence quarks, sea quarks and
gluons respectively. The spread of possible values for the
nuclear gluon distribution is truly remarkable. This uncertainty has
profound consequences for pQCD predictions at LHC energies where the average
parton $\xi$ will be much smaller than at RHIC.

\subsubsection{Higher Twist Corrections}

Nuclear corrections to the parton distributions deal with the effects
of nuclear binding on the long-distance behavior of a process. One
can also ask the question whether the hard process between partons is 
affected as well. Indeed it turns out that certain high-twist corrections become
important in collisions involving nuclei. Corrections beyond leading twist 
were first characterized in terms of new operators beyond parton distributions
that appear in the operator production expansion. Twist $t$ 
was defined as the dimension minus the spin of a local operator. The 
definition can be generalized 
to apply also to situations where an operator product expansion is not 
available. The leading twist operators in parton 
distributions, $\bar q \gamma^+ q$ and $F^{+\mu}F_{\mu}^{\,\,+}$, are
classified as twist $t=2$. We will use higher twist as a simple 
power counting scheme in terms of the large scale $Q$, such that a twist-$t$
contribution is suppressed by $1/Q^{t-2}$ compared to leading twist.
Jaffe's review \cite{Jaffe:1996zw} offers a discussion of both the
rigorous and the power counting definition of twist.

Higher twist effects are obviously important if the large scales $Q$
becomes to small (close to non-perturbative scales), or if
other enhancement effects weaken the power suppression. It was first pointed
out by Luo, Qiu and Sterman \cite{Luo:1992fz,Luo:1993ui,Luo:1994np} that 
in large nuclei with mass number $A$ some operators do not follow a 
classification in terms of an expansion in $\Lambda/Q \ll 1$, but rather
in the parameter
\begin{equation}
  \Lambda A^{1/3}/Q \sim \Lambda^2 L /Q >> \Lambda/Q \, .
\end{equation}
Here $\Lambda$ is a soft scale (of the order or $\Lambda_{\mathrm{QCD}}$
or the constituent quark mass) and $L \sim A^{1/3}/ \Lambda$ is the thickness 
of the nucleus. $L$ comes into play because in thick nuclear matter multiple 
hard scattering is possible and its probability increases with thickness. 
Multiple scattering should not modify the total cross section very much, but 
we expect some observables, e.g.\ transverse momentum spectra, to be 
significantly altered by multiple additional ``kicks'' that a scattered
particle experiences. The Cronin effect discussed below is a good example.

\begin{figure}[tb]
\begin{center}
\begin{minipage}[t]{12 cm}
\epsfig{file=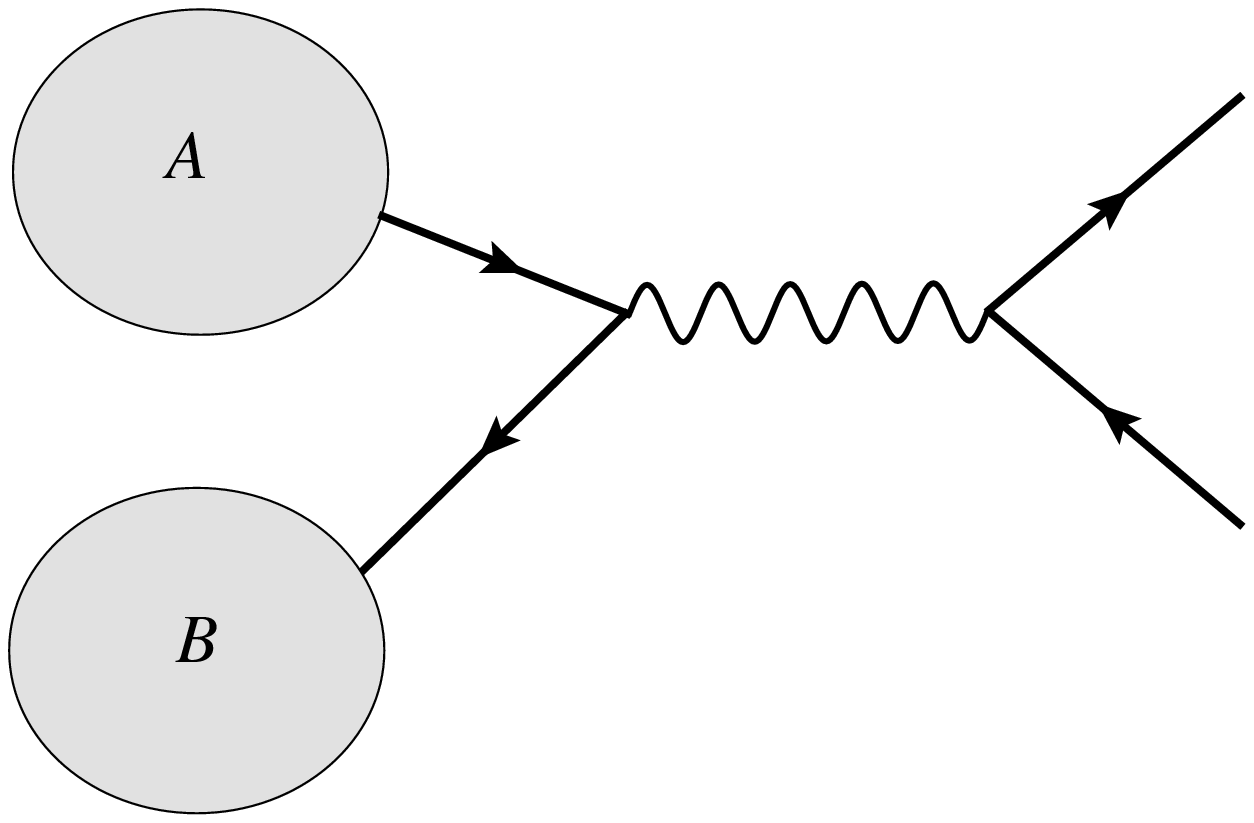,scale=0.40} \qquad
\epsfig{file=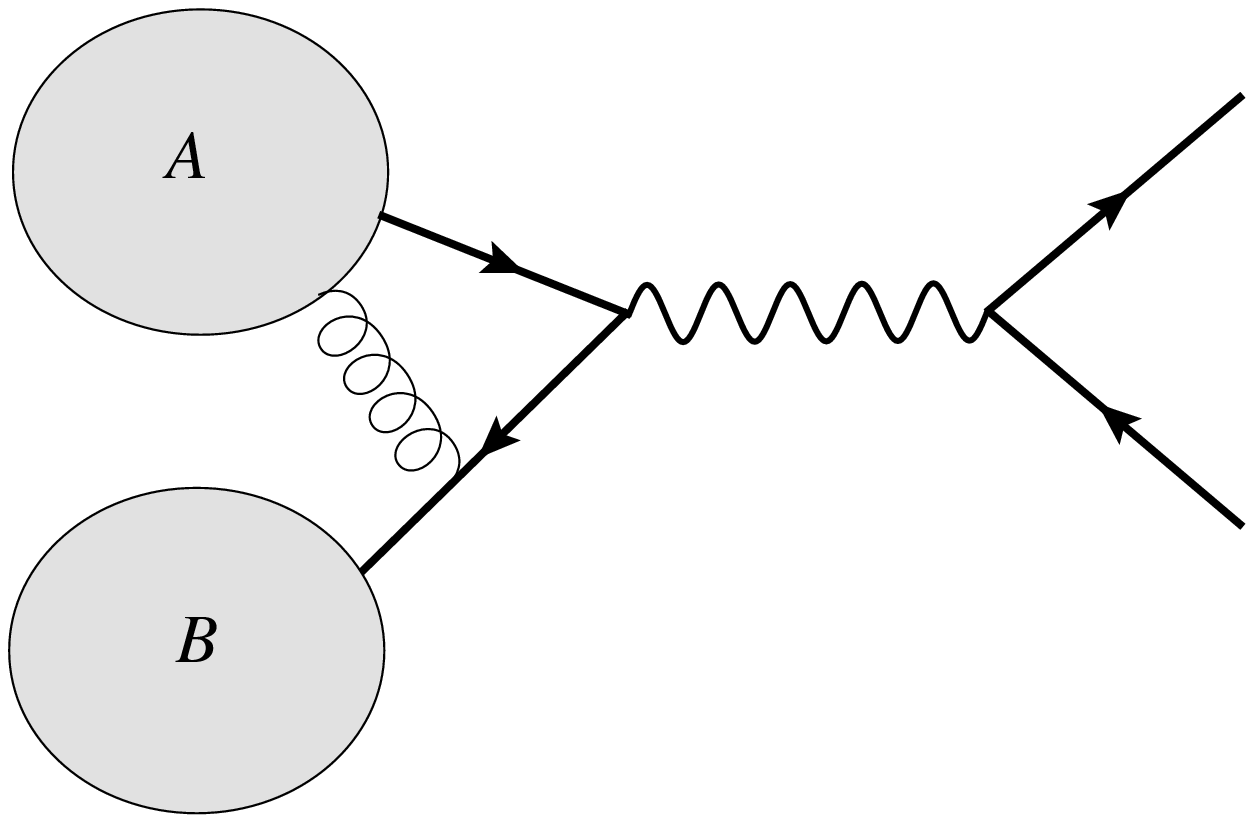,scale=0.40}
\end{minipage}
\begin{minipage}[t]{16.5 cm}
\caption{Left panel: Drell-Yan at leading twist and leading order in 
$\alpha_s$: quark-antiquark annihilation. Right panel: A nuclear enhanced
twist-4 correction in $A+A$ collisions that corresponds to double
scattering of the quark from the proton in the nucleus. First the
quark scatters off a soft gluon and then annihilates with an antiquark.
\label{fig:DY}}
\end{minipage}
\end{center}
\end{figure}

We want to review a simple example, the nuclear Drell-Yan process
$A+A \to l^+ + l^- + X$ \cite{Guo:1997mm,Fries:1999jj,Fries:2000da,
Fries:2002mu}. 
At leading order $\mathcal{O}(\alpha_s^0)$, and leading twist the virtual 
photon is produced through a simple quark-antiquark annihilation, 
$q+\bar q \to  l^+ + l^-$, see left panel in Fig.\ \ref{fig:DY}.
The corresponding cross section for dilepton pairs of mass $Q$ and 
(pair) transverse momentum $q_T$ is
\begin{equation}
  \label{eq:dy}
  \frac{d\sigma_{AB \to l^+ + l^-}}{dQ^2 dq_T^2} = \sigma_{\mathbf{DY}}
  \delta(q_T^2) \sum_q e_q^2 \int_{B_a}^1
  \left[ f_{q/A}(\xi_a) f_{\bar q/B}(\xi_b) + f_{\bar q/A}(\xi_a)
  f_{q/B}(\xi_b) \right] d\xi_b
\end{equation} 
where 
\begin{equation}
  \sigma_{\mathbf{DY}} =  \frac{4\pi}{3N_c}\frac{\alpha_{em}^2}{SQ^2}
\end{equation}
essentially is the cross section between partons $q$ and $\bar q$,
$N_c=3$, $e_q$ is the charge of quark $q$ in units of $e$, and 
$\xi_a = Q^2/(\xi_b S)$ is fixed with $B_a = Q^2/S$.
(\ref{eq:dy}) is a straight forward but nevertheless questionable result.
The $q_\perp$-spectrum is actually not well defined in the collinear
limit. Indeed the presence of two scales $q_T << Q$ presents  
additional problems. The safe way to discuss this result is by using moments 
in $q_\perp$-space. The lowest moment
is the cross section differential with respect to the mass squared
\begin{equation}
  \frac{d\sigma_{AB \to l^+ l^-}}{dQ^2} = \int_0^\infty
  \frac{d\sigma_{AB \to l^+ l^-}}{dQ^2 dq_\perp^2} dq_T^2 \, ,
\end{equation}
while the next moment can be used to define the average transverse momentum
squared
\begin{equation}
  \left\langle q_T^2 \right\rangle = \frac{
  \int_0^\infty \frac{d\sigma_{AB \to l^+ l^-}}{dQ^2 dq_T^2} q_T^2 dq_T^2
  }{\frac{d\sigma_{AB \to l^+ l^-}}{dQ^2}}.
\end{equation}
At leading order and leading twist $\langle q_\perp^2 \rangle = 0$ which
is also true for all higher moments.

The first nuclear enhanced higher twist correction corresponds to 
double scattering of one of the quarks off an additional gluon from
the other nucleus, e.g.\ $q+\bar q g \to  l^+ + l^-$, see right panel in Fig.\ 
\ref{fig:DY}. The cross section is
\begin{multline}
    \frac{d\sigma_{AB \to l^+ + l^-}^{(2)}}{dQ^2 dq_T^2} = -\delta'(q_T^2) 
    \sigma_{\mathbf{DY}} \frac{4\pi\alpha_s}{N_c}
  \sum_q e_q^2 \\ \times  \int_{B_a}^1 
  \left[ f_{q/A}(\xi_a) T_{\bar qg/B}(\xi_b) + f_{\bar q/A}(\xi_a)
  T_{qg/B}(\xi_b) + T_{qg/A}(\xi_a) f_{\bar q/B}(\xi_b) + T_{\bar qg/A}(\xi_a)
  f_{q/B}(\xi_b)   \right] d\xi_b
\end{multline}
Note that there is a derivative on the $\delta$-function.
The new matrix elements $T_{ab/A}$ measure a ``soft-hard'' two-parton density 
in the nucleus, e.g.\
\begin{multline}
  \label{eq:tsh}
  T_{qg/A}(\xi) = \int dy_4^- \frac{dy_3^-}{2\pi} \frac{dy_1^-}{2\pi}
  e^{i\xi P^+ y_1^-} \Theta(y_1^- - y_3^-) \Theta(-y_4^-)
 \left\langle A(P) \right| \bar q(0) \gamma^+ q(y_1^-) \\ \times
 F_a^{+\nu}(y_3^-) F_{a\,\nu}^{\>\>\>\>+}(y_4^-) \left|A(P) \right\rangle
\end{multline}
where $P^+$ is the large momentum component of a nucleon. Soft-hard in this 
particular case means that the quark or antiquark has a finite momentum
fraction $\xi$ while the gluon is very soft. Formally the soft-hard
matrix elements are limits of more general 2-parton distributions
$f^{(2)}_{qg/A} (\xi_q,\xi_g)$ with
\begin{equation}
  \label{eq:sh}
  T_{qg/A}(\xi_q) = \lim_{\xi_g\to 0} \xi_g f^{(2)}_{qg/A} (\xi_q,\xi_g) \, .
\end{equation}

Luo, Qiu and Sterman have classified the relevant twist-4 matrix elements that 
show nuclear enhancement. They all have a probabilistic interpretation
as 2-parton densities and they lead to soft-hard and hard-hard double
scattering on the parton level. The actual nuclear enhancement factor comes 
from unrestricted spatial integrations along the light cone. The coordinates 
associated with the two partons --- corresponding to $y_1^-/2$ and 
$(y_3^- + y_4^-)/2$ in Eq.\ (\ref{eq:tsh}) --- can be as far apart as the
nuclear matter extends along the light cone. Parametrically we have
\begin{equation}
  \label{eq:shmodel}
  T_{qg/A}(\xi) \sim A^{1/3} \Lambda^2
\end{equation}
where $\Lambda$ is a soft scale and $A$ denotes the mass number of nucleus 
$A$.

To arrive at infrared-safe results we again take moments. We note that 
$t=4$ double scattering does not make a contribution to the
integrated mass spectrum, $d\sigma^{(2))}/dQ^2 = 0$, or the total cross
section. However, it leads to non-vanishing transverse momentum, despite the
use of collinear factorization and the absence of radiation,
\begin{equation}
  \langle q_T^2 \rangle = \frac{4\pi\alpha_s}{N_c}
  \frac{\sum_q e_q^2 \int_{B_a}^1 
  \left[ f_{q/A}(\xi_q) T_{\bar qg/B}(\xi_b) + f_{\bar q/A}(\xi_a)
  T_{qg/B}(\xi_b) + T_{qg/A}(\xi_a) f_{\bar q/B}(\xi_b) + T_{\bar qg/A}(\xi_a)
  f_{q/B}(\xi_b) \right] d\xi_b}{\sum_q e_q^2 \int_{B_a}^1
  \left[ f_{q/A}(\xi_a) f_{\bar q/B}(\xi_b) + f_{\bar q/A}(\xi_a)
  f_{q/B}(\xi_b) \right]d\xi_b} \, .
\end{equation}
The $t=4$ matrix elements are universal functions that could in principle
be measured, but useful information is scarce. Most of the time
the soft-hard matrix elements are simply modeled using the shape of the
hard parton distribution
\begin{equation}
  \label{eq:shgluon}
  T_{qg/A}(\xi) = \lambda^2 A^{1/3} f_{q/A}(\xi)
\end{equation}
where $\lambda$ is a parameter with the dimension of energy which
parameterizes the strength of the soft gluon field.
For a symmetric situation with both nuclei being identical this leads
to the simple estimate
\begin{equation}
  \langle q_T^2 \rangle \approx \frac{4\pi\alpha_s}{N_c} 2\lambda^2
\end{equation}

Higher twist corrections for $t>4$, correspond to multiple scattering
beyond double scattering. It is possible to identify the operators with 
maximum nuclear enhancement $\sim A^{(t-2)/6}$ and they can be resummed in
certain situations. This is safe to do for Drell-Yan in $p+$A
collisions where the proton can be treated at leading twist 
\cite{Fries:2002mu,Majumder:2007hx}. The resulting effect is a 
diffusion of $q_T$ in transverse momentum space.
However, generally caution is necessary in nuclear collisions. Although 
the Drell-Yan process is rather simple, with no non-perturbative hadronic 
structure measured in the final state, factorization still breaks down 
beyond twist $t=4$ \cite{Doria:1980ak}.
In other words, while nuclear enhanced higher twist corrections
can be reliably calculated for Drell-Yan in $p+A$, there are true 
non-perturbative contributions that invalidate this expansion in $A+A$ 
collisions at the level of twist-6.

Nuclear enhanced higher twist corrections have been considered for several
observables, including deep-inelastic scattering on nuclei \cite{Guo:2001tz}, 
jets and dijets in electron-nucleus collisions \cite{Luo:1994np,Guo:1998rd}, 
Drell-Yan both at low and high $q_T$ \cite{Guo:1997mm,Fries:2002mu,Majumder:2007hx}, 
direct photon production \cite{Guo:1995zk},
and photon bremsstrahlung for jets \cite{Majumder:2007ne}.
Note that higher twist corrections can appear both as initial and final
state interactions. In fact, in most cases higher twist corrections could
lead to both effects and can not be put in one of those two categories. 
However, those
more general cases have not been considered in full detail, and we will
mostly assume here that higher twist corrections in the initial and final state
are independent of each other.
The most important applications to date for the scope of this article are
the Cronin effect (in the initial state) which we will discuss next, 
and medium-modified fragmentation functions (in the final
state) which will be reviewed in more detail
in the next subsection. We conclude by noting that there is a
patchwork of relevant and useful calculations on the topic of nuclear 
enhanced higher twist, but a lack of comprehensive and systematic studies.

\subsubsection{Cronin Effect}

The Cronin effect was one of the first nuclear modifications discovered
in experimental data \cite{Cronin:1974zm}. It was found that cross sections
of hadrons scale with a power of the atomic number $A$ that
is larger than 1 for intermediate transverse momenta $P_T \approx 1$ GeV/$c$.
This effect was found to not affect the total cross sections very much,
and to die out like a power law at larger $P_T$. This is reminiscent of
higher twist corrections and indeed these results can be interpreted in
the framework of higher twist. Intuitively, the Cronin effect
comes from multiple scatterings of partons on their way to the hard collision. 
These random kicks endow the parton with additional transverse momentum. 
This leads to a depletion of partons with very small (initial) 
intrinsic transverse momentum and an accumulation of partons at intermediate 
transverse momentum. At even larger values of $P_T$ the additional 
momentum kicks do not play a role and the effect decreases in importance.

The Cronin effect in its purest form can be studied in the case of
dilepton or photon production in $p+$A or A+A collisions. Then it is
guaranteed that deviations from the cross sections found in $p+p$ are 
initial state effects. We can simply refer to the discussion from the 
last subsection where we have established that higher twist corrections 
to the Drell-Yan process that correspond to
double scattering lead to an increase in the average 
transverse momentum squared $\langle q_T^2 \rangle$ which is proportional 
to $A^{1/3}$, and that a resummation of multiple scatterings leads to a 
Gaussian distribution of $q_T$ even at leading order in $\alpha_s$.

One can argue that these effects also increase $\langle P_T^2 \rangle$ 
in hadron production, although it is not always clear how to 
distinguish the effects of initial and final state interactions.
It is then quite common to refer to less rigorous but phenomenologically 
successful descriptions of the Cronin effect, see e.g.\ 
\cite{Accardi:2004be} for a review.
These models are usually built on the notion of an intrinsic transverse 
momentum of partons in hadrons or nuclei. The concept of intrinsic 
transverse momentum $k_T$ is not compatible with collinear factorization 
but has a long history as a phenomenological extension of the former. 
True schemes for $k_T$-factorization do exist, but only
for a handful of select processes, and they are technically more complex.
Nevertheless many features of the Cronin effect can be described by 
a model in which the average $k_T^2$ is enhanced in nuclei through
\begin{equation}
   \langle k_T^2 \rangle_{pA,AA} = \langle k_T^2 \rangle_{pp} + \delta k_T^2
\end{equation}
where $\delta k_T^2$ scales with the thickness of nucleus $A$. This is 
also known as $k_T$-smearing.

One possible implementation is through the use of $k_T$-dependent
``parton distributions''
\begin{equation}
  \tilde f_{a/A}(\xi,{k}_T,\mu) = \frac{1}{\pi \langle k_\perp^2\rangle}
  e^{-k_T^2/\langle k_T^2 \rangle} f_{a/A}(\xi,\mu)
\end{equation}
in which $\langle k_T^2 \rangle$ can be fitted to the system size,
or calculated from an underlying microscopic model, like a Glauber
\cite{Accardi:2003jh} or dipole model \cite{Johnson:2007kt}. Reference
\cite{Accardi:2004be} contains a compilation of parameters suitable 
for both RHIC and LHC energies.
As a side remark we note that both shadowing and the Cronin effect are
also natural consequences of gluon saturation and the mechanisms discussed
here should smoothly transition to their color glass counterparts for very
large center of mass energies \cite{JalilianMarian:2003mf,Kharzeev:2003wz}.

\subsubsection{Phenomenological Consequences of Initial State Effects}
\label{sec:initstate}

Initial state effects are considered background effects in heavy ion 
physics. They are sometimes called cold nuclear matter effects, although
the two terms are not synonymous. In fact, there are clearly final state
effects in cold nuclear matter as seen in hadron production in
$e+A$ collision by the HERMES experiment \cite{Airapetian:2007vu} and 
successfully described in terms of higher twist corrections 
\cite{Guo:2000nz}.
For hadron production and similar process in $A+A$ collisions a
factorization between initial and final state effects for hadron production
is not obvious. It is one of the big \emph{assumptions} of the hard
probes program in heavy ion physics that final state effects can be
factorized off and modeled separately from hard processes and initial state
effects.

\begin{figure}[tb]
\epsfysize=9.0cm
\begin{center}
\begin{minipage}[t]{13 cm}
\epsfig{file=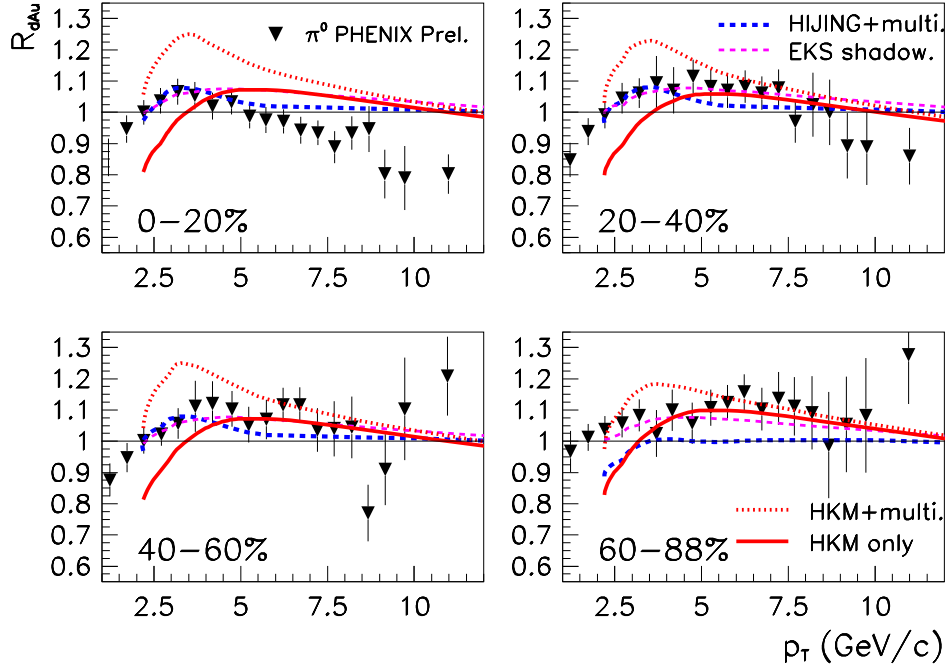,scale=0.7}
\end{minipage}
\begin{minipage}[t]{16.5 cm}
\caption{Cold nuclear matter effects for pion production in $d+$Au collisions 
at RHIC. The modification factor $R_{dAu}$ for $\pi^0$ is shown as a function 
of transverse momentum $P_T$. Data from the PHENIX experiment 
\cite{Buesching:2006ap} is compared to several calculations using different
sets of parton distributions with or without multiple scattering in the
initial state. See text for details.
Figure reprinted from \cite{Levai:2006yd} with permission from Elsevier. 
\label{fig:coldnuc}}
\end{minipage}
\end{center}
\end{figure}

One can analyze the effect of initial state interactions in nuclei by 
looking at hadron production in $p+$A and $d+$A collisions,
and by studying photons and dileptons for $p+$A, $d+$A \emph{and} 
A+A collisions.
While these analyses are not yet complete, the picture that starts to emerge 
is that for Au ions at RHIC energies the initial state effects do not change 
the yield of hadrons for high $P_T > 2$ GeV/$c$ by more than 20\%.

Fig.\ \ref{fig:coldnuc} shows calculations by Levai et al.\ 
\cite{Levai:2006yd} of the nuclear modification factor
\begin{equation}
  \label{eq:rdau}
  R_{dA} = \frac{dN^{dA}/dP_T}{\langle N_{\mathrm{coll}} \rangle
    dN^{pp}/dP_T}
\end{equation}
for pions in deuteron gold collisions compared to experimental data from
PHENIX \cite{Buesching:2006ap}. $\langle N_{\mathrm{coll}}\rangle$ is the
average number of binary nucleon-nucleon collisions expected for the given
centrality class.
The calculations use different sets of nuclear parton distributions
\cite{Eskola:1998df,Hirai:2001np}, with and without smearing through
multiple scattering, and are also compared to the transport model HIJING 
\cite{Li:2001xa}.
The modification factor is centered around 1 with very moderate deviations,
but we can clearly identify how the regions of anti-shadowing and the 
EMC effect map onto hadron-$P_T$ at midrapidity for the RHIC top energy
of 200 GeV. The calculations using HIJING and HKM nuclear parton distributions 
\cite{Hirai:2001np} together with multiple scattering are in reasonable
agreement with data except for the most central bin. However, the calculation
with only EKS modifications to parton distributions is doing equally
well. This suggests that higher-twist modifications to hard scattering and
modifications to parton distributions are not easily separated with the
available level of data accuracy.

This result can be confirmed by looking at the same modification 
factor $R_{AA}$ for direct photons in Au+Au collisions. 
Skipping ahead to Fig.\ \ref{fig:amyphot} we see that $R_{AA}$ for
photons is very close to 1 for $P_T > 4$ GeV/$c$ due to the absence
of final state effects. The existing, small deviations can be 
understood with the arsenal of initial state effects discussed in this
subsection. We conclude that initial state effects in A+A collisions
seem to be under control at RHIC energies.

Nevertheless there are some significant gaps in our understanding going 
forward to LHC.
We have already mentioned our poor knowledge of the nuclear gluon
distribution at smaller values of $\xi$. On a deeper level, it is a very 
difficult task to separate corrections to parton distributions
($\sim \log Q^2$) from higher twist corrections to hard processes
($\sim Q^2$) with data covering only a limited amount
of phase space. A separation into a contribution that follows 
DGLAP evolution and a power suppressed part has large uncertainties.
The use of a limited sample of $P_T$-spectra from RHIC for nuclear parton 
distribution fits bears the danger of introducing (erroneously) features 
into nuclear parton distributions which are non-universal.
The future Electron Ion Collider should be able to improve this situation
tremendously.

\subsection{\it Final State Effects and Energy Loss}

The main goal of the hard probes program in heavy ion
collisions is the determination of basic transport properties of the 
quark gluon plasma.
The idea that a hot medium formed in nuclear collisions should 
lead to energy loss of partons in the final state, and to a partial
quenching of high-$P_T$ hadrons, was proposed many years ago 
by Bjorken \cite{Bjorken:1982tu}. In the 1990s it was realized that the
most efficient process of energy loss is through
induced gluon bremsstrahlung. This topic was covered from several angles
in the 1990s in seminal works which estimated the effect on partons in 
perturbative plasma \cite{Thoma:1990fm}, for partons interacting 
perturbatively with static scattering centers (GW model) 
\cite{Gyulassy:1993hr, Wang:1994fx} and for multiple soft scatterings (BDMPS
model) \cite{Baier:1996kr,Baier:1996sk,Zakharov:1996fv,Zakharov:1997uu}.
Energy loss through elastic scattering had been calculated and was generally
found to be smaller than radiative energy loss for light quarks and 
gluons.

In this subsection we describe some of the underlying concepts and the most
important modern implementations of parton energy loss. We also comment
on some more recent developments including jet shapes and jet chemistry.
We will focus on light quarks and gluons. We would like to point the interested
reader to the review by Majumder and Van Leeuwen,
recently published in this journal \cite{Majumder:2010qh}, for complementary
information, and in particular for a detailed derivation of the higher twist 
energy loss formalism and for a discussion of heavy quark energy loss.

\subsubsection{Basic Phenomenology}

Partons of mass $m$ produced in hard QCD processes are typically off-shell, 
and the virtuality $\nu =\sqrt{p^2-m^2}$ is on average of the same 
order as the scale $Q$ of the momentum transfer in the hard process. The 
outgoing parton will radiate bremsstrahlung to get back to the mass shell, 
producing a parton shower and eventually a jet cone. This is an example for 
vacuum bremsstrahlung. Note that this picture is consistent with our earlier
discussion of hard processes where large angle radiation in the final state
would be counted as a higher order correction to the hard process while 
collinear radiation is resummed into fragmentation functions.

A particle that exchanges momentum with a medium will also change its 
virtuality with each interaction. It too will radiate bremsstrahlung to get 
back to the mass shell. This increased rate of radiation (or ``splitting'')
is an effective mechanism to carry away longitudinal momentum, and it acts
as a diffusion mechanism for transverse momentum (directions are relative 
to the original particle momentum).
This additional medium-induced bremsstrahlung obviously depends on the
density of the medium, or more precisely on the rate at which additional
virtuality can be transferred by the medium. This leads to the definition
of the transport coefficient
\begin{equation}
  \label{eq:qhat}
  \hat q = \frac{\langle k_T^2\rangle_L}{L} = \frac{\mu^2}{\lambda}
\end{equation}
which measures the average squared transverse momentum transferred to
the particle that propagates over a distance $L$, or equivalently
the average momentum transfer squared per interaction, $\mu^2$, divided
by the mean free path $\lambda$ of the particle.

It was realized early on that destructive interference is a key ingredient of 
these calculations. This is a well-known effect in QED, named after
Landau, Pomeranchuk and Migdal (LPM) \cite{Landau:1953um,Migdal:1956tc}. 
Let us consider the emission of a gluon $g$ from a quark $q$ that has
an initial energy $E$. If the relative transverse momentum between the
partons in the final state is $\mathbf{k}_T$ and the energy of the gluon
is $\omega$ then the formation time  
\begin{equation}
  t_f \sim \frac{\omega}{k_T^2}
\end{equation}
estimates when the final quark-gluon pair can be treated as two independent,
incoherent particles. If the mean free path $\lambda$ of the quark is of 
the order of the formation time or smaller, radiation is suppressed. 
In that situation the quark scatters coherently from $N_{\mathrm{coh}} 
\sim l_\mathrm{coh}/\lambda$ scatterers in the medium. For light, relativistic
partons the coherence length is given by the formation time $l_\mathrm{coh}=
\tau_f$.

Depending on the energy $\omega$ of the emitted gluon radiation one can 
qualitatively distinguish three domains for induced radiation in a
medium of finite length $L$ \cite{Baier:1996sk}:
\begin{itemize}
\item The incoherent regime for small $\omega$ in which the gluon
  radiation spectrum is independent of the length of the medium
  and the total energy loss $\Delta E$ would be proportional to the length $L$.
\item The completely coherent regime for large energies $\omega$ in which 
  the particle scatters coherently off the entire medium and the energy 
  loss $\Delta E$ is independent of the length $L$.
\item The LPM region in between the two extremes in which scatterings 
  off groups of $N_{\mathrm{coh}} \approx l_{\mathrm{coh}}/\lambda$ particles
  in the medium are coherent, and several or many of such interactions occur. 
  To determine the energy loss the differential gluon spectrum
  per unit length $x$, $\omega dI/d\omega dx \sim 1/\sqrt{\omega}$, has to 
  be integrated up to the limit $\omega_{\mathrm{cr}}$ which corresponds to 
  $l_{\mathrm{coh}} = x$, i.e.\ the boundary to the completely coherent 
  regime. For any given path length $x$ that critical value is, see Eq.\ 
  (\ref{eq:qhat}),
\begin{equation}
  \label{eq:omegacr}
  \omega_{\mathrm{cr}} = x \langle k_T^2\rangle_x  = x^2 \frac{\mu^2}{\lambda} \, .
\end{equation}
This leads to an energy loss rate 
\begin{equation}
  \label{eq:dedxest}
  \frac{dE}{dx} \sim - \hat q x
\end{equation}
and an energy loss $\Delta E \sim - \frac{1}{2}\hat q L^2$ over the entire length of 
the medium.
\end{itemize}
The LPM effect is expected to dominate the behavior of induced gluon radiation
in heavy ion collisions. The $L^2$-dependence is a characteristic signature 
of this effect.

In the following we will discuss several modern implementations of 
parton energy loss in more detail. They all differ in some of the underlying
approximations made.
\begin{itemize}
\item The Higher Twist formalism developed by Guo and Wang 
  \cite{Guo:2000nz,Wang:2001ifa}. It derives from the notion of higher 
  twist corrections for final state partons in $e+$A collisions.
\item The AMY formalism, based on the work by Arnold, Moore and Yaffe
  \cite{Arnold:2002ja,Jeon:2003gi}. It is based on hard thermal loop 
  resummation in a perturbative plasma.
\item The ASW formalism by Armesto, Salgado and Wiedemann 
  \cite{Salgado:2002cd,Salgado:2003gb} which resums multiple soft gluon 
  emission as in the BDMPS approach in a finite length medium using Poisson 
  statistics.
\item The GLV approach by Gyulassy, Levai and Vitev 
  \cite{Gyulassy:1999zd,Gyulassy:2000fs,Gyulassy:2000er} which considers
  hard scatterings off static scattering centers in an opacity expansion.
\end{itemize}

\subsubsection{The Higher Twist Formalism}

The systematic discussion of final state interactions of hard scattered
partons in a nuclear medium is dominated by several big questions. One of the
most fundamental ones is whether the final state interactions be factorized
off (a) the hard process, (b) the initial-state effects in nuclei, and (c) the
fragmentation into hadrons? There are ways to treat problem (c), or it can be
circumvented by looking at jets instead of hadrons, 
which is experimentally difficult at RHIC, but will be
routinely done at the LHC. Most of the QCD-inspired energy loss models
that we discuss here assume such a factorization. The Higher Twist (HT)
formalism eventually has to make the same assumption, but it takes guidance 
from a process in which such a factorization can actually be tested: 
semi-inclusive hadron production in deep inelastic scattering $e+$A 
off nuclei.

Guo and Wang were the first to write down a set of expanded evolution 
equations for
medium-modified fragmentation functions in $e+$A collisions 
\cite{Guo:2000nz,Wang:2001ifa}. They base their computation on the 
pioneering work of Qiu and Sterman on nuclear enhanced higher twist 
corrections discussed earlier.
Semi-inclusive hadron production, $e+A \to e+H+X$
is usually discussed by factorizing the cross
section $\sigma$ as a function of hadron momentum $l_H^\mu = (E_H, 
\mathbf{l}_H)$ and final lepton momentum $p_2^\mu = (E_2, 
\mathbf{p}_2)$ into a QED part called the leptonic tensor and a QCD part
called the hadronic tensor
\begin{equation}
  E_2E_H \frac{d\sigma}{d^2p_2 d^3 l_H} = \frac{\alpha_{\mathrm{em}}}{2\pi S
    Q^4} L_{\mu\nu} E_H \frac{dW^{\mu\nu}}{d^2 l_H} \, .
\end{equation}
This factorization is accurate to leading order in the electromagnetic 
coupling $\alpha_{\mathrm{em}}$. The leptonic tensor is
\begin{equation}
  L_{\mu\nu} = 2 p_1^\mu p_2^\nu + 2 p_1^\nu p_2^\mu - 2 p_1 \cdot p_2 \,
  g_{\mu\nu} \, ,
\end{equation}
and $Q^2 = -q^2$ measures the virtuality of the photon with momentum
$q^\mu = p_2^\mu -p_1^\mu$. We have labeled the initial momentum of the lepton
as $p_1^\mu$ and as usual we call the average momentum of a nucleon in the 
nucleus $P^\mu$ with a large light cone momentum fraction $P^+$.
It is common to choose the frame such that the photon momentum has a large 
$-$-component and no transverse components, $q^\mu =
(-Q^2/2q^-,q^-,\mathbf{0})$ in light cone notation.

\begin{figure}[tb]
\begin{center}
\begin{minipage}[t]{17 cm}
\epsfig{file=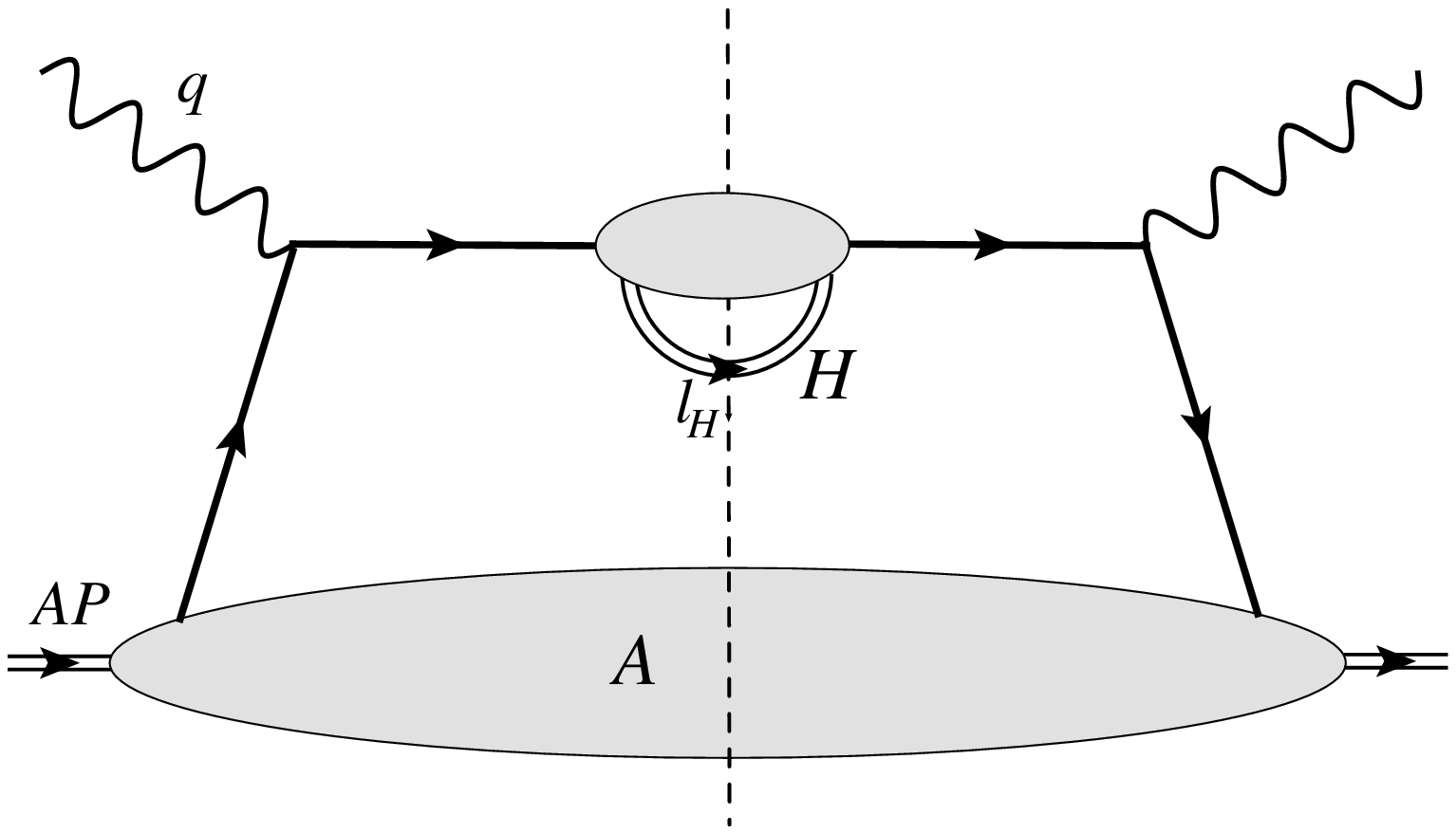,scale=0.5} \qquad\qquad
\epsfig{file=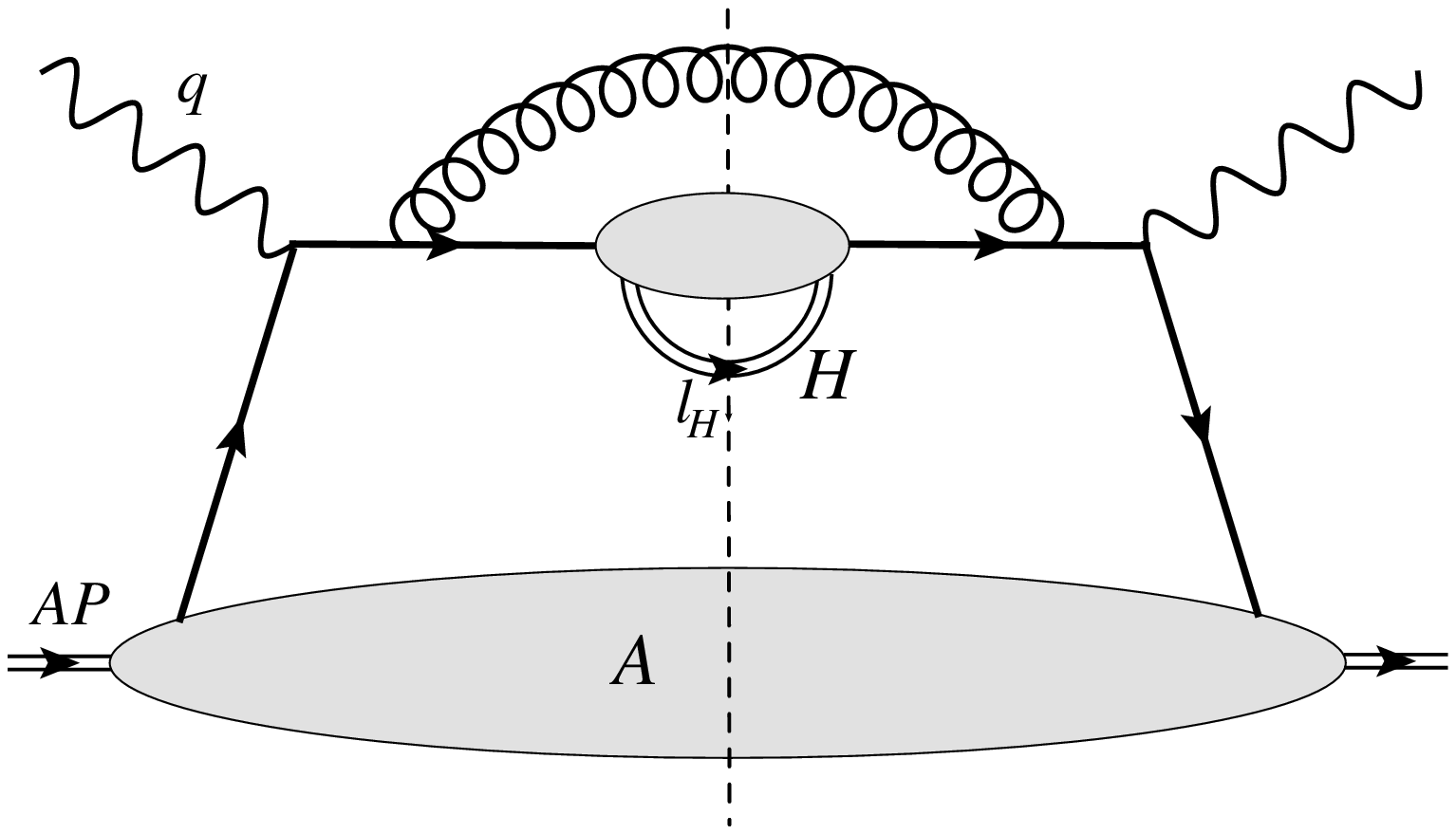,scale=0.5}
\end{minipage}
\begin{minipage}[t]{16.5 cm}
\caption{Left panel: Semi-exclusive deep-inelastic scattering at leading
order. Right panel: example for a real radiative correction that leads 
to the evolution equations for the fragmentation functions.
\label{fig:dis1}}
\end{minipage}
\end{center}
\end{figure}

The hadronic tensor measures the electromagnetic 
current (to which the photon couples) both in the amplitude and complex 
conjugated amplitude between initial nuclear states and the final 
hadronic states, $W^{\mu\nu} \sim \sum_X 
\langle A| j^\mu| H,X\rangle \langle H,X| j^\nu | A \rangle$. After
integrating the transverse degrees of freedom we can write it
as a function of only the longitudinal momentum fraction of the hadron
with respect to the photon momentum, $z_H = l_H^- /q^-$.
Leading-twist ($t=2$) collinear factorized QCD tells us that
\begin{equation}
  \label{eq:leadingtw}
  \frac{dW^{\mu\nu}_{(2)}}{d z_H} = \sum_{ab} \int d\xi f_{a/A} (\xi,Q) \int_{z_H}^1
  \frac{dz_b}{z_b} H^{\mu\nu}_{ab}(\xi,p,q,z_b) 
   D_{b/H} \left( \frac{z_H}{z_b},Q \right) 
\end{equation}
where $H^{\mu\nu}_{ab}$ describes the hard scattering of parton $a$ off the
virtual photon, producing a parton $b$ and maybe more unobserved final state
particles $x$, $a+\gamma^* \to b+x$.
While parton $a$ has a momentum fraction $\xi$ with respect to $P^\mu$,
parton $b$ has a momentum fraction $z_b = l_b^-/q^-$ with respect to the
photon. There is a clear separation between the initial and final state
long-distance processes described by the parton distribution $f$ and
the fragmentation function $D$ respectively. The leading order diagram
for this process is shown in the left panel of Fig.\ \ref{fig:dis1}

Collinear radiation off the final state parton $b$ leads to leading-twist
evolution equations for the fragmentation functions.  
The diagram in the right panel of Fig.\ \ref{fig:dis1} leads to a correction 
to Eq.\ (\ref{eq:leadingtw}) which can be written as
\begin{equation}  
  \frac{dW^{\mu\nu}_{(2')}}{d z_H} = \sum_{abc} \int d\xi f_{a/A} (\xi,Q) 
  \int \frac{dl_T^2}{l_T^2} \frac{\alpha_s}{2\pi}
  \int_{z_c}^1 \frac{dz_b}{z_b} H^{\mu\nu}_{ab}(\xi,p,q,z_b)
  \int_{z_h}^1 \frac{dz_c}{z_c} P_{b\to c}\left(\frac{z_c}{z_b}\right)
  D_{c/H} \left( \frac{z_h}{z_c},Q \right)
\end{equation}
with the familiar splitting functions $P_{b\to c}$ to a third parton $c$
which undergoes fragmentation. The leading collinear term can be
resummed and leads to evolution equations which are completely analogous
to the DGLAP equations (\ref{eq:dglap}) for parton distributions\footnote{The 
equations in \cite{Guo:2000nz,Wang:2001ifa} often omit the integral
over $z_b$ since the hard parton scattering tensor $H$
contains a $\delta(z_b -1)$ function at leading order in $\alpha_s$ only, which
has been canceled with the integral.}.

\begin{figure}[tb]
\begin{center}
\begin{minipage}[t]{17 cm}
\epsfig{file=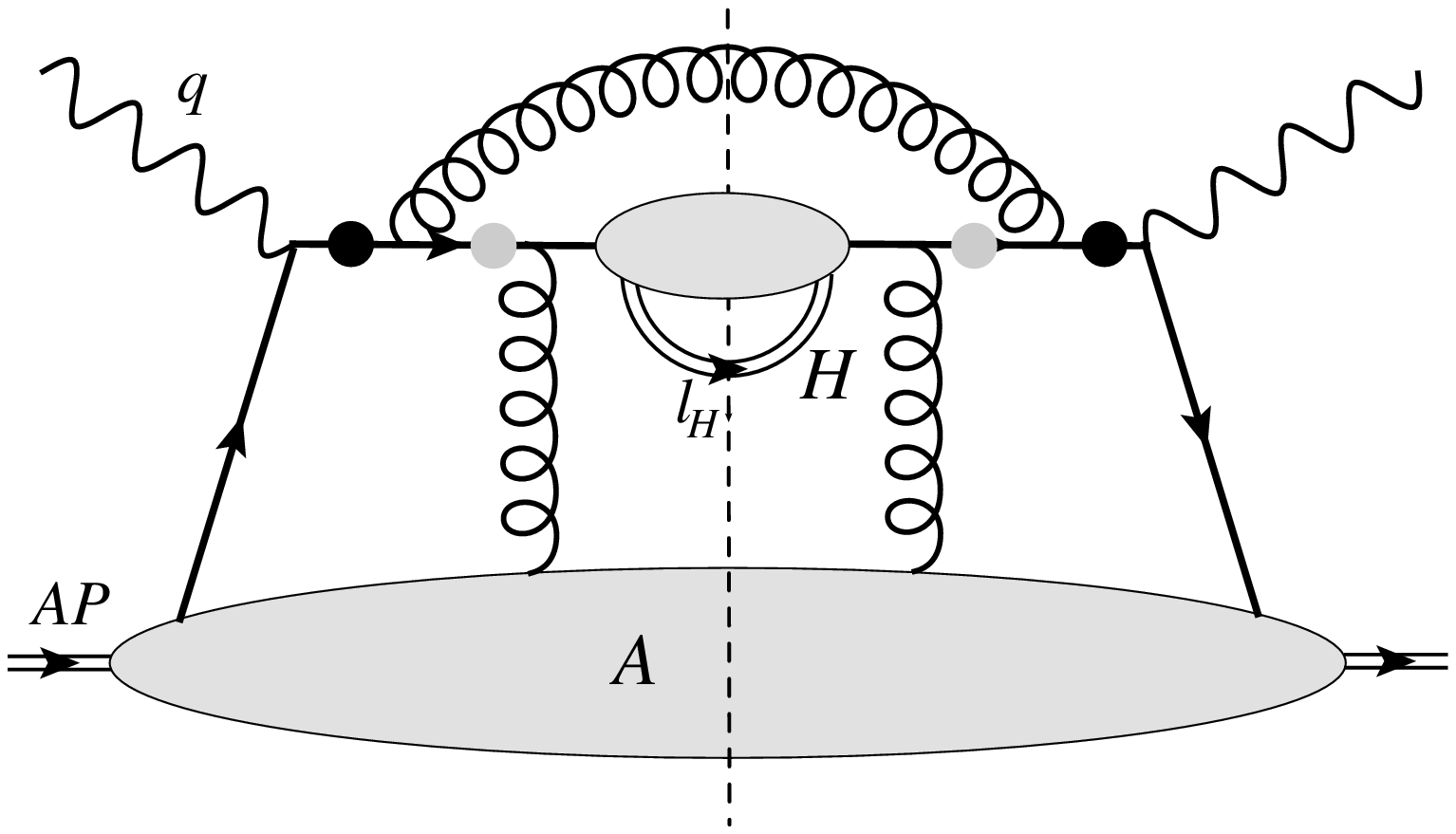,scale=0.5} \qquad\qquad
\epsfig{file=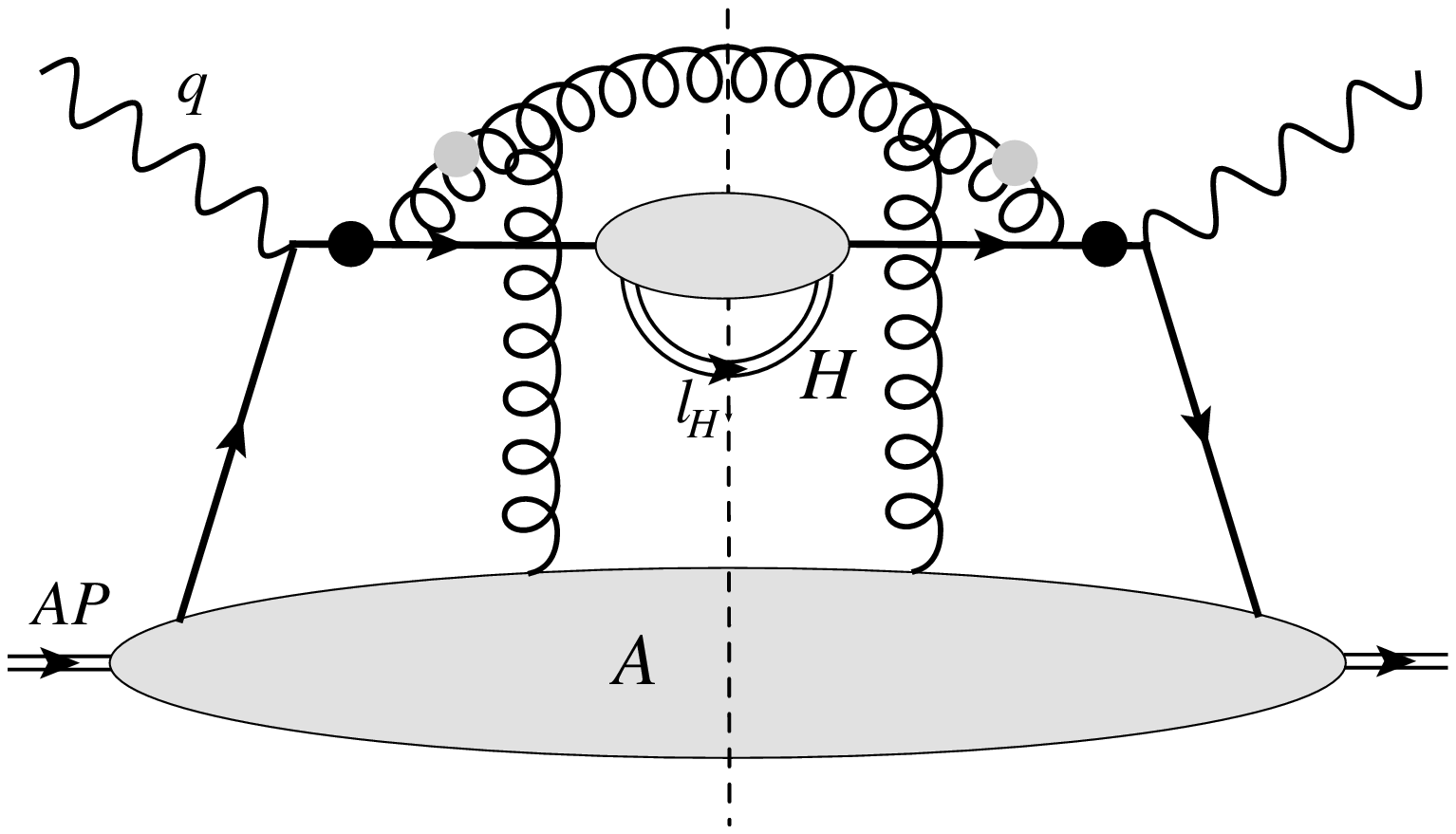,scale=0.5}
\end{minipage}
\begin{minipage}[t]{16.5 cm}
\caption{Two examples of twist-4 diagrams that contribute to the
evolution equations of medium-modified fragmentation functions.
Propagators with poles that make the gluon from the nucleus soft 
(momentum $\sim \xi_D$) are indicated by a grey circle, propagators whose poles
make the gluon harder (momentum $\sim \xi_L$) are shown with a solid black
circle.
There are many more diagrams including those with different final state cuts
and interference diagrams in which the gluon from the nucleus couples 
to different particles in amplitude and complex conjugated amplitude.
\label{fig:dis2}}
\end{minipage}
\end{center}
\end{figure}

Guo and Wang showed that the hadronic tensor at the level of nuclear enhanced 
twist-4 receives contributions from diagrams like the ones shown in 
Fig.\ \ref{fig:dis2} where parton $b$ or $c$, or the radiated parton could 
scatter off an additional medium parton $d$. 
They computed the result from those diagrams and found
\begin{multline}
  \frac{dW^{\mu\nu}_{(4')}}{d z_H} = \sum_{abcd} \int d\xi 
  \int \frac{dl_T^2}{l_T^4} \frac{\alpha_s}{2\pi}
  \int_{z_c}^1 \frac{dz_b}{z_b} H^{\mu\nu}_{ab}(\xi,p,q,z_b)
  \int_{z_h}^1 \frac{dz_c}{z_c} \left(\frac{z_c}{z_b}\right)
  D_{c/H} \left( \frac{z_H}{z_c} \right)   \\ \times
  \left[ P_{b\to c}\left(\frac{z_c}{z_b}\right) T_{ag/A}(\xi,\xi_L) + 
  \delta\left( \frac{z_c}{z_b}-1 \right) \Delta T_{ag/A}(\xi,l_T^2) \right] \, .
\end{multline}
For simplicity of notation we have assumed here that the second parton $d$ 
is a gluon that does not change the identity of the parton it couples to.
The other cases can be treated accordingly.  
The twist-4 matrix elements are similar to the soft-hard matrix
elements introduced in Eq.\ (\ref{eq:tsh}). If parton $a$ is a quark
we have
\begin{multline}
  \label{eq:tgw}
  T_{qg/A}(\xi,\xi_L) = \int dy_4^- {dy_3^-} \frac{dy_1^-}{2\pi}
  e^{i(\xi+\xi_L) P^+ y_1^-} \left( 1- e^{-i\xi_L P^+ y_4^-} \right) 
  \left( 1- e^{-i\xi_L P^+ (y_1^- -y_3^-)} \right) \Theta(y_1^- - y_3^-) \Theta(-y_4^-)
  \\ \times 
  \left\langle A(P) \right| \bar q(0) \gamma^+ q(y_1^-)
   F_a^{+\nu}(y_3^-) F_{a\,\nu}^{\>\>\>\>+}(y_4^-) \left|A(P) \right\rangle \, .
\end{multline}
The expression also contains a derived matrix element
\begin{equation}
  \Delta T_{qg/A}(\xi,l_T^2) = \int_0^1 \frac{dz}{1-z} \left[ 2 T_{qg/A}(\xi,
  \xi_L)|_{z=1} - (1+z^2) T_{qg/A}(\xi,\xi_L) \right]
\end{equation}
which remains after the virtual corrections have canceled the singularity
at $z=1$.

We note that the normalization of the matrix elements, that is following 
Guo and Wang here, differs by a factor $2\pi$ from Eq.\ (\ref{eq:tsh}). 
We have also gone beyond the soft-hard matrix elements by allowing parton 
$d$ to have a non-vanishing momentum fraction 
\begin{equation}
  \xi_L = \frac{l_T^2}{2P^+ q^- z(1-z)}
\end{equation}
with $z=z_c/z_b$. It is the structure $(1-\exp(\ldots))(1-\exp(\ldots))$
in (\ref{eq:tgw}) that exhibits interference and will eventually lead to the
LPM effect in the Higher Twist
formalism. One can easily see how this interference emerges in 
the calculation. The momentum
of parton $d$ is fixed by the poles of partons $b$/$c$ and there 
are two kinematic possibilities shown in Fig.\ \ref{fig:dis2}. Either
the momentum of parton $d$ is very soft with a momentum fraction
$\xi_D \sim k_T^2/Q^2$ which vanishes when its intrinsic transverse momentum 
$\mathbf{k}_T$ is set to zero. The phase $\exp(-i\xi_D P^+ y^-)$ then reduces
to unity. The other pole sets the momentum fraction to $x_L$, and 
interestingly the amplitudes for both poles exhibit a relative minus sign.
This interference is common when higher twist corrections are considered
together with radiative corrections. A discussion of this soft-hard
interference in the context of Drell-Yan can be found in 
\cite{Guo:1997mm,Fries:1999jj,Fries:2000da,Fries:2002dn}.

The collinear radiation corrections at twist-4 can be resummed in
\emph{modified} evolution equations for new fragmentation functions
$\tilde D$ in cold nuclei just as in the twist-2 (DGLAP) case. The resulting
set of equations takes exactly the same form as in Eq.\ (\ref{eq:dglap}),
with $D \to \tilde D$ and new, medium-dependent splitting functions
\begin{equation}
  \tilde P_{ab} (z) =  P_{ab}(z) + \Delta P_{ab}(z)
\end{equation}
where $P_{ab}$ is the usual vacuum splitting function and $\Delta P_{ab}$
is the twist-4 correction. For the case of $q\to q+g$ splitting induced
by a gluon we have
\begin{equation}
  \label{eq:modsplit}
  \Delta P_{q\to q}(z) = \frac{2\pi \alpha_s C_A}{N_c} \frac{1}{l_T^2}
  \left[ \left( \frac{1+z^2}{1-z}\right)_+ 
  \frac{T_{qg/A}(\xi,\xi_L)}{f_{q/A}(\xi)} 
  + \delta(z-1) \frac{\Delta T_{qg/A}(\xi,l_T^2)}{f_{q/A}(\xi)} \right] \, .
\end{equation}
The other modified splitting functions are discussed in Ref.\
\cite{Wang:2001ifa} for scattering off gluons and in Ref.\ \cite{Schafer:2007xh}
for scattering off quarks.
Even though this result is technically correct and very useful we can also 
see its limitations.
The splitting functions, and as a consequence the modified fragmentation
functions $\tilde D$, are no longer universal as they depend on 
the underlying process through the matrix elements $T$. In fact they
are no longer just functions of a single momentum fraction $z$ but also 
depend in a non-trivial way on $\xi$. On
the other hand, the breaking of universality encodes the 
medium effects that we are after.

The new medium-modified fragmentation functions allow us to write hadron 
production in $e+$A collisions in a very simple way analogous to 
Eq.\ (\ref{eq:leadingtw}) as
\begin{equation}
  %\label{eq:leadingtw}
  \frac{dW^{\mu\nu}_{(4)}}{d z_H} = \sum_{ab} \int d\xi f_{a/A} (\xi,Q) \int_{z_H}^1
  \frac{dz_b}{z_b} H^{\mu\nu}_{ab}(\xi,p,q,z_b) 
   \tilde D_{b/H} \left( \frac{z_H}{z_b},Q \right)  \, .
\end{equation}
The dependence of the $\tilde D$ on other quantities is usually suppressed in
the notation. We have now come to a point where one has to introduce  a
certain amount of modeling since a rigorous solution would include a
simultaneous fit of the $\tilde D_{ab/A}$ and the $T_{ab/A}$ in the same
environment (because of the loss of universality) with the evolution equations 
as constraints. This is too complex a task given the available data.

The twist-4 matrix elements are modeled similar to the 
less general soft-hard matrix elements from Eq.\ (\ref{eq:sh}). It seems
safe to assume that $T_{ab/A}$ can be factorized into a product of two
parton distributions for partons $a$ and $b$ resp. Guo and Wang model
the interference effect by introducing a massless parameter for the
radius $R_A$ of the nucleus, $x_A = 1/(MR_A)$, where $M$ is the mass of a 
nucleon. They suggest
\begin{equation}
  T_{ab/A}(\xi,\xi_L) \approx \frac{C}{x_A} \left(1-e^{-\xi_L^2/x_A^2}\right)
  \left[ f_{a/A}(\xi) \, \xi_L f_{b/A}(\xi_L) + f_{a/A}(\xi+\xi_L) \, 
  \kappa_b \right]
\end{equation}
where $C$ is a normalization constant and 
$\kappa_b \approx \lim_{x\to 0} x f_{b/A}(x)$ formally is the value of the parton 
density for parton $b$ when it is very soft, i.e.\
with momentum fraction of order $x_D$. Obviously $\kappa_g$ is proportional
to the gluon density $\lambda^2$ introduced in Eq.\ (\ref{eq:shgluon}). 
Note that the dependence on the size of the system is hidden in $x_A$ and
the leading size dependence is $\sim A^{1/3}$.
On the other hand the formation time of the radiation is $\sim 1/(x_L P^+)$ 
and the factor $1-e^{-\xi_L^2/x_A^2}$ leads to LPM suppression unless
$\xi_L \gg x_A$.
In Ref.\ \cite{Wang:2002ri} the authors suggest an even simpler model that 
drops the second term 
\begin{equation}
  T_{ab/A}(\xi,\xi_L) \approx \frac{\tilde C_b}{x_A}
  \left(1-e^{-\xi_L^2/x_A^2} \right) f_{a/A}(\xi) \, .
\end{equation}

Even if the set of twist-4 matrix elements were perfectly known there is
still the task to extract the medium-modified fragmentation functions
from the set of evolution equations. Guo and Wang originally suggested the
first iteration
\begin{equation}
  \label{eq:gwmedium}
  \tilde D_{a/H}(z,\mu) =  D_{a/H}(z,\mu) + \frac{\alpha_s}{2\pi}
  \int_0^{\mu^2} \frac{dl_T^2}{l_T^2} \int_z^1 \frac{dy}{y} \sum_b
  \Delta P_{a\to b}(y) D_{b/H}\left( \frac{z}{y} \right)
\end{equation}
as an approximate solution. Note that the $\Delta P_{a\to b}$ carry all 
the information about the medium through Eq.\ (\ref{eq:modsplit}) 
and its counterparts. Deng and Wang recently showed how to solve 
modified evolution equations numerically \cite{Wang:2009qb}.

Using the first iteration the energy loss of a parton can be calculated as the shift in the
momentum fraction due to the term $\Delta P_{a\to b}$ in the equation above
\begin{equation}
  \langle \Delta y \rangle = 1- \frac{\alpha_s}{2\pi}
  \int_0^{\mu^2} \frac{dl_T^2}{l_T^2} \int_z^1 {dy} \sum_b
  \Delta P_{a\to b}(y) \, .
\end{equation}
For quarks one finds \cite{Guo:2000nz,Wang:2002ri}
\begin{equation}
  \langle \Delta y \rangle = \tilde C_g \frac{\alpha_s^2 C_A}{N_c}
  \frac{6x_B}{Q^2} \frac{1}{x_A^2} (- \ln 2x_B) 
\end{equation}
which is proportional to the nuclear size squared, $\langle \Delta y \rangle
\propto A^{2/3}$, as expected for the LPM regime. $x_B = Q^2/(2P^+q^-)$ is
the Bjorken variable in deep inelastic scattering. In Ref.\ \cite{Wang:2002ri}
the authors can explain the observed suppression of semi-inclusive hadron 
production in the HERMES experiment \cite{Airapetian:2003mi} very well using
the medium modified fragmentation functions described above.
One can go one step further and try to interpret the medium modification as
a simple rescaling of the vacuum fragmentation functions. One uses an ansatz
\cite{Wang:2002ri,Wang:1996yh}
\begin{equation}
  \label{eq:htmmf}
   \tilde D_{a/H}(z) \approx \frac{1}{1-\Delta z} D_{a/H} \left(
   \frac{z}{1-\Delta z} \right)
\end{equation}
where $\Delta z$ is the typical energy loss for parton $a$. This formula can
only be a satisfying approximation in a limited region with $z+\Delta z < 1$.
At this level the medium-modified fragmentation functions have been
cast in a very general form and one can try to apply the general concepts
to systems other than $e+$A. In particular one can extract the
stopping power $\langle \Delta E/L \rangle \approx E \Delta z/L$ for
a particle with initial energy $E$ from data in heavy ion collisions.
Using the techniques here (and including the dilution of the medium through
the longitudinal expansion in nuclear collisions) E.\ Wang and X.N.\ Wang 
concluded in 2002 that the differential energy loss extracted 
from data in Au+Au collisions at RHIC was about 15 times larger 
than the one for cold nuclear matter measured at HERMES \cite{Wang:2002ri}.

Beyond the first iteration, Deng and Wang have systematically studied the
evolution of gluon induced parton-to-parton fragmentation functions (starting
from $D_{a/a}=\delta(z-1)$ and $D_{a/b}=0$, $b\ne a$ initial conditions at a low
scale). They also investigate the energy and medium thickness dependence and
apply their techniques to pion fragmentation with parameterized vacuum 
fragmentation functions as input. Modified evolution
suppresses the fragmentation functions at intermediate and large values of $z$
as expected \cite{Wang:2009qb}.

In recent years progress was made on the HT formalism (in its original
meaning for deep-inelastic scattering) by considering medium
modifications for double fragmentation \cite{Majumder:2004pt}, elastic
energy loss \cite{Majumder:2008zg} and through successful resummation of
multiple scatterings per photon or gluon emission \cite{Majumder:2007ne,
Majumder:2009ge}.

\subsubsection{The AMY Formalism}

The particular merit of the formalism widely know as AMY, after Arnold, Moore and Yaffe,
is its complete internal consistency in a very high temperature regime. The
basic picture is the following.
Partons propagating through a plasma with temperature $T$, themselves 
having momenta of order $T$ or larger but small virtualities, interact
perturbatively with quarks and gluons in the plasma
with thermal masses $\sim gT$. They pick up transverse momentum of the same 
order $gT$, and then radiate gluons (for quarks) or split into quark-antiquark 
pairs or two gluons (for gluons), again at typical transverse momentum scales 
$gT$. This leads to formation times that are rather long, of the order 
$T/(gT)^2 \sim 1/(g^2 T)$.
The shortcomings are the requirements of small initial off-shellness of the
parton, an unlikely condition for a parton emerging from a hard process,
and the condition of very small coupling $g \ll 1$ to justify thermal
perturbation theory, which requires very large temperatures $T \gg T_c$. 
The exact temperature from which on such calculations start to be reliable 
is a matter of debate. 
One must also assume that the temperature does not change during 
the formation time of radiation which is questionable for rapidly evolving 
fireballs. Nevertheless, the rigor of the formalism has made AMY
an appealing choice in the canon of energy loss calculations.

The AMY formalism grew out of a computation of the complete leading order, 
hard thermal loop (HTL) resummed perturbative photon and gluon emission
rates. Arnold, Moore and Yaffe introduced, for the first time, the correct 
treatment of collinear emission in a finite temperature medium, which 
must take into account the LPM effect due to the long formation times
$\sim 1/(g^2 T)$ \cite{Arnold:2001ba,Arnold:2001ms,Arnold:2002ja}. 
The applications of these results to photon production in quark gluon 
plasma had been mentioned in a previous section. The results for gluon 
radiation off a parton with typical momentum $T \gg gT$ in the medium can 
be used to calculate its rate of energy loss. For quarks of momentum
$p$ radiating gluons of momentum $k$ the rate is 
\cite{Arnold:2002ja,Jeon:2003gi}
\begin{equation}
  \label{eq:amyrate}  \frac{\partial \Gamma_{q\to qg}}{\partial k\partial t}(p,p-k) = 
  \frac{C_R \alpha_s}{4 p^7} n_B(k) n_F(p-k)
  \frac{1+(1-x)^2}{x^3(1-x)^2} \int \frac{d^2 h}{(2\pi)^2} 2\mathbf{h} \cdot
  \mathrm{Re} \, \mathbf{F}(\mathbf{h},p,k)
\end{equation}
where $n_B$ and $n_F$ are the boson and fermion occupation factors for the
gluon and the quark in the final state, $x = k/p$ is the momentum fraction 
of the gluon, and $\mathbf{h}=\mathbf{p} \times \mathbf{k}$ is 
a useful measure for the non-collinearity of the final state.
$\mathbf{F} \propto \mathbf{h}$ is a function defined by the integral equation
\begin{multline}
   2\mathbf{h} = i \delta E(h,p,k) \mathbf{F}(\mathbf{h}) + 
   \frac{\alpha_s}{2\pi} \int d^2 q_\perp V({q}_\perp) 
   \times \left[ (2 C_R - C_A)\left(\mathbf{F}(\mathbf{h}) 
   - \mathbf{F}(\mathbf{h}-k\mathbf{q}_\perp) \right) \right. \\ + \left. 
   C_A \left(\mathbf{F}(\mathbf{h}) - \mathbf{F}(\mathbf{h}+p\mathbf{q}_\perp) 
   \right) +
   C_A \left(\mathbf{F}(\mathbf{h}) - \mathbf{F}(\mathbf{h}-(p-k)
   \mathbf{q}_\perp) \right) \right] \, .
\end{multline}
We have dropped the second and third argument in $\mathbf{F}(\mathbf{h},p,k)$
for brevity, in all cases above they are equal to the
initial momentum $p$ and the radiated momentum $k$ resp. It is the function
$\mathbf{F}$ that encodes important properties of the medium via
the potentials
\begin{equation}
  V({q}_\perp) = \frac{m_D^2}{q_\perp^2(q_\perp^2-m_D^2)} , \qquad
  m_D^2 = \frac{g^2 T^2}{6} (2N_c + N_f)
\end{equation}
as a function of momentum transfer $\mathbf{q}_\perp$ and temperature $T$.
\begin{equation}
  \delta E(h,p,k) = \frac{h^2}{2pk(p-k)} + \frac{m_k^2}{2k} +
  \frac{m_{p-k}^2}{2(p-k)} - \frac{m_p^2}{2p}
\end{equation}
is the energy difference between final and initial state. The masses
$m$ are $m_D/\sqrt{2}$ for gluons and $gT/\sqrt{3}$ for quarks in the
respective channels with momenta $p$, $k$ and $p-k$.

The rates for other processes can be obtained from Eq.\ (\ref{eq:amyrate})
by replacing the splitting functions, adjusting the Bose or Fermi
factors, and putting the correct color factor $C_R$. The missing splitting
functions needed are
\begin{align}
  N_f \frac{x^2+(1-x)^2}{x^2(1-x)^2} & \quad \mathrm{for} \, g\to q\bar q 
  \, , \\
  \frac{1+x^4+(1-x)^4}{x^2(1-x)^3} & \quad \mathrm{for} \, g\to gg \, .
\end{align}

The rates for different processes as a function of time $t$ can be implemented 
in coupled rate equations for quark, antiquark, and gluon momentum 
distributions $f_q$, $f_{\bar q}$ and $f_g$ resp.
\begin{align}
  \label{eq':amyrate2}
  \frac{f_q(p)}{\partial t} & = \int_{-\infty}^{\infty} dk \left[ 
  \frac{\partial\Gamma_{q\to qg}(p+k,k)}{\partial k\partial t} f_q(p+k) 
  - \frac{\partial\Gamma_{q\to qg}(p,k)}{\partial k\partial t} f_q(p) 
  + \frac{\partial\Gamma_{g\to q \bar q}(p+k,k)}{\partial k\partial t} f_g(p+k) \right] \, ,\\
  \frac{f_g(p)}{\partial t} & = \int_{-\infty}^{\infty} dk \left[ 
  \frac{\partial\Gamma_{q\to qg}(p+k,p)}{\partial k\partial t} \left( f_q(p+k) + f_{\bar q}(p+k)\right)
  - \frac{\partial\Gamma_{g\to q\bar q}(p,k)}{\partial k\partial t} f_g(p)
  \nonumber \right.  \\ & \qquad \qquad
  \left. + \frac{\partial\Gamma_{g\to gg}(p+k,k)}{\partial k\partial t} f_g(p+k)
  - \frac{\partial\Gamma_{g\to gg}(p,k)}{\partial k\partial t} f_g(p+k) 
  \Theta(2k-p) \right] \, .
\end{align}
The equation for the antiquark distribution is analogous to the equation
for the quark distribution. Note that the emitted momentum $k$ can be positive
or negative, which means that a parton can in principle also acquire momentum
from the medium. Of course, a parton with momentum much larger than typical 
thermal momenta will still lose momentum on average.
Final quark and gluon spectra can be subjected to vacuum fragmentation
to compute final hadron spectra.

\subsubsection{The GLV Formalism}

The GLV energy loss model by Gyulassy, Levai and Vitev 
\cite{Gyulassy:1999zd,Gyulassy:2000fs,Gyulassy:2000er} is based on the 
earlier Gyulassy-Wang model. It
describes the medium as an ensemble of static scattering centers with
Yukawa potentials exhibiting a screening mass $\mu$, which also set
the typical scale of transverse momentum transfer. 
The scattering amplitude is then expanded in terms of the opacity
$L/\lambda$ where $L$ is the length of the medium and $\lambda$ is the 
mean free path.
The leading, zeroth order term, for a parton with energy $E$ corresponds 
to vacuum radiation with a spectrum
\begin{equation}
  \label{eq:glv0}
  \frac{dI^{(0)}}{dx dk^2} = \frac{C_R \alpha_s}{2\pi}
  \left( 2-2x+x^2 \right) E \frac{1}{k^2}
\end{equation}
as a function of the gluon longitudinal momentum fraction $x$ and
transverse momentum $\mathbf{k}$. $C_R$ is the color Casimir factor
in the appropriate representation, $C_F =4/3$ for quarks and $C_A = 3$ for 
gluons.

At the next order in opacity one considers the interference
of vacuum radiation and a single medium-induced radiation with
momentum transfer $\mathbf{q}$
\cite{Gyulassy:2000er}
\begin{equation}
  \label{eq:glv1}
  \frac{dI^{(1)}}{dx dk^2} = \frac{C_R \alpha_s}{2\pi}
  \left( 2-2x+x^2 \right) \frac{L}{\lambda} E \frac{1}{k^2} \, 
  \int d^2 q \frac{\mu^2}{\pi(q^2+\mu^2)^2}
  \frac{\mathbf{k}\cdot\mathbf{q} (\mathbf{k-q})^2 L^2}{16 x^2 E^2 + 
  (\mathbf{k-q})^4 L^2} \, .
\end{equation}
The integrals can be evaluated analytically away from the extreme
cases $x \to 0$ and $x \to 1$. Then maximally loose kinematic 
constraints $0 < k < \infty$, $0 < q < \infty$ can be assumed and
the total energy loss at first order in opacity is
\begin{equation}
  \Delta E ^{(1)} = \frac{C_R\alpha_s}{4} \frac{\mu^2}{\lambda} L^2
  \ln \frac{E}{\mu} \, .
\end{equation}
It exhibits the characteristic $L^2$-dependence.
In contrast the zeroth order gluon spectrum (\ref{eq:glv0}) will lead to 
a ``vacuum quenching''
\begin{equation}
  \Delta E^{(0)} = \frac{4 C_R \alpha_s}{3\pi } E \ln \frac{E}{\mu}
\end{equation}
where $\mu$ in this case is chosen to play the role of a lower cutoff 
for the transverse momentum $k$. 

Higher orders in the opacity $(L/\lambda)^n$ can be treated numerically 
\cite{Gyulassy:2000er}. Since the number of emitted gluons is finite,
fluctuations around a given mean value are important. They can be taken 
into account through Poisson statistics \cite{Gyulassy:2001nm},
leading to a probability distribution $P(\epsilon)$ for the fractional
energy loss $\epsilon = \Delta E/E$. The probabilities for emission of
$n$ gluons are
\begin{equation}
  \label{eq:poisson1}
  P_n(\epsilon) = \frac{e^{-\bar N_g}}{n!} \prod_{i=1}^{n} \int dx_i 
  \frac{dI}{dx_i} \delta\left( \sum_i x_i - \epsilon \right) 
\end{equation}
where $\bar N_g$ is the average number of gluons and $dI/dx$ is the gluon 
energy spectrum to the desired order in opacity, e.g.\ derived from Eq.\ 
(\ref{eq:glv1}) to first order. The total probability distribution is 
\begin{equation}
  \label{eq:poisson2}
  P(\epsilon) = \sum_{n=1}^\infty P_n (\epsilon) \, .
\end{equation}
It can be used to define a medium-modified fragmentation function
\begin{equation}
  \label{eq:mmffasw}
  D^{\mathbf{GLV}}_{a/H}(z,Q) = \int_0^1 d\epsilon 
  \frac{P(\epsilon)}{1-\epsilon} D_{a/H}\left(\frac{z}{1-\epsilon},Q\right)
  \, 
\end{equation}
analogous to Eq. (\ref{eq:htmmf}).
Recall that we tacitly assume that parton energy loss and the actual
hadronization of the parton are factorizable, and hadronization itself 
happens outside of the medium.
There have also been attempts to include elastic scattering consistently
in the GLV approach \cite{Gyulassy:2002yv,Wicks:2005gt}.

\subsubsection{The ASW Formalism}

The energy loss model due to Armesto, Salgado and Wiedemann
\cite{Salgado:2002cd,Salgado:2003gb} assumes a Poisson-like distribution 
of gluon emissions as described in Eqs.\ (\ref{eq:poisson1}) and 
(\ref{eq:poisson2}). The proponents of the model present a resummed
version of the so-called quenching weight $P$.
They add the explicit possibility $P_0$ of zero gluon emissions, i.e.\ a 
finite probability that a parton escapes unquenched.
For their main result they use the gluon radiation spectrum $dI/dx$ from the 
BDMPS approach assuming finite propagation length $L$
\cite{Baier:1996kr,Baier:1996sk,Zakharov:1996fv,Zakharov:1997uu,
Wiedemann:2000tf}. They also present results for a resummation of 
gluon spectra in a GLV-like opacity expansion \cite{Wiedemann:2000za},
but we will not discuss that latter option in detail.

In the case of BDMPS soft scattering they introduce a characteristic
gluon frequency 
\begin{equation}
  \omega_c = \frac{1}{2} \hat q L^2
\end{equation}
and a dimensionless quantity 
\begin{equation}
  R = \omega_c L  \, .
\end{equation}
Note that $\omega_c$ is close to the definition in Eq.\ (\ref{eq:omegacr}). 
$R$ is introduced to enforce the kinematic constraint $k_\perp < \omega$. 
From its definition we can infer that $R\to \infty$ corresponds to an 
infinitely large medium if $\omega_c$ is finite. It is also the limit in which
the previous BDMPS result is recovered.

One can perform a numerical resummation of the Poisson sum $P$ 
from Eq.\ (\ref{eq:poisson2}) for $n >0$. The total probability is 
then written as a sum
\begin{equation}
  P(\Delta E) = p_0(\omega_c,R) \delta(\Delta E) + p (\Delta E,\omega_c,R)
  \, 
\end{equation}
which contains a discrete probability $p_0$ for zero energy loss and
the resummed probability for finite energy loss ($n>0$). The unphysical case 
$\Delta E > E$ can be dealt with by either renormalizing the total probability
to unity (``reweighted'') or by introducing a second $\delta$-function 
at $\Delta E = E$ which accumulates the probability for total loss
of the jet (``non-reweighted'') \cite{Wiedemann:2000tf,Dainese:2004te}.
The uncertainty in the treatment of the case $\Delta E >0$ leads to a 
rather large uncertainty in describing the data, see e.g.\ Fig.\ 
\ref{fig:daineseraa}.
The authors of the ASW model provide a Fortran code which computes both
the discrete and the continuous part of the quenching weight as functions
of $\Delta E/\omega_c$ and $R$ \cite{aswquenching}.
As in the GLV case the quenching weights can be used to define modified
fragmentation functions using Eq.\ (\ref{eq:mmffasw}), although also in the
ASW case the true, non-perturbative fragmentation process itself is 
considered independent and not affected by the medium.

For applications in heavy ion collisions we need a way to go beyond the
simple assumptions of a homogeneous medium in which $\hat q$ is constant
along the propagation path of the parton. In fact in a realistic fireball
$\hat q = (\mathbf{x},t)$ where $\mathbf{x}$ is the position in the fireball
and $t$ the time. The extraction of $\hat q = (\mathbf{x},t)$, or at least
a spatially and time-averaged version $\langle \hat q\rangle$ of it, is
actually the goal of the hard probes program.
One can deduce $\omega_c$ and $R$ from the two lowest moments of
$\hat q$ integrated over the path of a jet particle,
\begin{equation}
  I_n = \int_0^\infty \hat q\left(\mathbf{x}(t), t\right) 
  |\mathbf{x}(t) - \mathbf{x}_0|^n  dt \, .
\end{equation}
Here $\mathbf{x}(t)$ is the trajectory of the particle originating from a
point $\mathbf{x}_0$ at $t=0$ (the integrals can also be shifted by a
finite formation time). Then we have
\begin{equation}
  \omega_c = I_1 \, , \qquad R = \frac{2I_1^2}{I_0} \, .
\end{equation}

Due to the sheer impossibility to extract detailed space-time information
on $\hat q$ it has become standard to assume that
$\hat q$ is locally proportional to a quantity whose distribution and
time-evolution is approximately known. A popular choice is the $3/4$th power
of the local energy density $\epsilon$
\begin{equation}
  \label{eq:bdmpsqhat}
  \hat q (\mathbf{x},t) = 2 K \epsilon^{3/4}(\mathbf{x},t) \, ,
\end{equation}
or the entropy density $s$. This is the parametric dependence expected
from dimensionality arguments for a fully thermalized quark gluon plasma 
\cite{Baier:2002tc}. $K$ is then treated as a fit parameter and we
expect it to be close to unity for a weakly coupled plasma and larger for
a strongly coupled system. Other choices for modeling the shape of $\hat q$
found in the literature are the temperature $T(\mathbf{x},t)$ of an equilibrated
plasma, and the density of the number of participant nucleons or 
number of binary nucleon-nucleon collisions, which are then usually
treated as time-independent medium distributions.

\subsubsection{Final State Effects: Other Developments}

We want to end the discussion of final state effects in nuclear collisions
by briefly touching upon two special topics. We have mostly focused
our attention on the production of hadrons since this is the dominant
mode of measurement at RHIC. At LHC, the calorimetric measurements of 
jets will become much more important as their energy grows much
above the background event. The advent of fast and reliable jet algorithms
for high-multiplicity environments \cite{Cacciari:2005hq,Salam:2007xv,
Cacciari:2008gp} has added to the excitement.

One possible way of modeling jets is through advanced Monte Carlo simulations
of medium induced gluon radiation that does not just focus on the leading
parton but tracks the evolution of the entire parton shower. Monte Carlo jet
quenching modules like PYQUEN \cite{Lokhtin:2005px}, Q-PYTHIA 
\cite{Armesto:2009fj}, JEWEL \cite{Zapp:2008gi} , YaJEM \cite{Renk:2008pp} 
and MARTINI \cite{Schenke:2009gb} have made progress in that direction. 
Another approach is the study of jet shapes in heavy ion environments
that track the flow of energy through cones of given radius 
\cite{Vitev:2008rz}. These studies will be much more flexible and
comprehensive than leading hadron studies as they give much better answers
to the question of ``energy loss'' which is, of course, rather a 
redistribution of energy than a real loss.

Another interesting field that has emerged in recent years is the study
of hadron chemistry in jets. A heavy ion environment not only redistributes
energies of energetic particles, it can also lead to significant changes
in the relative abundances of hadrons. This can either happen through
profound changes in the way hadronization works in high multiplicity 
environments \cite{Sapeta:2007ad}, see also the discussion on quark
recombination in the next section, or through the exchange of particles
with the quark gluon plasma which leads to a phenomenon termed jet 
conversion \cite{Ko:2007zz,Liu:2008zb,Liu:2008bw,Liu:2008kj}.
Jet conversions would increase the number of protons and kaons relative
to pions in nuclear collisions vs $p+p$ collisions. In the former case
constantly occurring conversions between quarks and gluons wash 
out the different
color factors for their respective suppression, leading to equal
suppression of particles from light quark and gluon fragmentation. In
the later case the small sample of high-$p_T$ strange quarks is 
tremendously enhanced relative to up and down quarks when quenched in 
a chemically equilibrated quark gluon plasma. In Ref.\ \cite{Liu:2008zb}
the authors predicted a factor 2 increase in the $R_{AA}$ of kaons
vs pions which seems to bee seen in preliminary STAR data \cite{Ruan:2010qg}.

Conversions have been particularly well understood in the case of photons
\cite{Fries:2002kt,Fries:2005zh,Turbide:2005fk} and dileptons 
\cite{Srivastava:2002ic,Turbide:2006mc}. Both induced photon bremsstrahlung
\cite{Arnold:2001ba,Zakharov:1996fv} and elastic annihilation and Compton
scattering with the medium ($q+\bar q \to \gamma +g$ and $q+g\to \gamma+q$
resp.) lead to photon yields that are comparable with other sources
(thermal, hard direct, vacuum bremsstrahlung) at intermediate transverse
momenta of a few GeV/$c$. Both the evolution equations of the Higher
Twist formalism and the rate equation of AMY can accommodate more channels
and ``flavor'' changing processes in a straight forward manner to study
these effects.

\subsection{The Perturbative Approach: Critique and Challenges}

Despite the large amount of effort put into the development of a
perturbative description of hadron production in heavy ion collisions, there
are uncertainties remaining about the exact nature of jet-medium interactions
in the kinematic and temperature regimes important at RHIC. We will discuss in 
more detail in Sec.\ \ref{sec:results} below that the four approaches
described here generally fare well in describing RHIC data, but they can
reach very different quantitative conclusions about the quenching strength 
$\hat q$. 
This should not come as a big surprise since the approaches differ in some
of their basic assumptions, and there are large uncertainties in
modeling hard probes beyond the calculation of an energy loss rate for
a quark or gluon. 

Currently the big picture can be summarized as follows:
perturbative calculations under various assumptions are compatible with RHIC
data, but the constraints are insufficient to rule out any of the models. The
experimental constraints
are also insufficient to completely exclude non-perturbative mechanisms of jet quenching.
Calculations using the AdS/CFT correspondence to model strongly interacting
QCD \cite{Herzog:2006gh,Liu:2006ug,Gubser:2006bz} can describe the
same basic phenomenology. Most likely this challenge to perturbative QCD can only
be answered at LHC. The extrapolation of jet quenching to larger jet energies
is significantly different in strong coupling and perturbative scenarios 
\cite{Horowitz:2007su}. It is also possible to imagine a small regime of
strong non-perturbative quenching around $T_c$ together with perturbative
quenching at  higher temperatures. Such mixed scenarios might be hard to
distinguish experimentally. One such picture was recently explored by Liao and Shuryak 
\cite{Liao:2008dk}. They found that a ``shell''-like quenching profile in
which quenching is enhanced around $T_c$ can give better simultaneous 
fits to single hadron suppression and elliptic flow.

\begin{table}[tb]
\begin{center}
\begin{minipage}{14cm}
\caption{pQCD-based energy loss models: This table summarizes some of the key 
assumptions of the four perturbative calculations discussed here. The models
differ with respect to the medium (thermalized, perturbative), kinematics,
scales ($E$ = energy of the parton, $k_T$ = transverse momentum of the
emitted gluon, $\mu$ = typical transverse momentum picked up from the medium,
$T$ = temperature, $\Lambda$ = typical momentum scale of a (non-thermalized)
medium, $x$ = typical momentum fraction of the emitted gluon), and the
treatment of the resummation.}
\label{tab:eloss}
\end{minipage}
\end{center}
\renewcommand{\arraystretch}{3}
\begin{center}
\begin{tabular}{|c|c|c|c|} 
\hline
Model & Assumptions about the Medium & Scales & Resummation \\ \hline\hline
GLV & \begin{minipage}[b]{8cm}static scattering centers (Yukawa), opacity
  expansion\end{minipage} & $E \gg k_T \sim \mu$,
$x \ll 1$ & Poisson
\\ \hline 
ASW & \begin{minipage}[b]{8cm}static scattering centers, multiple soft scattering (harmonic oscillator
approximation)\end{minipage}  & $E \gg k_T \sim \mu$, $x \ll 1$ & Poisson
\\ \hline
HT & \begin{minipage}[b]{8cm} arbitrary matrix element at scale $\Lambda$ (thermalized or non-thermalized medium) \end{minipage}  &
$E \gg k_T \gg \Lambda \sim \mu$ &  DGLAP
\\ \hline
AMY & \begin{minipage}[b]{8cm}perturbative, thermal, $g << 1$ (asymptotically large $T$)\end{minipage}  & $ E > T \gg
gT \sim \mu $ & Fokker-Planck
\\ \hline
\end{tabular}
\end{center}
\end{table}

Additional uncertainties come from a variety of issues regarding the
details of the phenomenological modeling:
\begin{itemize}
\item {\it The initial state:} Jets will interact with their
  environment before a quark gluon plasma is fully formed. The time
  dependence of $\hat q$ during the first fm/$c$ of the collisions is highly
  uncertain, in particular if initial interactions are dominated by coherent gluon
  fields \cite{Gelis:2010nm,Lappi:2010ek}.  Some calculations set an ad-hoc
  start time or use different extrapolations to small times.
  One estimate of uncertainties can be found in \cite{Armesto:2009zi} and is
  discussed in more detail in Fig.\ \ref{fig:armesto} below.
\item {\it Fireball evolution:} Wildly different fireball parameterizations used in
  jet quenching calculations can be found in the literature up to this
  day. The correct longitudinal and transverse expansion with the correct
  cooling rate have to be taken into account. Recently comparisons of
  different calculations using the same underlying fireball calculated from
  hydrodynamics have become available \cite{Bass:2008rv}. However, as we will discuss in
  detail in the Sec.\ \ref{sec:hydro}, uncertainties remain in hydrodynamic
  calculations as well which are transferred to hard probes when hydrodynamics 
  is used as a background.
 \item {\it The hadronic phase:} The models referenced here deal with parton
  energy loss in a partonic medium. Clearly a jet will also interact with a
  surrounding hadronic medium. Some models could in principle deal with this
  situation (e.g.\ the higher twist approach only needs the jet to be
  dominated by a sufficiently high energy parton), others have to fail (like
  the AMY approach). But none of them addresses the question of fully formed
  high energy hadrons in a jet interacting with a hadronic medium.
  Shower simulations with full space-time evolution might be helpful to
  constrain at least the size of the problem from the partonic side.
\item {\it Event-by-event fluctuations:}
  Not much is known from experiment about spatial fluctuations in the
  fireball, but clearly we should expect a fireball to exhibit a certain
  degree of inhomogeneity as suggested by many models of initial 
  nucleus-nucleus interactions. Compared to quenching in a
  smooth, average fireball, event-by-event quenching can lead to considerably
  different results for hadron suppression and elliptic flow \cite{Rodriguez:2010di}.
\item {\it Back-reaction from the medium:}
  While the medium modifies jets, jets on the other hand modify the
  surrounding medium by transferring energy and momentum. The heating
  of the medium can be considerable \cite{Fochler:2010wn}, and a variety of
  shock phenomena can occur
  \cite{CasalderreySolana:2004qm,Neufeld:2008fi,Betz:2008js}. 
  Clearly, a comprehensive approach will consider both the jet and
  the medium as variable and would not fix one or the other as a background or
  boundary condition. In particular, if part of the energy (or maybe most of the energy
  for some jets) thermalizes, this most likely proceeds through
  non-perturbative channels which are not included in either of the models
  discussed here.
\end{itemize}
There are no systematic studies of all uncertainties together. A pessimistic 
estimate of their compounded effect would be a factor 3-5 uncertainty in the extraction
of $\hat q$ from RHIC data.

Let us finally revisit the four approaches to calculate leading particle
energy loss discussed here. Why do they lead to similar qualitative,
but sometimes quite different quantitative results? We have discussed
the underlying assumptions of each model in its respective section.
We summarize them once more in Tab.\ \ref{tab:eloss} in terms of the 
different ways the medium is modeled, the hierarchy of scales and the
way resummation is handled. Two key points are the different assumptions
about the transverse momentum $k_T$ of the emitted gluon vs the transverse 
momentum $\mu$ picked up from the medium, and the philosophy of multiple
soft emissions vs an opacity expansion (with single, somewhat larger
transverse kicks). The full solution of the problem is quite complex, even in
a fully perturbative medium. We refer the interested reader to the recent
assessment by Arnold \cite{Arnold:2009mr} and the review by Majumder and
Van Leeuwen \cite{Majumder:2010qh}.
It might be a while until more comprehensive calculations are available, but
efforts in this direction are under way, e.g.\ within the TECHQM and JET
collaborations. Simplistic assumption have to be improved and narrow 
kinematic regimes have to be widened.

%%%%%%%%%%%%%%%%%%%%%%%%%%%%%%%%%%%%%%%%%%%%%%%%%%%%%%%%%%%%%%%%%%%%%%%

\section{Success of Hydrodynamic Models at RHIC
\label{sec:hydro}}

Since Landau~\cite{BeLa56} and Bjorken~\cite{Bjorken:1982qr} proposed 
the idea to apply hydrodynamics to the production of
particles in high energy collisions, it has evolved into one of 
the most useful phenomenological models for our understanding of 
high energy heavy ion collisions.   
Starting with Landau's hydrodynamic description and Bjorken's scaling solution, 
various kinds of implementations of hydrodynamics were developed to understand 
experimental data from the Alternating Gradient Synchrotron (AGS) and 
and the Super Proton Synchrotron (SPS).   
Hydrodynamics and hydro-inspired models (e.g.\ the Blast-Wave Model 
\cite{Schnedermann:1993ws}) give us reasonable explanations for a 
large amount of experimental data: single particle spectra (with respect 
to transverse momentum $P_T$ and pseudorapidity $\eta$), two particle 
correlations, collective flow, and electromagnetic probes.  
However they did not appear to work for some aspects of anisotropic flow, e.g.\ 
directed flow $v_1$ and anisotropic flow $v_2$. 
For example, the rapidity dependence of directed flow of charged pions is 
different from that of protons \cite{Appelshauser:1997dg,Alt:2003ab}. 
The charged pion $v_1$ decreases with rapidity, but the proton $v_1$ increases 
(see Figs.~18 and 19 in Ref.~\cite{Alt:2003ab}). 
This difference between pions and protons is difficult to understand 
from collectivity arguments which is one of the crucial features of 
hydrodynamics. To explain this interesting behavior in detail, hydrodynamics
may be too simplistic and additional effects may play a role.  
Therefore transport models in which more complicated dynamics of the
underlying theory are included fare better for directed flow. 
Another hint comes from the fact that hydrodynamic models routinely
overestimate the size of elliptic flow at SPS energies \cite{Alt:2003ab}.
This suggests that the system does not completely equilibrate. Hence 
the validity of hydrodynamics at SPS energies is not very clear.
This led to the pre-RHIC view that hydrodynamic models were rather 
simple-minded phenomenological tools.  

All of this changed dramatically after the first experimental results from
RHIC came out in 2000. Hydrodynamic models could naturally explain the 
unexpectedly large elliptic flow at RHIC compared to that at SPS 
\cite{Kolb:2000fha,Huovinen:2001cy}. The success of hydrodynamics at RHIC, 
together with observations of jet quenching in the medium many times larger
than that in cold nuclear matter, indications for the existence of a
Color Glass Condensate and the success story of recombination models, 
are cited as the main results from the RHIC program, as evidence for
production of a quark gluon plasma with deconfined partons, and for
the conjecture that this QGP is strongly coupled (a ``sQGP'') and in fact 
may be the most perfect liquid ever seen \cite{white_papers,QGP}. 

In the wake of this success hydrodynamics has become the most promising
tool to describe most of the expansion and cooling process of the bulk
of the matter created in heavy ion collisions at RHIC.
However, at the same time limitations of hydrodynamic models, old ones and
novel ones, were found: e.g.\ the failure to describe experimental data on 
two-particle correlations, or elliptic flow as a function of pseudorapidity. 
This suggests that the assumption of perfect fluidity is not valid for 
the entire fireball at RHIC. At least in the hadronic phase and in the
final state viscous effects can not be neglected \cite{Hirano:2005wx}.  
Since the perfect fluidity hypothesis is tested through precision analyses 
of elliptic flow \cite{Bhalerao:2005mm} one needs to overcome this obstacles
with hybrid models that couple hydrodynamics to hadron-based transport
models and through the development of viscous hydrodynamic codes.

Both of those avenues have undergone remarkable developments in recent years.
In particular our understanding of second order viscous, relativistic 
hydrodynamics has increased tremendously over the past few years.
While the extraction of the equation of state for quark gluon plasma and
the phase transition used to be at the forefront of hydrodynamic modeling,
the extraction of transport coefficients, like shear and bulk viscosity,
of the hot and dense matter created at RHIC has been added to the main goals. 
In this section we will review the basic principles and numerical 
concepts of hydrodynamics together with some recent progress. We will 
then discuss applications of hydrodynamic models at RHIC.

\subsection{\it Basics of Relativistic Hydrodynamics}

\subsubsection{The Framework of Ideal Hydrodynamics} 
%(ideal hydro, viscous hydro)
Let us start with the ideal relativistic hydrodynamic equations of motion. 
In ideal hydrodynamics thermalization is perfect and there is a well-defined
local rest frame for each fluid cell. In that local rest frame, the 
energy-momentum tensor of a volume element of the fluid 
(where Pascal's law works) is given by 
\be
\tilde{T}^{\mu\nu}(x)=
\left (
\begin{array}{cccc}
\epsilon (x) & 0 & 0 & 0\\
0 & p(x) & 0 & 0 \\
0 & 0 & p(x) & 0 \\
0 & 0 & 0 & p(x) \\
\end{array}
\right )
\ee
where $\epsilon(x)$ and $p(x)$ are energy density and pressure, respectively,
as functions of the space-time point $x^\mu$ of the fluid cell. 
Introducing the four-velocity $u^\mu(x)$ of each fluid cell in the
lab frame we can boost the local energy-momentum tensor into the
laboratory frame which gives us the well-known result  
\be 
  T^{\mu \nu}(x)= [ \epsilon (x) + p(x) ] u^\mu (x) u^\nu  (x) - p(x) g^{\mu
    \nu} \, . 
\ee
The motion of the fluid is simply described by the equations
of energy and momentum conservation, 
\be
\partial_{\mu}T^{\mu \nu}(x) = 0 \, , 
\label{eq-hydro}
\ee
from which the entropy current conservation, 
\be
\partial_\mu (su^\mu)=0
\label{eq-entropy}
\ee
is derived. 
Other conserved currents $j^\mu(x)$ besides energy and momentum ---
most importantly the baryon current $j_B^\mu(x)$ --- can be taken into 
account by imposing the conservation laws
\be
  \partial_{\mu}j^\mu(x) = 0 \, .
  \label{eq-net_baryon}
\ee
In ideal hydrodynamics each of these conserved currents can be written
as
\begin{equation}
  j^\mu(x) = n(x) u^\mu(x)
\end{equation}
where $n(x)$ is the density of the corresponding conserved charge in the
local rest frame of a fluid cell.
(\ref{eq-hydro}) and (\ref{eq-net_baryon}) are the equations of motion
that need to be solved. An equation of state (EoS) $p=p(\epsilon)$ is the
last equation needed to close the system of equations. It is the only
place where the underlying dynamics of the system comes into play.
In practical applications the only current usually considered in heavy 
ion physics is the net-baryon current $j_B^\mu$. Even that current is often 
negligible at RHIC and LHC. 

The equation of state gives us direct information about the QCD phase diagram. 
This direct link to the main goal of the RHIC program is one of the
outstanding features of hydrodynamic models. 
Most hydrodynamic computations use an EoS with a first-order 
phase transition based on the Bag Model. 
However, recent lattice QCD results suggest that the phase transition 
at low baryon densities is rather a smooth crossover \cite{Aoki:2006we,Petreczky:2009at}. 
Since then parameterizations of the EoS from lattice QCD have become
fashionable. 
We will describe the recent developments in this area 
in more detail in Sec.\ \ref{sssec-hydrodnamic}.

\subsubsection{Dissipative Corrections}

When we start to include effects of dissipation into relativistic 
hydrodynamics, we are confronted with a rather complicated situation.  
One of the difficulties is that a naive introduction of viscosities causes
the first order theory (i.e.\ first order in gradients) to exhibit 
acausalities. Heat might propagate instantaneously because of the
parabolic character of the equations. This problem is unique to the
relativistic generalization of viscous hydrodynamics.
In order to avoid this problem, second order terms in heat flow and 
viscosities have to be included in the expression for the entropy
\cite{Muller,Israel,IsSt1,IsSt2,GO1,GO2,Ott}, but the systematic treatment 
of these second order terms is difficult.
Although there is remarkable progress toward the construction of a 
fully consistent relativistic viscous hydrodynamic theory there 
are still ongoing discussions about the correct formulation of 
the equations of motion and about the appropriate numerical procedures 
\cite{Teaney:2009qa}.

The basic tenet that has to be given up in dissipative hydrodynamics
is the assumption of a uniquely defined local rest frame. Away from 
equilibrium the vectors defining the flows of energy, momentum and
conserved currents might be misaligned. We can still define a local rest
frame by just choosing a velocity $u^\mu(x)$ in the lab frame. Then
the energy-momentum tensor and the conserved current take the more
general shape
\begin{align}
    T^{\mu \nu}(x)& = [ \epsilon(x) + p(x)+\Pi(x) ] u^\mu(x) u^\nu(x)  - 
      [p(x)+\Pi(x)] g^{\mu \nu} + 2 W^{(\mu} u^{\nu)} + \pi^{\mu\nu} \, , \\
    j^\mu (x) &= n(x) u^\mu + V^\mu \, , 
\end{align}
where $V^\mu$ and $W^\mu$ are corrections to the flow of conserved
charge and energy that are orthogonal to $u^\mu$ and $T^{\mu\nu} u_\nu$ resp.,
$\pi^{\mu\nu}$ (with the orthogonality conditions 
$u_\mu \pi^{\mu\nu}=\pi^{\mu\nu}u_\nu =0$) is the symmetric, trace-less shear 
stress tensor, and $\Pi$ is the bulk pressure. $(\ldots )$ around indices
indicate symmetrization.
Usually $u^\mu$ is chosen to define one of two standard frames: the Eckart 
frame where the velocity is given by the physical flow of net charge 
(then $V^\mu = 0$), or the Landau frame where the velocity is given by
the energy flow (then $W^\mu = 0$).  
We refer the reader to the article by Muronga and Rischke for further
details \cite{Muronga:2004sf}.

At first order the new structures have to be proportional to gradients
in the velocity field $u^\mu$, and only three proportionality constants appear
in these relations: the shear viscosity $\eta$, the bulk viscosity $\zeta$
and the heat conductivity $\kappa$. These transport coefficients are the
fundamentally new quantities in dissipative hydrodynamics. With the usual
definitions the first order relations in the Landau frame are \cite{Muronga:2004sf}
\begin{align}
  \Pi &= -\zeta \nabla_\mu u^\mu   \\
  q^\mu &= - \kappa \frac{nT^2}{e+p} \nabla^\mu \frac{\mu}{T} \\
  \pi^{\mu\nu} &= 2 \eta \nabla^{<\mu} u^{\nu>} \, .
\end{align}
Here $q^\mu = -(\epsilon+p)/n \, V^\mu$ is the heat flow in the
Landau frame, $\nabla^\mu = (g^{\mu\nu} -u^\mu u^\nu) \partial_\nu$ is the covariant
derivative orthogonalized to the flow vector, $T$ and $\mu$ are temperature
and chemical potential for the conserved current resp., and $\langle \cdots \rangle$
refers to a symmetrization of indices with the trace subtracted. 
The entropy current $S^\mu$ receives additional contributions beyond the
equilibrium term $s u^\mu$ and one can show that with all
three transport coefficients positive the entropy is strictly non-decreasing,
$\partial_\mu S^\mu \ge 0$.

At second order many more new parameters, related to relaxation phenomena,
appear.  Currently, most viscous hydrodynamic calculations use the relativistic 
dissipative equations of motion that were derived phenomenologically 
by Israel and Stewart \cite{IsSt2} and variations of those, while some use the
method by \"Ottinger and Grmela \cite{GO1,GO2,Ott}, see e.g.\ \cite{Dusling:2007gi}.
Recently, a second-order viscous hydrodynamics from AdS/CFT 
correspondence was derived \cite{Baier:2007ix}, as well as a set of 
generalized Israel-Stewart equations from kinetic theory via Grad's 14-momentum 
expansion which have several new terms \cite{Betz:2009zz}. 
On the other hand however, a stable first-order relativistic dissipative 
hydrodynamic scheme was proposed on the basis of 
renormalization-group methods \cite{Tsumura:2007ji,Tsumura:2007wu}.
The discussion surrounding the appropriate (second order)
equations of motion is ongoing and a review is beyond the scope of 
this phenomenological overview. We refer the reader to the original articles 
cited in this subsection for further guidance.

In heavy ion physics, most of the focus in viscous hydrodynamics has been
on the shear viscosity, and its ratio with the entropy density, $\eta/s$.
Interesting initial investigations of the effects of bulk viscosity have begun 
\cite{Fries:2008ts,Song:2009rh}, while heat conductivity still has not been 
investigated systematically in connection with RHIC data.

\subsubsection{Numerical Calculation}

In most ideal hydrodynamic models employed at RHIC numerical computations
are carried out in the Eulerian formalism. For shock-capturing schemes the
SHASTA \cite{SHASTA} and RHHLE \cite{Schneider:1993gd} algorithms were 
developed. The SHASTA algorithm is the most widely spread numerical
implementation for both ideal and viscous hydrodynamics. 
On the other hand, the NEXSPHERIO code \cite{Andrade:2006yh} is based on 
smoothed particle hydrodynamics (SPH) \cite{Aguiar:2000hw}. In order 
to describe the flow of a fluid at high energy but low baryon number  
entropy is taken as the SPH base. 
Lagrangian hydrodynamics is also used in hydro codes for RHIC 
physics \cite{Nonaka:2006yn}. 
Lagrangian hydrodynamics has several advantages over Eulerian hydrodynamics 
when applied to ultra-relativistic nuclear collisions. 
At high energies, the initial distribution of the energy is strongly
localized in longitudinal direction due to the Lorentz contraction of the
two nuclei in the lab frame. 
In order to handle this situation appropriately a fine resolution is required in
Eulerian hydrodynamics, and computational costs become large. 
On the other hand, in Lagrangian hydrodynamics the grid moves along with
the expansion of the fluid. Therefore, one can perform the calculation 
at all stages on those lattice points which were prepared for the 
initial conditions. 
Another merit of Lagrangian hydrodynamics is the fact that it enables 
us to derive the physical information directly because it traces the 
flux of the currents. For example, the path of a volume element of fluid 
in the $T$-$\mu_B$ plane, spanned by temperature and baryon chemical potential,
can be easily followed.  This allows us to discuss directly how the transition between 
the QGP phase and the hadronic phase affects physical phenomena. 

%%%%%%%%%%%%%%%%%%%%%%%%%%%%%%%%%%%%
% viscous hydrodynamics 
For relativistic viscous hydrodynamics the numerics
becomes more complicated in terms of numerical viscosity and 
the stability of the calculation.  
Currently several numerical calculations have been implemented and 
first quantitative comparisons with experimental data become available. 
Most numerical algorithms used in viscous hydrodynamics are
based on SHASTA \cite{Song:2008si, Chaudhuri:2009hj}, but other
numerical procedures have also been explored, e.g.\ SPH \cite{Denicol:2009zz} 
and the discretization method in Ref.\ \cite{Dusling:2007gi}. 
We have only begun to explore the problems related to numerically solving causal
relativistic dissipative fluid dynamics; see e.g.\ Ref.\ \cite{Molnar:2009tx}
for a test procedure applying an algorithm to the Riemann problem.

%%%%%%%%%%%%%%%%%%%%%%%%%%%%%
%%%%%%%%%%%%%%%%%%%%%%%%%%%%%
\subsection{\it Applications to RHIC Physics}
%%%%%%%%%%%%%%%%%%%%%%%%%%%%%

\subsubsection{Hydrodynamics for Heavy Ion Collisions}

%%%%%%%%%%%%%%%%%%%%%%%%%%%%%%%%%%%%%%%%%%%%
% figure mean PT as a function of particle mass 
%%%%%%%%%%%%%%%%%%%%%%%%%%%%%%%%%%%%%%%%%%%%
\begin{figure}[tb]
\begin{center}
\begin{minipage}{14cm}
\begin{center}
\epsfig{file=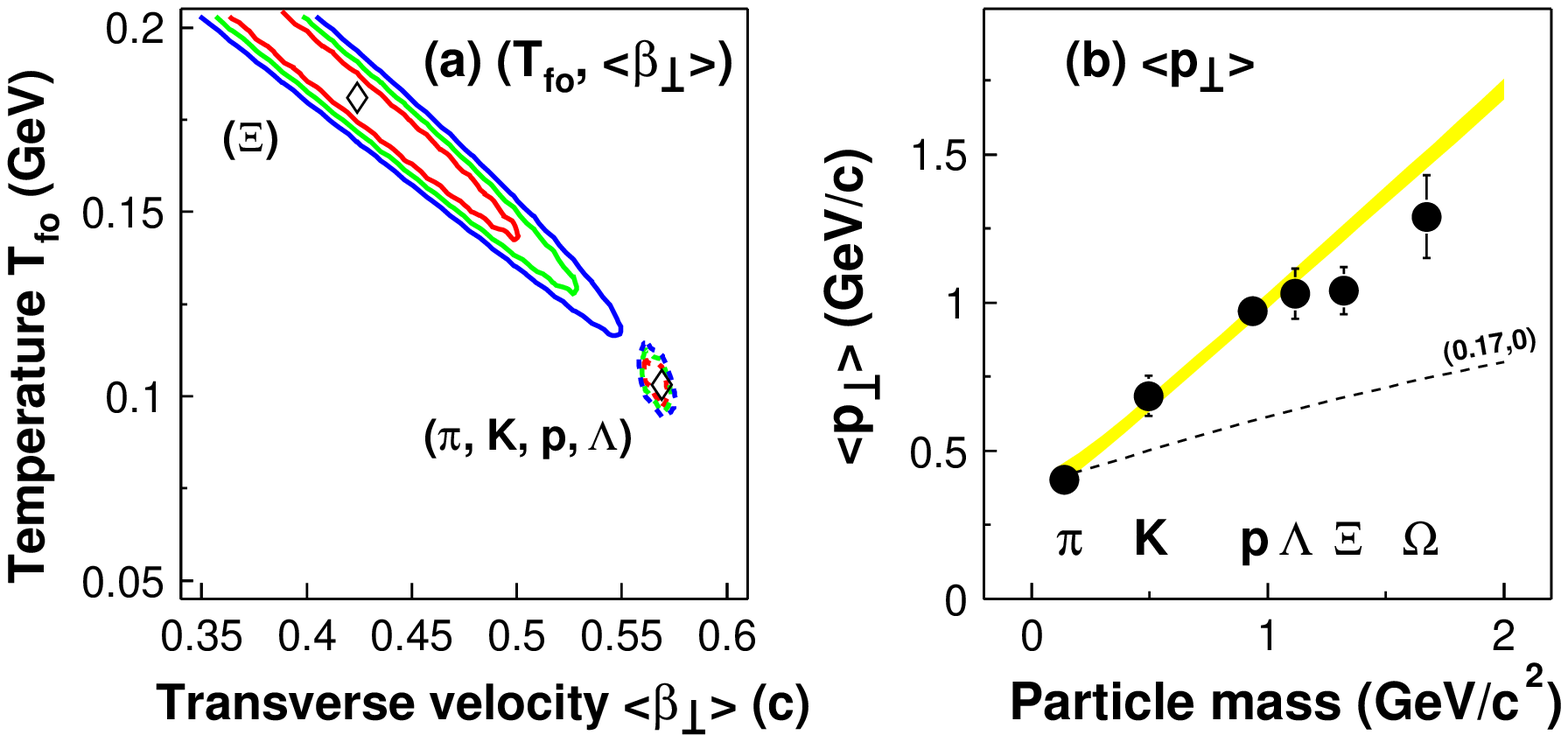,scale=0.7,bb=290
  10 535 250,clip=}
\end{center}
\caption{Mean transverse momentum $\langle P_T \rangle$ as a function 
of particle mass in central Au+Au collisions at $\sqrt{s_{NN}} = 130$ GeV.
Data are from STAR, yellow band from a hydrodynamically inspired fit,
and the dashed line represents the expected behavior from a
hydrodynamic fit at the chemical freeze-out temperature of 170 MeV at which
the radial flow is still too small.
Figure reprinted from \cite{Adams:2003fy} with permission from the American
Physical Society. 
\label{fig_meanpt_star}}
\end{minipage}

\end{center}
\end{figure}
%%%%%%%%%%%%%%%%%%%%%%%%%%%%%%%%%%%%%%%%%%%%

Let us first review why RHIC data strongly suggests a long hydrodynamic 
expansion and cooling phase of the bulk matter created in nuclear collisions.
We can see the clear signals of bulk collectivity in experimental data
most convincingly in data on the mean transverse momentum 
$\langle P_T \rangle $ as a function of observed particle masses and in
measurements of elliptic flow $v_2$ at RHIC.
In Fig.\ \ref{fig_meanpt_star} we see data on $\langle P_T \rangle$ for
several hadron species measured in Au+Au collisions at $\sqrt{s_{NN}}=130 $
GeV at RHIC, together with a band resulting from the hydrodynamically 
inspired fit to the $\pi$, $K$, $p$, and $\Lambda$ data \cite{Adams:2003fy}.
The observed mass dependence clearly shows that radial flow exists and that
it dominates the motion of bulk particles. Small discrepancies between
data and theory exist and will be discussed in Sec.\ \ref{sec:results}.

The second remarkable success of hydrodynamics is the explanation of the
large elliptic flow $v_2$ as a function of $P_T$ seen at RHIC, and
the characteristic dependence on mass, as shown in Fig.\ \ref{fig_v2_pt}.
To realize such strong elliptic flow in a leading order parton cascade model,  
unrealistic large cross sections are needed \cite{Molnar:2004yh},  
which clearly supports a hydrodynamic interpretation.  
The hydrodynamical analyses also indicate that thermalization of the matter
at RHIC is achieved very rapidly. Most estimates put the equilibration
time between 0.6 and 1.0 fm/$c$  after the collision \cite{white_papers}.
$v_2$ is a wonderful quantity to analyze with hydrodynamic models,
even deviations from the expected equilibrium values can 
provide information on viscosities \cite{Bhalerao:2005mm,Drescher:2007cd}.

Hydrodynamics assumes thermal equilibrium and very short
mean free paths of particles. We know that these assumptions are broken
at least at very early times and at the latest times in collisions. They
might also be broken at the boundaries of the system where densities are
smaller, leading to corona effects. This means that hydrodynamic calculations
have to be combined with some calculations or parameterizations of the
initial state, and one has to be careful about the treatment of the 
freeze-out, the conversion of hydro fluid cells back into particles.
We can further infer from the $P_T$-dependence of physical observables 
such as elliptic flow that thermalization of particles starts to
be incomplete at high $P_T$. At RHIC hydrodynamic models generally work 
very well up to $P_T \sim 2$  GeV/$c$. Above that threshold novel effects, like quark 
recombination, come into play at intermediate values of $P_T$. At the 
largest values, above 6 GeV/$c$, perturbative QCD production dominates. 
Of course, the hydrodynamic regime comprises about 99\% of the particles 
produced in a typical Au+Au collision at RHIC.

%%%%%%%%%%%%%%%%%%%%%%%%%%%%%
\subsubsection{Initial Conditions}
\label{sec:init}

The hydrodynamic equations of motion need initial conditions for
all their dynamic variables, which are then evolved forward in
time by the solutions.
These initial conditions are outside of the framework of hydrodynamic models
and have to be determined by other means. Physically, they are
determined by the processes during the initial collision of the nuclei
and the approach to equilibrium which is eventually reached at a time 
$\tau_0$. As we already discussed this equilibration time has been 
estimated to be rather short. However, in practice it should be
a parameter since the precise equilibration mechanisms are still 
under debate and calculations of the initial conditions at the start 
of the equilibrated plasma phase have been elusive so far 
\cite{Baier:2000sb,re-early-therm}.

Historically, parameterized initial conditions for entropy densities 
(or alternatively the energy densities) and the net baryon densities have
been used \cite{Nonaka:2006yn,Kolb:2001qz,Hirano:2002ds,Teaney:2000cw}.  
In the transverse plane these distributions are usually parameterized 
based on Glauber-type models of nuclear collisions. In longitudinal 
direction often initial distributions inspired by Bjorken's scaling solution
are used. Then few parameters remain in these initial conditions, 
such as the maximum values of the energy or entropy density, and net 
baryon density. They are usually fixed by comparison with experimental data on
single particle rapidity and transverse momentum spectra.

%%%%%%%%%%%%%%%%%%%%%%%%%%%%%%%%%%%%%%%%%%%%%%%%%%%%%%%
\begin{figure}[tb]
%\begin{center}
\begin{minipage}[h]{10cm}
\epsfig{file=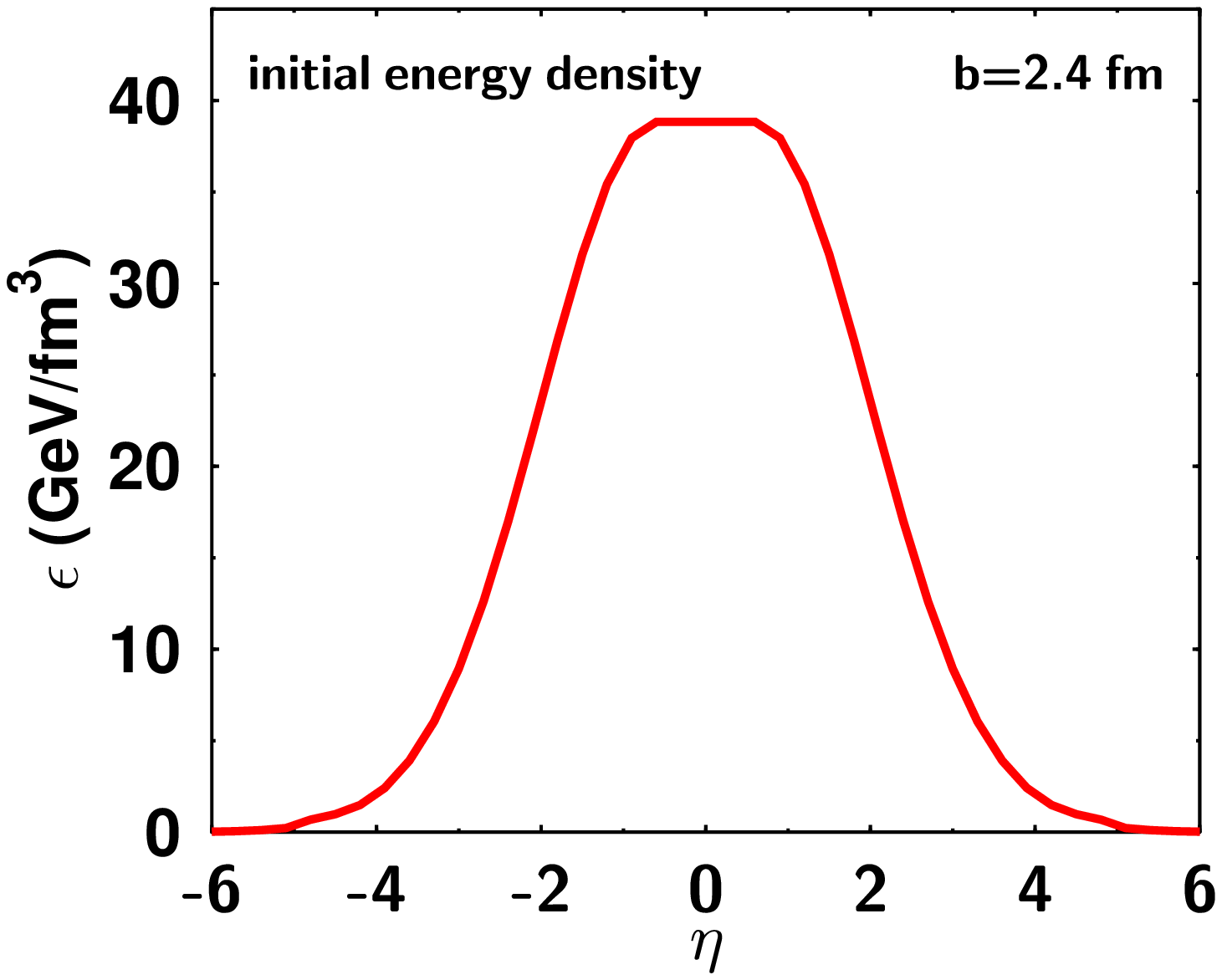,scale=0.5}
\end{minipage}
\hspace{-1cm}
\begin{minipage}[h]{10cm}
\epsfig{file=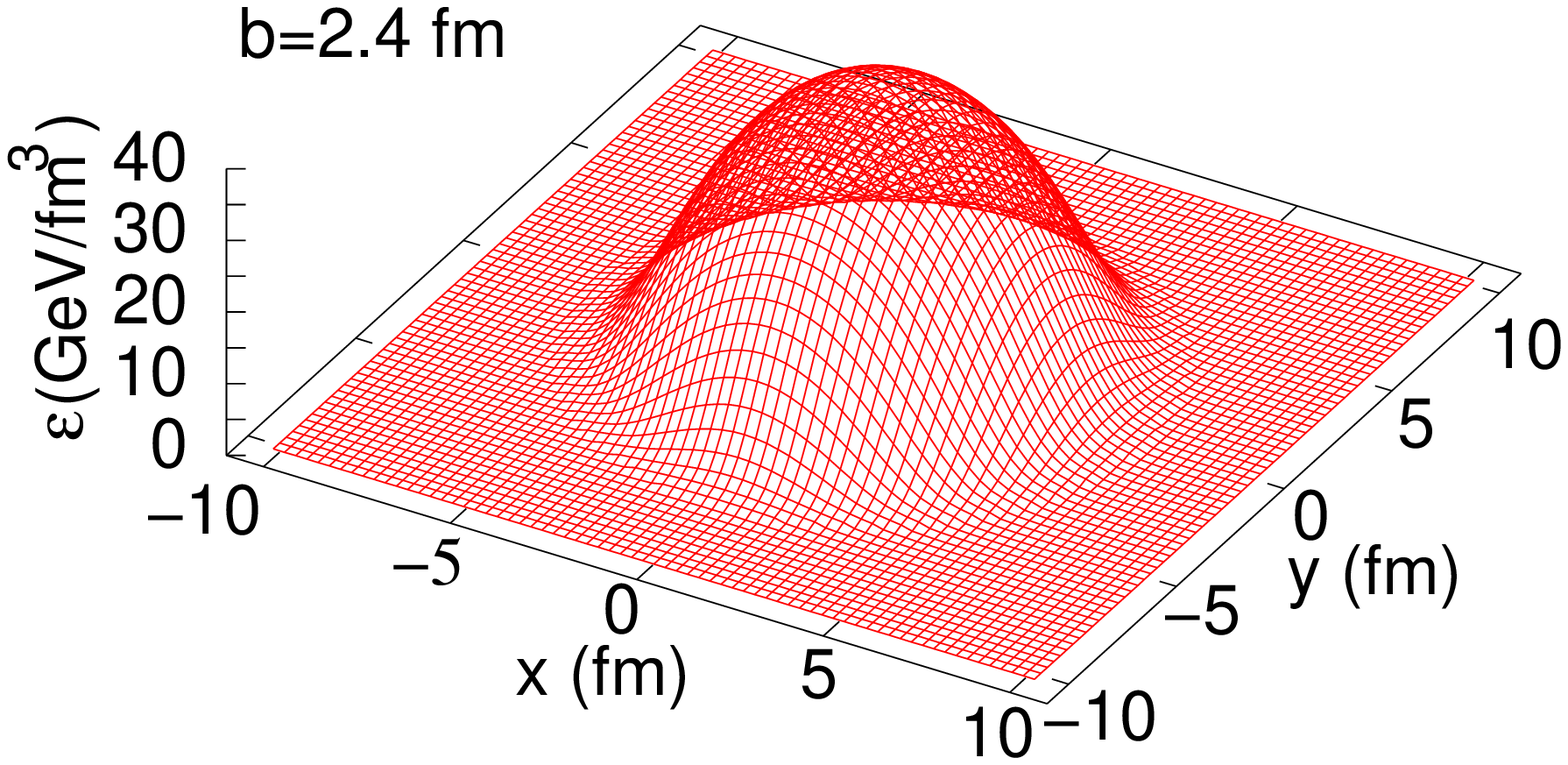,scale=0.5}
\end{minipage}
\caption{Initial energy density $\epsilon(\tau_0,\eta,x,y)$ from a Glauber 
model in longitudinal direction (left panel) and in transverse direction 
(right panel)) for central Au+Au collisions. Figures taken from
\cite{Nonaka:2006yn}. 
\label{fig:initial1}}
%\end{center}
\end{figure}
%%%%%%%%%%%%%%%%%%%%%%%%%%%%%%%%%%%%%%%%%%%%%%%%%%%%%%%

Let us discuss a straight forward set of energy density and net baryon 
density distributions that give reasonable results. We assume a factorization
into a longitudinal profile $H(\eta)$ and a transverse profile $W(x,y;b)$
which are given by  
\begin{eqnarray}
  \epsilon(x,y,\eta)& = & \epsilon_{\rm max}W(x,y;b)H(\eta), \nonumber   \\
  n_B(x,y,\eta)& = & n_{B{\rm max}}W(x,y;b)H(\eta) \, .
\end{eqnarray}
Here the maximum values $\epsilon_{\rm max}$ and $n_{B{\rm max}}$ are
parameters, $b = |\mathbf{b}|$ denotes the impact parameter of the 
collision, and $\eta$ is the space-time rapidity.
The longitudinal distribution can be parameterized by  
\begin{equation}
  H(\eta)=\exp \left [ -(|\eta|-\eta_0)/2 \sigma^2_\eta  \right ] 
  \theta (|\eta|-\eta_0) 
  +  \theta (\eta_0-|\eta|) , 
\end{equation}
where parameters $\eta_0$ and $\sigma_\eta$ can also be determined 
by comparison with experimental data on single particle 
distributions.
The function $W(x,y;b)$ in the transverse plane is determined by a
superposition of the density of wounded nucleons --- characteristic for
soft particle production processes --- and the density of binary collision 
 --- characteristic for hard particle production processes 
\cite{Kolb:2001qz}
\begin{equation}
  W(x,y;b)=w\frac{d^2N_{\rm BC}}{ds^2} +(1-w) 
  \frac{d^2N_{\rm WN}}{ds^2} \, .
\end{equation}
%This function is normalized by $W(0,0;0)$.    
The density of wounded nucleons is given by 
\begin{equation}
\frac{d^2N_{\rm WN}}{ds^2}
  = T_A(\vba) \left( 1 - e^{-T_B(\vbb) \sigma}\right)
  + T_B(\vbb) \left( 1 - e^{-T_A(\vba) \sigma}\right) 
\end{equation}
where $\vba=\vs + \mathbf{b}$, $\vbb=\vs-\mathbf{b}$, and 
$\sigma \approx 42$ mbarn is the total nucleon-nucleon cross section 
at $\sqrt{s_{NN}}=200$ GeV \cite{Adler:2003ii}.    
$T_A$ is the nuclear thickness function of nucleus $A$, 
\begin{equation}
T_A(\vs)=\int dz \rho_A(z, \vs), 
\end{equation}
where the nuclear density $\rho_A(z, \vs)$ can e.g.\ be taken to be of
Woods-Saxon shape  
\begin{equation}
\rho_A(r) = \rho_0 \frac{1}{1 + e^{(r-R_A)/a}}.
\label{Eq-Woods-Saxon}
\end{equation}
In Eq.~(\ref{Eq-Woods-Saxon}) appropriate parameters $a$, $R_A$, and $\rho_0$ 
are 0.54 fm, 6.38 fm and 0.1688 fm$^{-3}$, respectively \cite{Adler:2003ii}.
On the other hand, the distribution of the number of binary 
collisions is given by  
\begin{equation}
\frac{d^2N_{\rm BC}}{ds^2} = \sigma T_A(\vba) T_B(\vbb). 
\end{equation}
The weight factor $w$ can be set to 0.6, consistent with experimental data.
Figure \ref{fig:initial1} shows examples of initial longitudinal 
and transverse profiles.
Obviously the parameters in initial conditions are truly additional
uncertainties and make quantitative predictions with hydrodynamic models 
more difficult.

As a straight forward solution one can choose to set the initial 
longitudinal flow to Bjorken's scaling solution \cite{Bjorken:1982qr}, 
and one can set the initial transverse flow to zero. This simplest 
initial flow profile will serve as the basis for all further investigation 
here. 
The possibility of an initial transverse flow was discussed e.g.\
by Kolb and Rapp \cite{Kolb:2002ve}. The initial flow improves the 
results for $P_T$-spectra and reduces the anisotropy. Utilizing a 
parameterized evolution model it has been pointed out that a Landau-type 
initial condition with complete longitudinal compression and vanishing
initial flow is favorable for the description of the Hanbury Brown-Twiss (HBT) 
correlation radii \cite{Renk:2004yv,Renk:2003gn}. 
This suggests that HBT analyses may be a sensitive tool for the determination 
of the initial longitudinal flow.
As we will discuss in detail in the result section, hydrodynamic calculations
during the early RHIC years did show very bad agreement with experimental 
data, especially for the ratio of $R_{\rm out}/R_{\rm side}$, leading to the
notion of an HBT puzzle \cite{HeKo02}.

Let us come back to the the apparent early thermalization times found
at RHIC. Usually it is said that small initial times $\tau_0$ are 
needed to describe the elliptic flow data as elliptic flow builds up 
at the earliest stage of expansion when the eccentricity of the fireball is
largest \cite{Kolb:2000fha,Huovinen:2001cy}. 
However, we note that with suitable sets of initial conditions and 
freeze-out temperatures in fact a larger initial proper time 
is also compatible with data. Luzum and Romatschke show that three very
different sets of initial and freeze-out temperature ($T_i$,$T_f$) ---
$(0.29, 0.14)$ GeV with $\tau_0=2$ fm, $(0.36, 0.15)$ GeV  
with $\tau_0=1$ fm, and $(0.43, 0.16)$ GeV with $\tau_0=0.5$ fm ---
provide almost identical differential elliptic flow in a viscous hydrodynamic 
calculation \cite{Luzum:2008cw}.
This suggests that better constraints on initial conditions 
are indispensable to avoid wrong conclusions from comparisons of hydrodynamic 
calculations with experimental data.

%%%%%%%%%%%%%%%%%%%%%%%%%%%%%%%%%%%%%%%%%%%%%%%%%%%%%%%
\begin{figure}[tb]
%\begin{center}
\begin{minipage}[h]{10cm}
\epsfig{file=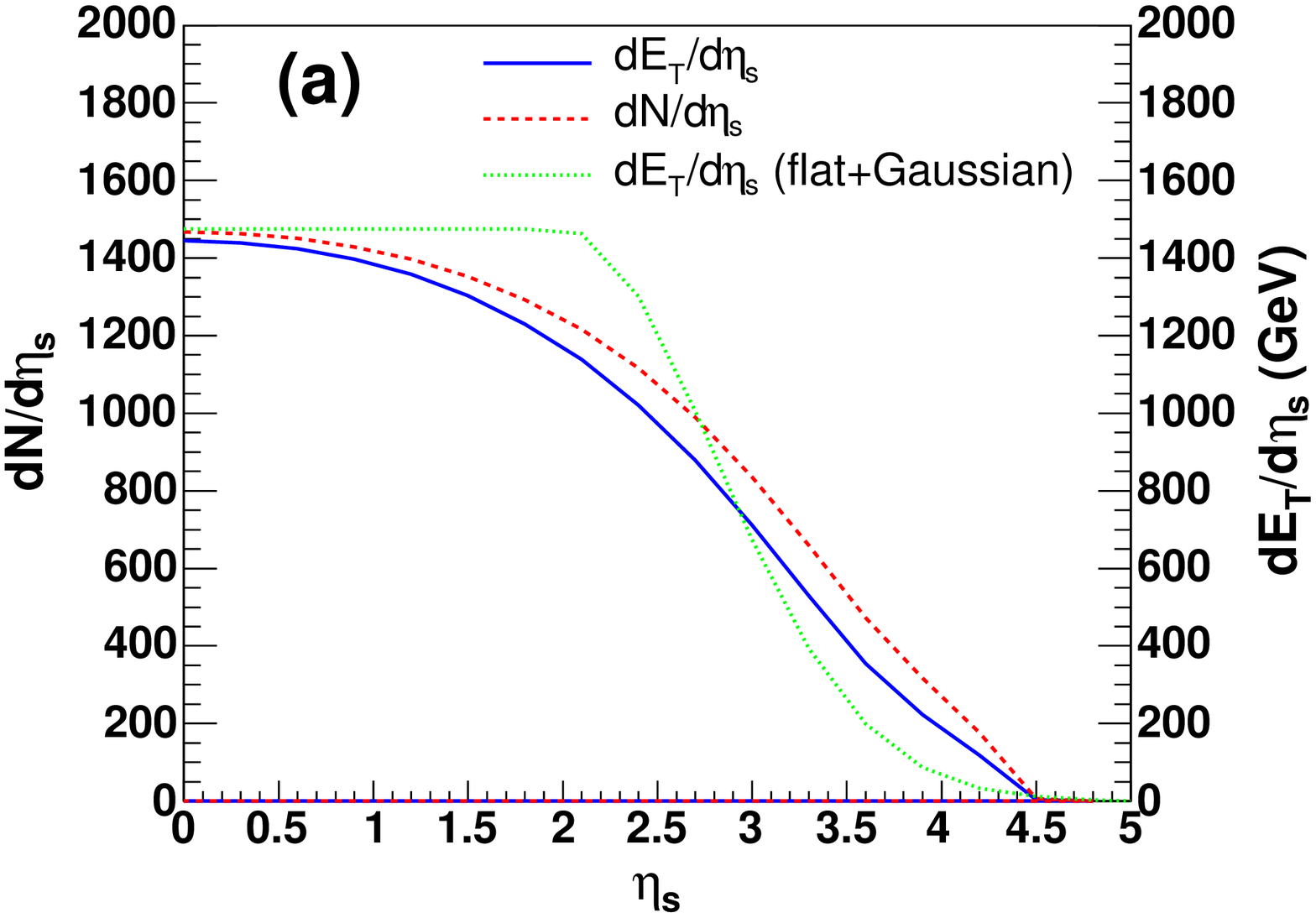,scale=0.4}
\end{minipage}
\begin{minipage}[h]{10cm}
\epsfig{file=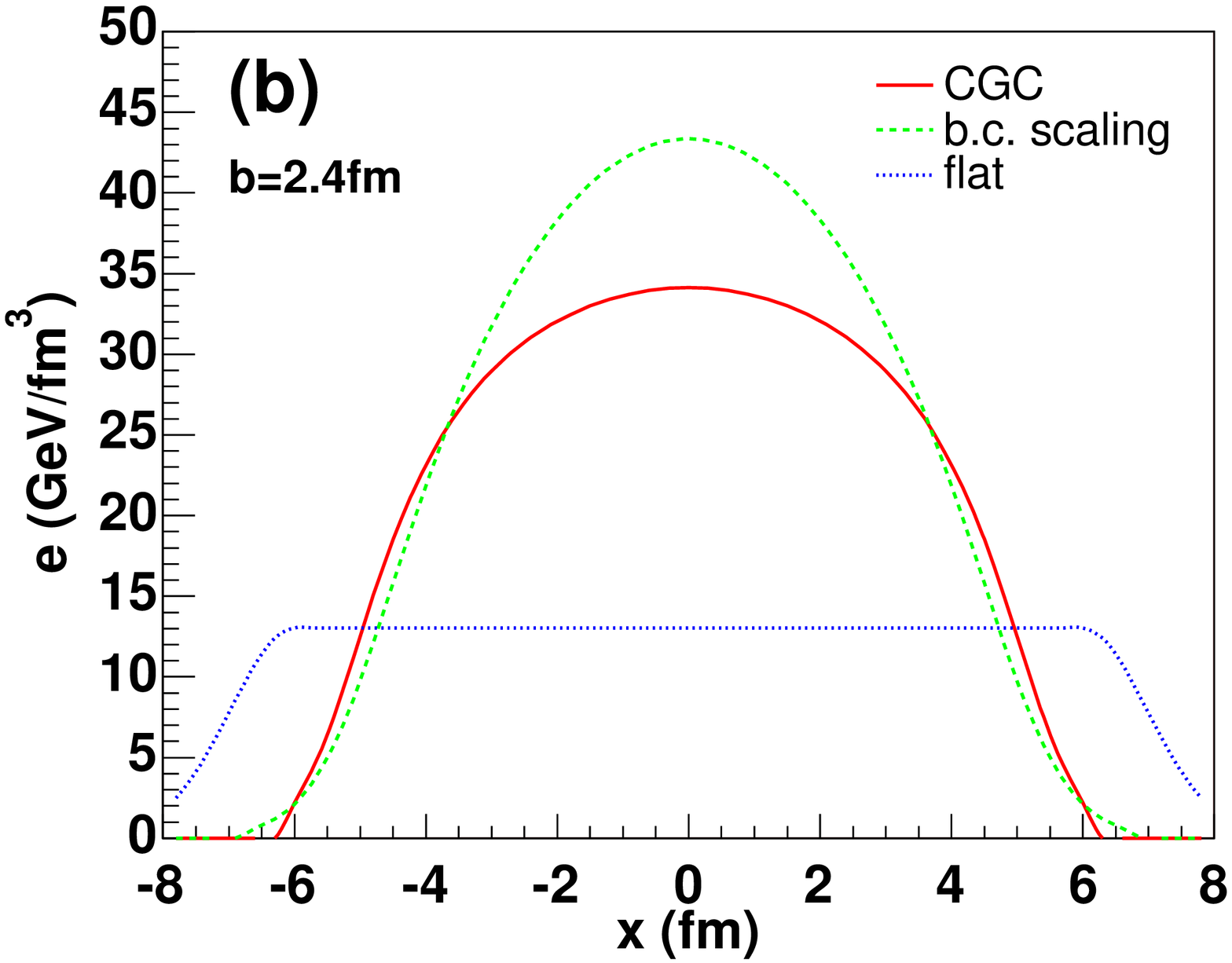,scale=0.4}
\end{minipage}
\caption{Initial energy density from the Color Glass Condensate in 
longitudinal direction (left panel) and in the transverse plane 
(right panel) for central Au+Au collisions at $\sqrt{s_{NN}} =  200$ GeV.
Figures reprinted from \cite{Hirano:2004en} with permission from Elsevier. 
\label{fig:initial2}}
%\end{center}
\end{figure}
%%%%%%%%%%%%%%%%%%%%%%%%%%%%%%%%%%%%%%%%%%%%%%%%%%%%%%%

%%%%%%%%%%%%%%%%%%%%%%%%%%%%%%%%%%%%%%%%%%%%
\begin{figure}[tb]
\begin{center}
\rotatebox{-90}{\epsfig{file=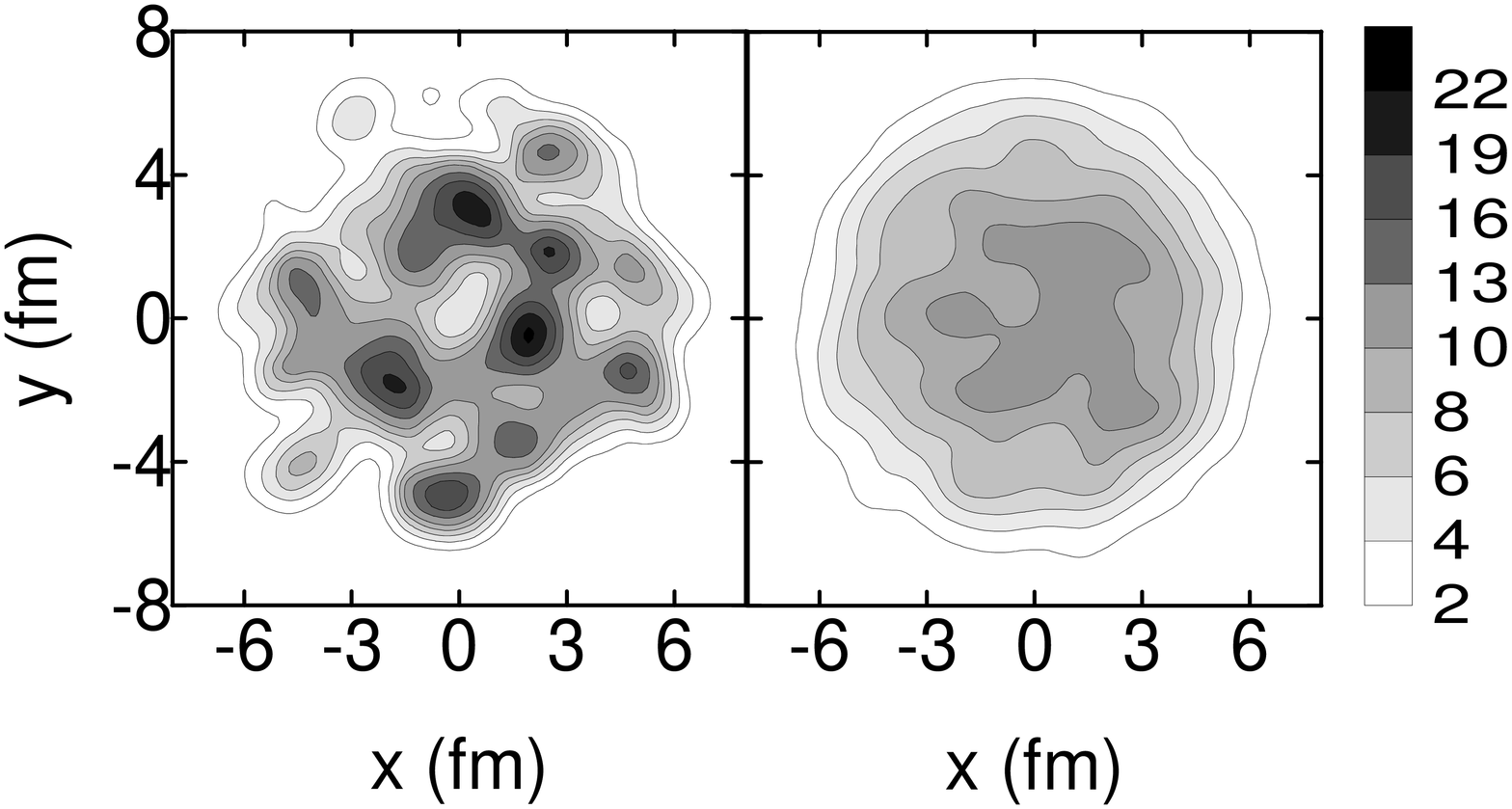, scale=0.5}}
\caption{Examples of initial energy density profiles in the transverse plane
for central Au+Au collisions given by NeXus at mid-rapidity. The 
energy density is plotted in units of GeV/fm$^3$. Left panel:
one random event. Right panel: average over 30 random events 
(similar to the smooth initial conditions often used).
Figure reprinted from \cite{Socolowski:2004hw} with permission from the
American Physical Society.
\label{fig:initial3}}
\end{center}
\end{figure}
%%%%%%%%%%%%%%%%%%%%%%%%%%%%%%%%%%%%%%%%%%%%

Other approaches for generating realistic initial conditions beyond the
Glauber-based parameterizations are available. 
Color Glass Condensate-inspired initial conditions are becoming increasingly
popular, see Fig. \ref{fig:initial2}. They feature larger eccentricities 
than Glauber-based models which has significant implications on elliptic 
flow \cite{Hirano:2004en}.
In that case additional dissipation during the early quark gluon plasma 
stage is needed in order to achieve agreement with experiments 
\cite{Hirano:2004en, Hirano:2005xf}.
Others models are the string rope model \cite{Csernai:2009zz} and the
pQCD + saturation model \cite{Eskola:2005ue}. In the latter the initial 
time $\tau_0$ is given by the inverse of the saturation scale, 
which is set to a very short $\tau=$0.18 (0.10) fm at RHIC (LHC).
More recently there is a push to implement effects of 
event-by-event fluctuations in initial conditions. In the
NEXSPHERIO hydro model these events are created by the event generator
NeXus \cite{Hama:2004rr}. Fig.\ \ref{fig:initial3} shows an example
from Ref.\ \cite{Socolowski:2004hw}. 
We will discuss the implications of fluctuations in more detail later.

%%%%%%%%%%%%%%%%%%%%%%%%%%%%%
\subsubsection{QCD and Hydrodynamics}
\label{sssec-hydrodnamic}

Recent lattice QCD calculations show that the phase transition at vanishing 
or low chemical potential is a crossover 
\cite{Aoki:2006we,Petreczky:2009at,Borsanyi:2010cj}. 
However the Bag Model-based equation of state with a first-order phase transition has dominated calculations in the early RHIC years. 
There are several simulations of lattice QCD at high temperature and 
low density. However, at high net baryon density the investigation 
with lattice QCD becomes difficult. Because of the sign problem, we can 
not apply the usual Monte Carlo methods for finite density lattice QCD.

%%%%%%%%%%%%%%%%%%%%%%%%%%%%%%%%%%%%%%%%%%%%
\begin{figure}[h]
\begin{center}
\epsfig{file=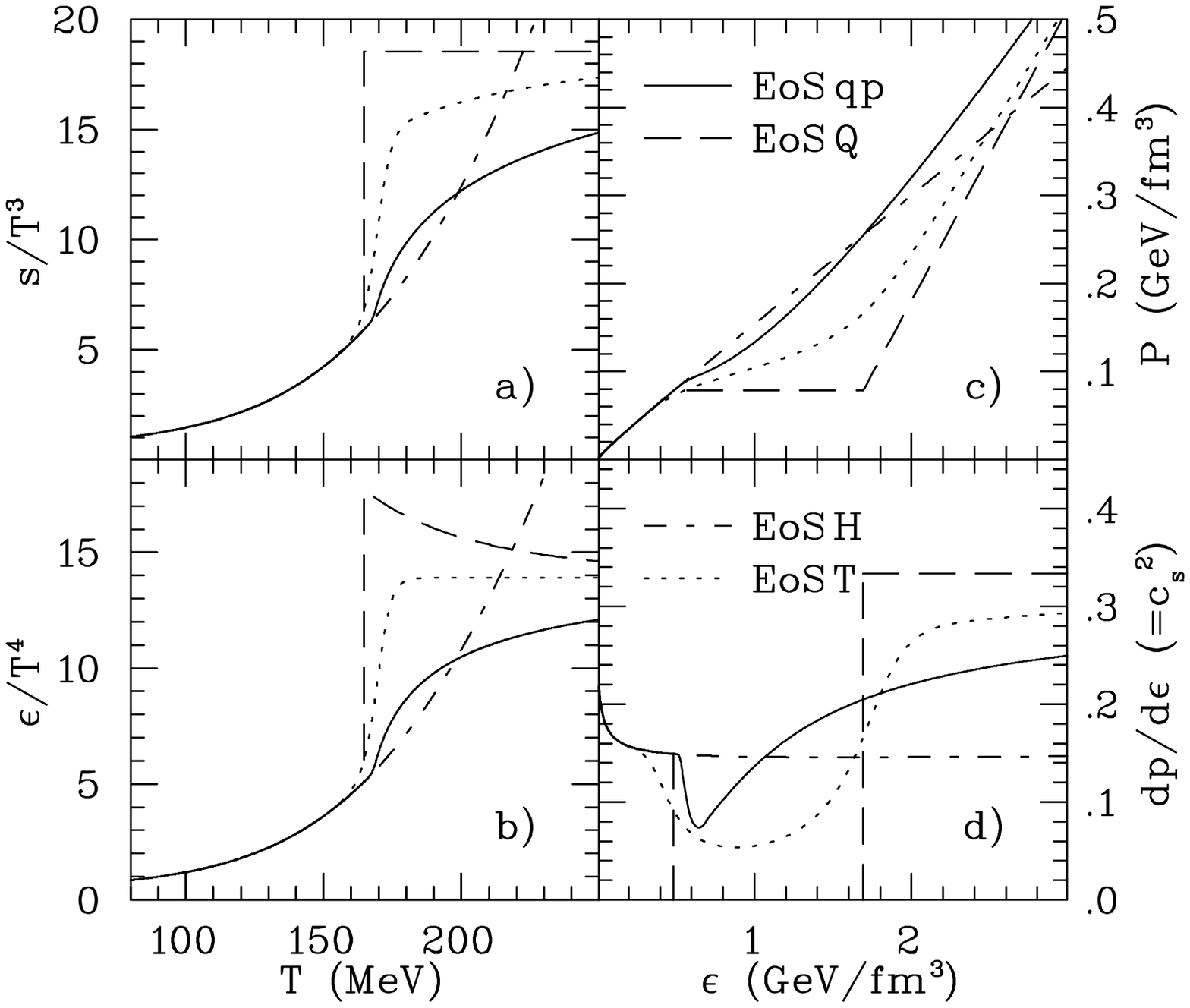, scale=0.6}
\caption{(a) The entropy density $s$ divided by $T^3$, and (b) 
the energy density $\epsilon$ divided by $T^4$ both as functions of 
temperature; (c) the pressure $p$, and (d) the velocity of sound squared 
$c_s^2 = dp/d\epsilon$ both as functions of $\epsilon$. In all four
panels four equations of state are shown: (i) qp = quasiparticle EoS, 
(ii) Q = ideal parton gas with first order phase transition, (iii) H  =
hadron resonance gas, and (iv) T = ansatz with crossover. 
Figure reprinted from \cite{Huovinen:2005gy} with permission from Elsevier.
\label{fig:eos}}
\end{center}
\end{figure}
%%%%%%%%%%%%%%%%%%%%%%%%%%%%%%%%%%%%%%%%%%%%

A comprehensive study of the dependence of hydrodynamic expansion on
the equation of state was carried out by Huovinen \cite{Huovinen:2005gy}.
Figure \ref{fig:eos} shows different equations of state used in Ref.\
\cite{Huovinen:2005gy}. Suitable parameters for the initial conditions and 
freeze-out temperature were chosen for each EoS so that the hydrodynamic 
calculation reproduces the experimental data reasonably. Huovinen found 
that the closest fit to the data of $v_2$ as a function of $P_T$ 
was obtained for a strong first-order phase transition while the results for 
the lattice-inspired EoS fit the data badly. 
Similar results were also obtained in more detailed analyses 
\cite{Huovinen:2009yb}. 
The current discrepancy between lattice results (crossover) and the hydrodynamic 
analysis of data (first order) is unsatisfactory.  Dependences on parameters, 
e.g.\ flow in the initial conditions, and the freeze-out procedure obscure the 
observation of the equation of state in hydrodynamic calculations.
Nevertheless, hydrodynamics offers the most direct tool to investigate the equation
of state and we have to move to a comprehensive analysis including
viscosities, hadronic re-interactions and an honest assessment of uncertainties
from initial conditions to solve this problem in the near future.

One of the most fascinating topics in connection with the QCD phase diagram 
is the QCD critical point (QCP) and its possible manifestation in 
relativistic heavy ion collisions \cite{Stephanov:1998dy}. 
The possible existence of this point and its location in the QCD phase 
diagram have been attracting a great deal of interests.
Recent studies based on effective theories show a wide range of possible 
locations of the critical point in the phase diagram \cite{Stephanov:2007fk}. 
In addition, a recent study on the curvature of the critical surface in
lattice QCD shows that the existence of the QCD critical point is less than
certain \cite{deForcrand:2008zi}. 
A solid experimental result seems the best way to settle the question of
the existence of the QCP because a unanimous conclusion from lattice QCD 
and effective theories appears to be difficult.  
For quantitative tests of the existence and location of the QCP we need to run
hydrodynamics with equations of state with a critical point included, and we need
to identify appropriate physical observables which show clear signatures of 
a critical point \cite{Nonaka:2005vr,Asakawa:2008ti}. The upcoming lower energy scan
at RHIC which will produce QGP at higher baryon chemical potential will
provide a suitable data sample.

Relativistic viscous hydrodynamic models need more information besides the
equation of state in order to run: shear and bulk viscosities, heat conductivity, and
relaxation times. On the flip side one should argue that there is a
unique chance at RHIC to extract quantitative values for some
of these transport coefficients from data.
Estimates for some transport coefficients are available 
from lattice QCD, finite temperature QCD perturbation theory and 
effective theories. For strongly coupled quark gluon plasma lattice QCD 
should be a reliable tool. However the determination of transport coefficients 
in lattice QCD with the Green-Kubo formula is not easy.
Pioneering results for shear and bulk viscosities with large
uncertainties have been obtained in \cite{Nakamura:2004sy, Meyer:2007ic}.
For weakly coupled QGP perturbative results have been obtained by
Arnold, Moore and Yaffe \cite{Arnold:2000dr,Arnold:2003zc,Arnold:2006fz}.
Microscopic transport models have helped to estimate transport coefficients 
of a hadron gas \cite{Muronga:2003tb, Muroya:2004pu, Demir:2008tr} and a
parton gas \cite{Xu:2007ns,Xu:2007jv}.
The results seem to indicate that $\eta/s$ from a hadron gas is larger 
than the KSS bound $1/(4\pi)$ \cite{Baier:2007ix} which is proposed by 
the AdS/CFT correspondence. On the other hand, shear viscosities from
radiative parton transport can be surprisingly lower than leading order
perturbative results and close to the AdS/CFT bound. We have accumulated 
a few predictions in Fig.\ \ref{fig_eta-s}. Even though there is still little
trust in quantitative numbers the preliminary conclusion is that the origin of the 
perfect fluid signatures at RHIC should be in the partonic phase 
\cite{Demir:2008tr}.

There are also attempts to compare hydrodynamic calculations to partonic
transport models.  Early parton cascades with leading order dynamics 
\cite{Molnar:2001ux} had suggested that large cross sections are needed to 
account for the elliptic flow observed at RHIC. Thermalization of the parton matter
in the initial stage would be extremely slow. More recent parton cascades
including radiative corrections have led to results that are compatible with
large elliptic flow and rapid thermalization \cite{Xu:2004mz,Xu:2007aa,Xu:2008av}.
More studies comparing hydrodynamics and transport models in detail are needed as
they provide necessary mutual checks.

%%%%%%%%%%%%%%%%%%%%%%%%%%%%%
\subsubsection{The Freeze-out Process}

Currently two separate freeze-out processes are believed to occur in 
heavy ion collisions. One is the chemical freeze-out at which the ratios 
of hadrons are fixed, and the other is the kinetic freeze-out at which
the particles stop to interact.
Chemical freeze-out temperature and chemical potentials are extracted with  
the statistical model on the basis of the grand-canonical formalism. 
Surprisingly, statistical models are in excellent agreement with experimental 
data for hadron ratios in a wide range of collision energy from AGS to 
RHIC \cite{rev-particle-production}. 
From the values of chemical freeze-out temperatures around 170 MeV
we can infer that the chemical freeze-out process takes place just after 
the QCD phase transition \cite{rev-particle-production}.
At the kinetic freeze-out temperature the mean-free path of particles
has grown to be of the order of the system size and a hydrodynamic 
description which requires zero mean-free path is clearly no longer
applicable. A first naive guess for the kinetic freeze-out temperature 
$T_f$ is the pion mass ($\sim140$ MeV). Practically the value of 
the kinetic freeze-out temperature in a hydrodynamic model is determined 
from comparison with data for $P_T$-spectra.

The task at the end of a hydro calculation is the population of fluid cells
of a given temperature and flow with particles.
For calculations of single particle spectra, the simple assumption of
a sudden freeze-out process at a certain proper time for each fluid cell 
is sufficient, neglecting the reverse  process from particles to the
hydrodynamic medium.
Under this assumption the Cooper-Frye formula \cite{CoFr74} is widely used.
Here we start from a simple case \cite{BlOl90}:    
Suppose that a number of particles $N(\tau)$  exists in the   
enclosed volume $\Omega$ that is bounded by a closed surface 
$S(\tau)$ at time. $N(\tau)$ is given by  
\begin{equation}
N(\tau)=\int_{\Omega(\tau)} d^3r \, n(\vecr,\tau)\, , 
\end{equation}
where $n(\vecr, \tau)$ is particle number density.
At time $\tau + \delta \tau$, the number of particle $N(\tau)$ 
has changed to  
\begin{equation}
\frac{dN}{d \tau} = \int_{\Omega(\tau)} d^3r \frac{\partial n}{\partial \tau} 
+ \frac{1}{d \tau} \int_{d \Omega}d^3r \, n(\vecr,\tau), 
\label{Eq-num1}
\end{equation}
where the volume has changed to $\Omega + d \Omega$.      
Utilizing current conservation, $\frac{\partial n}{\partial \tau} + 
\nabla \cdot \vj$ ($\vj$ is the current of particles),    
Eq.\ (\ref{Eq-num1}) is rewritten as  
\begin{equation}
\frac{dN}{d\tau}=-\int_{S(\tau)}d^2s \, \vj \cdot \vn+ \int _{S(\tau)} d^2s 
\frac{d\zeta}{d \tau}n,    
\label{Eq-num2}
\end{equation}
where $\vn$ is the normal vector of surface element $d^2s$ and 
$d\zeta$ is the distance between the surface of $\Omega(\tau)$  
and that of $\Omega(\tau)+d \Omega$.  
In Eq.\ (\ref{Eq-num2}), $dN/d\tau$ is the number of particles 
which cross the surface $S(\tau)$ during $d\tau$.  
Then the total number of particles through the hypersurface $\Sigma$,
which is the set of surfaces $\{S(\tau)\}$, is
\begin{equation}
N=\int _\Sigma j^\mu d\sigma_\mu, 
\label{Eq-num3}
\end{equation}
where $j^0=n$, $d\sigma_0=d^3r$, $d \vsigma =d\tau d^2s \, \vn$.
If we write
\begin{equation}
j^\mu=\frac{d^3P}{E}\frac{g_h}{(2\pi)^3}
\frac{1}{\exp[(P_\nu u^\nu-\mu_f)/T_f]\pm 1} P^\mu
\end{equation}
for the current $j^\mu$ in Eq.\ (\ref{Eq-num3}), we obtain the   
Cooper-Frye formula \cite{CoFr74}  
\begin{equation}
E\frac{dN}{d^3P} =  
\sum_h\frac{g_h}{(2\pi)^3} \int_\Sigma d\sigma_\mu 
P^\mu \frac{1}{\exp[(P_\nu u^\nu-\mu_f)/T_f]\pm 1}, 
\label{Eq-CF}
\end{equation}
where $g_h$ is a degeneracy factor of hadrons and $T_f$ and $\mu_f$ are 
the freeze-out temperature and chemical potential.  
In other words we obtain $d\sigma_\mu$ by estimating 
the normal vector on the freeze-out hypersurface $\Sigma$. 
Using Eq.\ (\ref{Eq-CF}), we can then calculate all particle distributions
after freeze-out.

%%%%%%%%%%%%%%%%%%%%%%%%%%%

\begin{table}[tb]
\begin{center}
\begin{minipage}{14cm}
\caption{Calculations with relativistic ideal hydrodynamical models at RHIC
energies. See text for acronyms.}
\label{table-ideal-hydro}
\end{minipage}
\end{center}
\begin{tabular}{|l|p{2.0cm}|c|p{2cm}|c|p{4cm}|} 
\hline
Ref. & IC & EoS & Hydro Exp. & hadronization \& FO & observables \\ \hline \hline
Kolb \cite{Kolb:2000fha} & Glauber & 1st order & 2d & KF & elliptic flow \\ \hline 
Huovinen  \cite{Huovinen:2001cy} & Glauber & 1st order & 2d & KF & radial, elliptic flow \\ \hline
Kolb \cite{Kolb:2001qz} & Glauber & 1st order & 2d & KF & centrality dependence of multiplicity and flow \\ \hline 
Hirano \cite{Hirano:2002ds} & Glauber & 1st order & 3d & CF and KF & flow and HBT \\ \hline
Teaney \cite{Teaney:2000cw} & Glauber     & 1st order & 2d & hadron cascade & flow, QGP signature \\ \hline 
Nonaka \cite{Nonaka:2006yn} & Glauber & 1st order & 3d & hadron cascade & spectra, flow \\ \hline
Kolb \cite{Kolb:2002ve} & Glauber, initial $v_T$ & 1st order & 2d & CF and KF & $P_T$ spectra, $v_2$ \\ \hline
Eskola \cite{Eskola:2005ue} & pQCD + saturation &   1st order   &   cylindrical symmetry    & KF &  low and high $P_T$ spectra \\ \hline
Hirano \cite{Hirano:2004en} & CGC & 1st order & 3d, \hspace{3mm}hydro+jet &  CF and KF & $R_{AA}$, $I_{AA}$      \\ \hline
Hirano \cite{Hirano:2005xf} & CGC & 1st order & 3d & hadron cascade & $v_2$ \\ \hline 
Andrade \cite{Andrade:2006yh} &  NEXUS & 1st order & 3d & KF & $v_2$ \\ \hline 
Hirano \cite{Hirano:2009ah} & CGC + fluctuation & 1st order & 3d & hadron cascade & $v_2$ \\ \hline 
Huovinen \cite{Huovinen:2005gy} & Glauber & comparison & 2d & KF & hadron spectra, $v_2$ \\ \hline 
Socolowski \cite{Socolowski:2004hw} & NeXus & 1st order & 3d & CE  & HBT \\ \hline 
Hirano \cite{Hirano:2003pw} & Glauber & 1st order & 3d, hydro+jet & CF and KF & hadron spectra, $v_2$ \\ \hline
\end{tabular}
\end{table}
%\end{center}
%%%%%%%%%%%%%%%%%%%%%%%%%%%%%

More realistic models have been 
investigated. One of them is the Continuous Emission Model (CEM) 
in which particles are emitted continuously from the whole expanding 
volume of the system at different temperatures and different times
\cite{Socolowski:2004hw}.  
In the early days of hydrodynamics only kinetic freeze-out was implemented.
Indeed, at lower collision energies such as at AGS and SPS, the differences 
between chemical freeze-out and kinetic freeze-out points are not large. 
However, at RHIC a significant difference between kinetic freeze-out 
temperatures from hydro-inspired models and the chemical freeze-out 
from the statistical model appears \cite{Adams:2003xp}. 
This phenomenon also manifests itself through the failure to get the
correct absolute normalization of some $P_T$-spectra, e.g.\ the proton
in hydrodynamic calculations, with only a kinetic freeze-out 
\cite{Heinz:2001xi}. 
Hence a consistent modeling of separate freeze-out processes 
via modified equations of state was introduced
\cite{Hirano:2002ds,ArGrHaSo01,Arbex:2001vx,Te02}. 

It turns out that some experimental data is still not understood in 
a satisfying way even with two separate freeze-out procedures. 
For example, mean transverse momentum $\langle P_T \rangle$ as a 
function of particle mass does deviate from the linear scaling law, which 
suggests significant final state interactions in the hadronic phase 
\cite{Nonaka:2006yn}. To explain these effect, and to account for the
apparently large viscosities in the hadronic phase, as discussed before,
hydro+cascade hybrid models were introduced.
They use a hydrodynamic computation of the expansion and cooling of hot 
QCD bulk matter, and then couple the output consistently to a hadron-based 
transport model for the final-state interactions.
Pioneering work on hydro+cascade hybrid models was done by Bass et al.\
\cite{Bass:2000ib} using UrQMD. Similar investigations were carried out 
in Refs.\ \cite{Teaney:2000cw, Hirano:2005xf}.
The improvements introduced by these hybrid models are discussed in more
detail in the results section.

%%%%%%%%%%%%%%%%%%%%%%%%%%%%%
% 
\begin{table}[tb]
\begin{center}
\begin{minipage}{14cm}
\caption{Calculations with relativistic viscous hydrodynamical models at RHIC
energies. See text for acronyms.}
\label{table-viscous-hydro}
\end{minipage}
\end{center}
\begin{tabular}{|l|p{2.0cm}|p{2cm}|p{2cm}|p{3cm}|p{3.6cm}|} 
\hline
Ref. & IC & EoS & Hydro Exp. & hadronization \& FO & observables \\ \hline \hline
Song\cite{Song:2008si} & Glauber & EoS dependence & 2d, I-S & KF (direct $\pi$) & $v_2$ \\ \hline 
Dusling\cite{Dusling:2007gi}&Glauber & ideal QGP gas & 2d, O-G & KF, viscous corrections & $v_2$ \\ \hline 
Romatschke\cite{Romatschke:2007mq} &Glauber &   semi-realistic EoS \cite{Laine:2006cp} & 2d, I-S&  KF &multiplicity, $v_2$ \\ \hline        
Luzum\cite{Luzum:2008cw} & Glauber and CGC &   semi-realistic EoS \cite{Laine:2006cp}
&  2d, conformal relativistic viscous hydro\cite{Baier:2007ix}  &KF, viscous corrections &  multiplicity, $v_T$, 
$v_2$ \\ \hline 
Chaudhuri\cite{Chaudhuri:2009hj} & Glauber & lattice  +HRG EoS 
& 2d, I-S &KF & $v_2$  \\ \hline
\end{tabular}
\end{table}

%%%%%%%%%%%%%%%%%%%%%%%%%%%%%%%%%%%%%%%%%%%%%%%%%%%%%%%
\begin{figure}[tb]
%\begin{center}
\begin{minipage}{8.5cm}
\epsfig{file=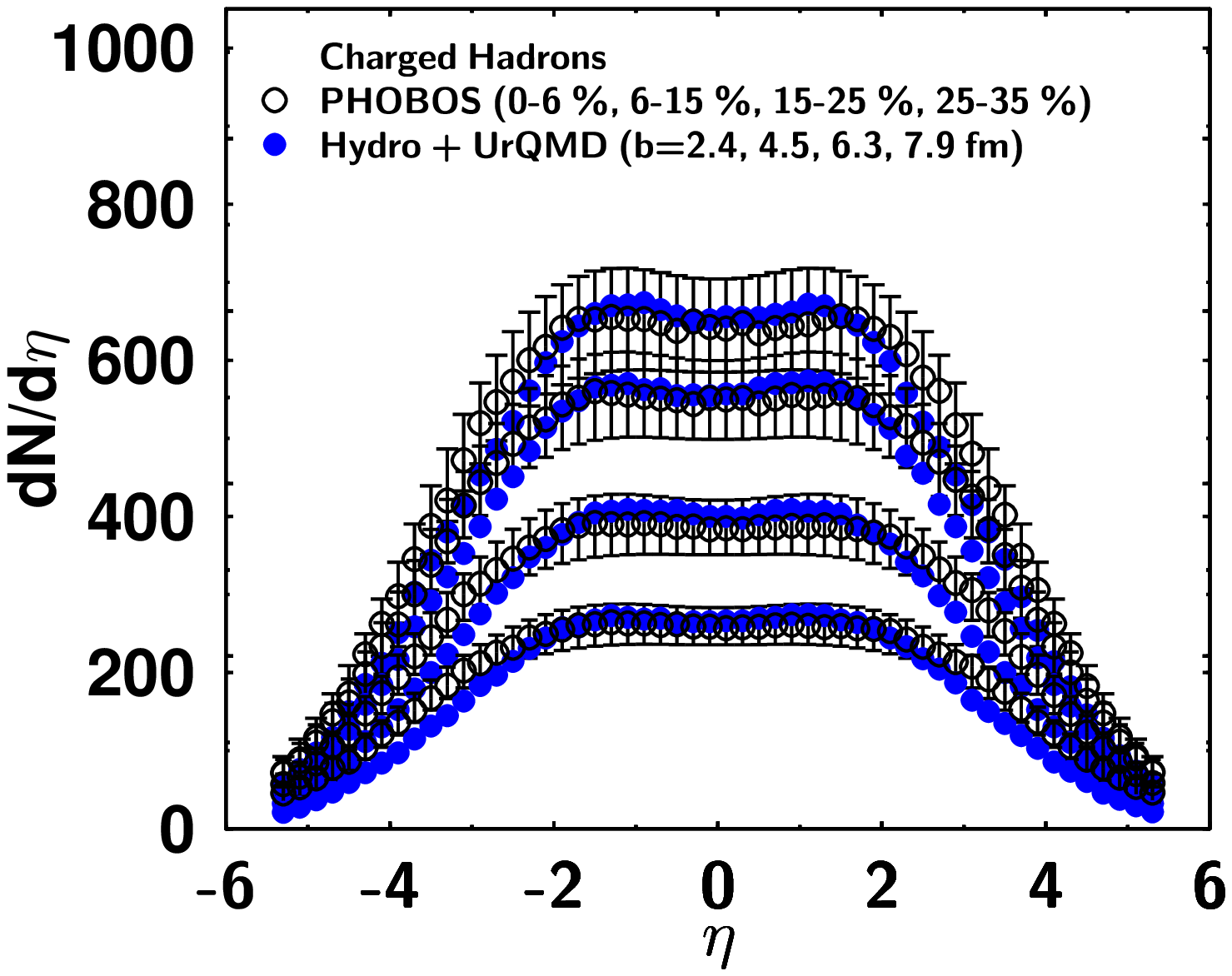,scale=0.5}
\caption{Centrality dependence of pseudorapidity distributions of charged 
particles from the hydro+UrQMD hybrid model \cite{Nonaka:2006yn} compared 
wit PHOBOS data for $\sqrt{s_{NN}}=200$ GeV Au+Au collisions \cite{PHOBOS_psuedo}. 
Impact parameters in the calculation with the corresponding centrality
bins in parentheses are 2.4 (0-6\%), 4.5 (6-10\%), 6.3 (10-15\%), and 7.9 fm 
(25-35\%). Figure taken from \cite{Nonaka:2006yn}.}
\label{fig_rap}
\end{minipage}
\hspace{0.5cm}
\begin{minipage}{8.5cm}
\vspace{-1.5cm}
\epsfig{file=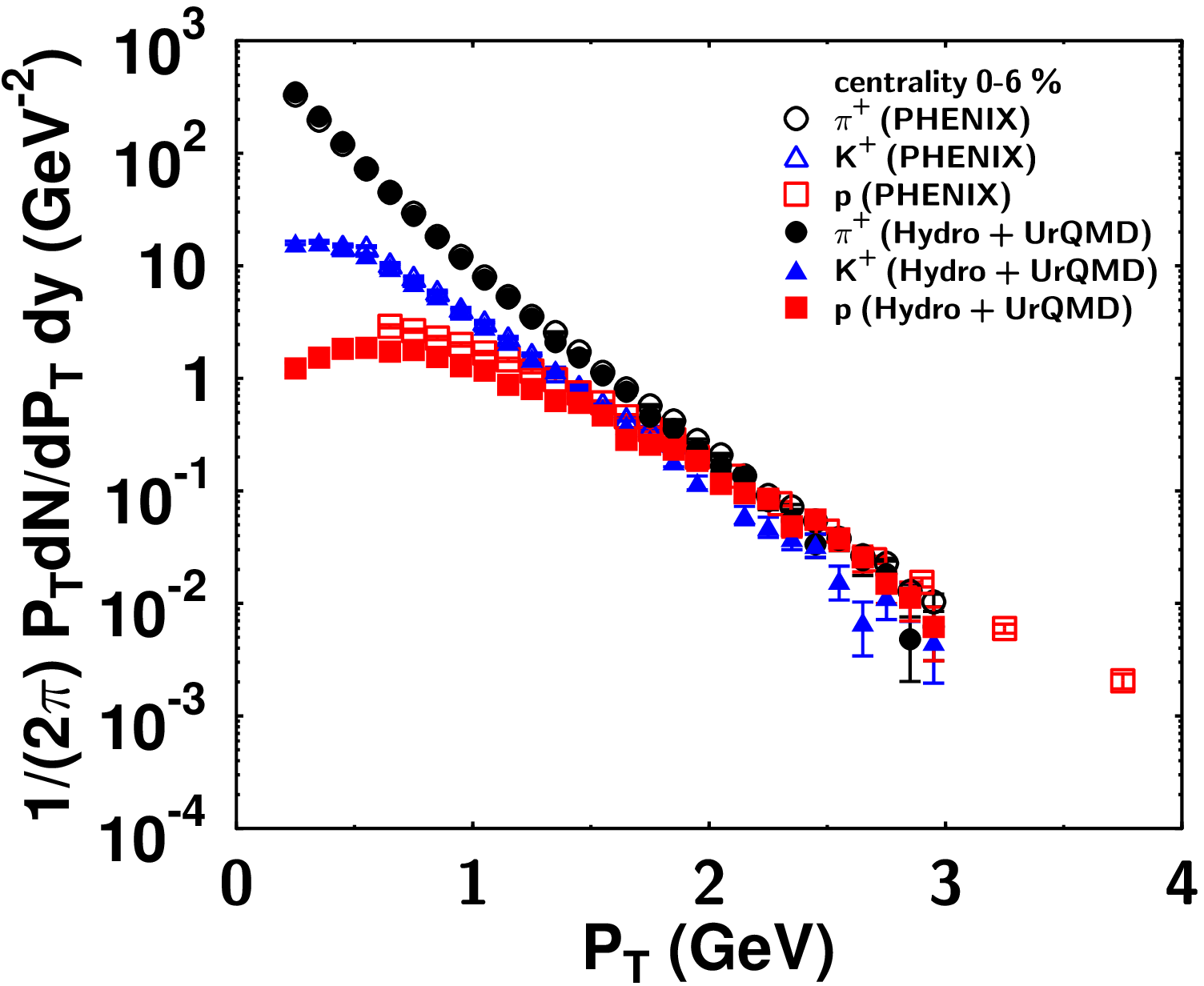,scale=0.5}
\caption{$P_T$-spectra for $\pi^+$, $K^+$ and $p$ at central
 $\sqrt{s_{NN}}=200$ GeV Au+Au 
collisions with PHENIX data \cite{PHENIX_PT}. The points are not renormalized.
Figure taken from \cite{Nonaka:2006yn}.}
\label{fig_Pt}
\end{minipage}
%\end{center}
\end{figure}
%%%%%%%%%%%%%%%%%%%%%%%%%%%%%%%%%%%%%%%%%%%%%%%%%%%%%%%

\subsection{\it New Developments}

Through its success at RHIC hydrodynamics has positioned itself as one of 
the most useful phenomenological models for heavy ion collisions. A
large amount of studies working with hydrodynamics have been
carried out in the RHIC era. Historically, most of them have been using ideal
hydrodynamics, but obviously viscous hydrodynamics
will be a major focus point for the near future, as both
mathematical and numerical issues are settled.
The thorough vetting of hydrodynamic models has also become
a top issue. Different existing codes have to be tested against each
other using identical parameters and initial conditions.
One such effort is under way in the ``Theory-Experiment Collaboration for 
Hot QCD Matter'' (TECHQM) \cite{TECHQM}.

Hydrodynamic models can be easily adopted for 
descriptions of the entire process of heavy ion collisions
that include initial collisions, thermalization, bulk dynamics,
hadronization, freeze-out, final state interactions and even hard and
electromagnetic probes. Such comprehensive approaches are called
multi-module modeling. There are ongoing efforts to construction 
such models \cite{Nonaka:2006yn,Hirano:2005xf,Petersen:2008dd,Werner:2009zza}.
The coupling of hydrodynamics and hadronic transport together 
to form a hybrid model is only the first step. Event generators
for the initial state will soon become standard. It has also 
become necessary to conduct jet quenching studies using realistic input 
from hydrodynamically evolved fireballs \cite{Bass:2008rv}.

%%%%%%%%%%%%%%%%%%%%%%%%%%%%%%%%%%%%
% relativistic viscous hydrodynamic 
%To close out this section about hydrodynamics at RHIC
We have compiled a list of relativistic ideal hydrodynamic models which are 
mentioned in this work in Tab.~\ref{table-ideal-hydro}. We reference each
work and quote the initial conditions, freeze-out treatment, and equation
of state that have been used, and the observables that have been calculated. 
The acronyms KF, CF and CE stand for kinetic freeze-out, chemical freeze-out 
and continuous emission, respectively. 
%%%%%%%%%%%%%%%%%%%%%%%%%%%%%%%%%%

For relativistic viscous hydrodynamics, the number of phenomenological
studies is much smaller. Quantitative discussions comparing to RHIC data have
been carried out, but need to be weighed with caution. Many points need
further study.
%%%%%%%%%%%%%%%%%%%%%%%%%%%%%%%%%%
In Tab.~\ref{table-viscous-hydro}, we list the relativistic viscous
hydrodynamic studies which compare results to experimental data from RHIC. 
Here I-S (G-O)  stands for relativistic viscous hydrodynamics proposed by
Israel and Stewart \cite{Israel,IsSt1,IsSt2}  (Grmela and  \"Ottinger \cite{GO1,GO2,Ott}). 
The major difference between ideal hydrodynamics 
(Tab.\ \ref{table-ideal-hydro}) and viscous hydrodynamics 
(Tab.\ \ref{table-viscous-hydro}) is that the former all agree on the set of
equations to be solved, while the latter solve a variety of different second
order schemes. While all of those should only exhibit small deviations from first order
hydrodynamics, this strengthens the point that we have to apply due caution
when analyzing RHIC data.

%%%%%%%%%%%%%%%%%%%%%%%%%%%%%%%%%%%%%%%%%%%%%%%%%%%%%%%
\begin{figure}[tb]
\begin{minipage}{8.5cm}
        \epsfig{file=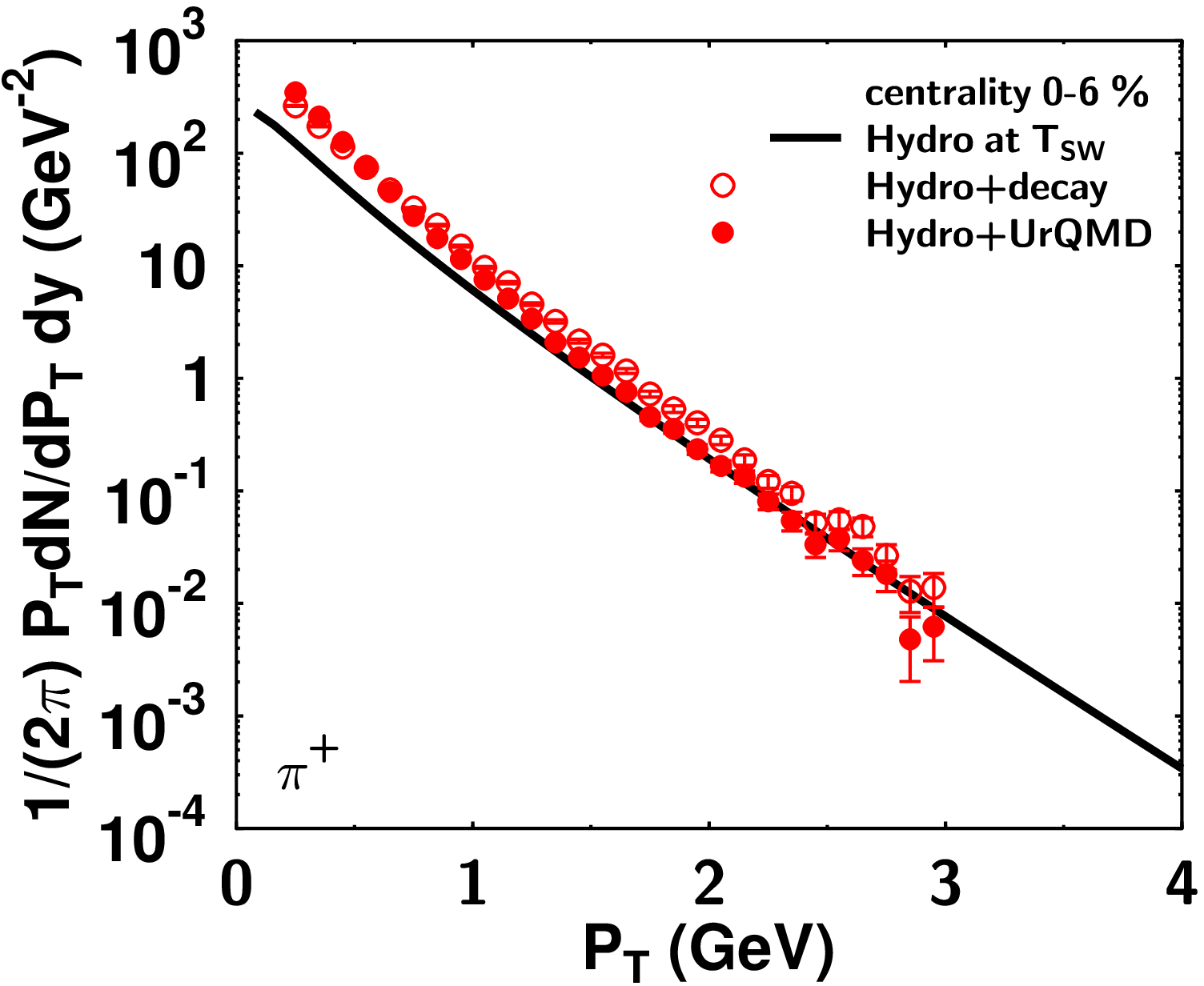,scale=0.5}
        \caption{$P_T$-spectra of $\pi^+$ from \cite{Nonaka:2006yn}.
Shown are the spectra for hydrodynamics at 
the switching temperature of 150 MeV, result from hydro + decay only, and 
hydro + full UrQMD, in central collision.
Figure taken from \cite{Nonaka:2006yn}.}        \label{fig_pt2}
\end{minipage}
\hspace{0.5cm}
\begin{minipage}{8.5cm}
\vspace{-1.0cm}
\epsfig{file=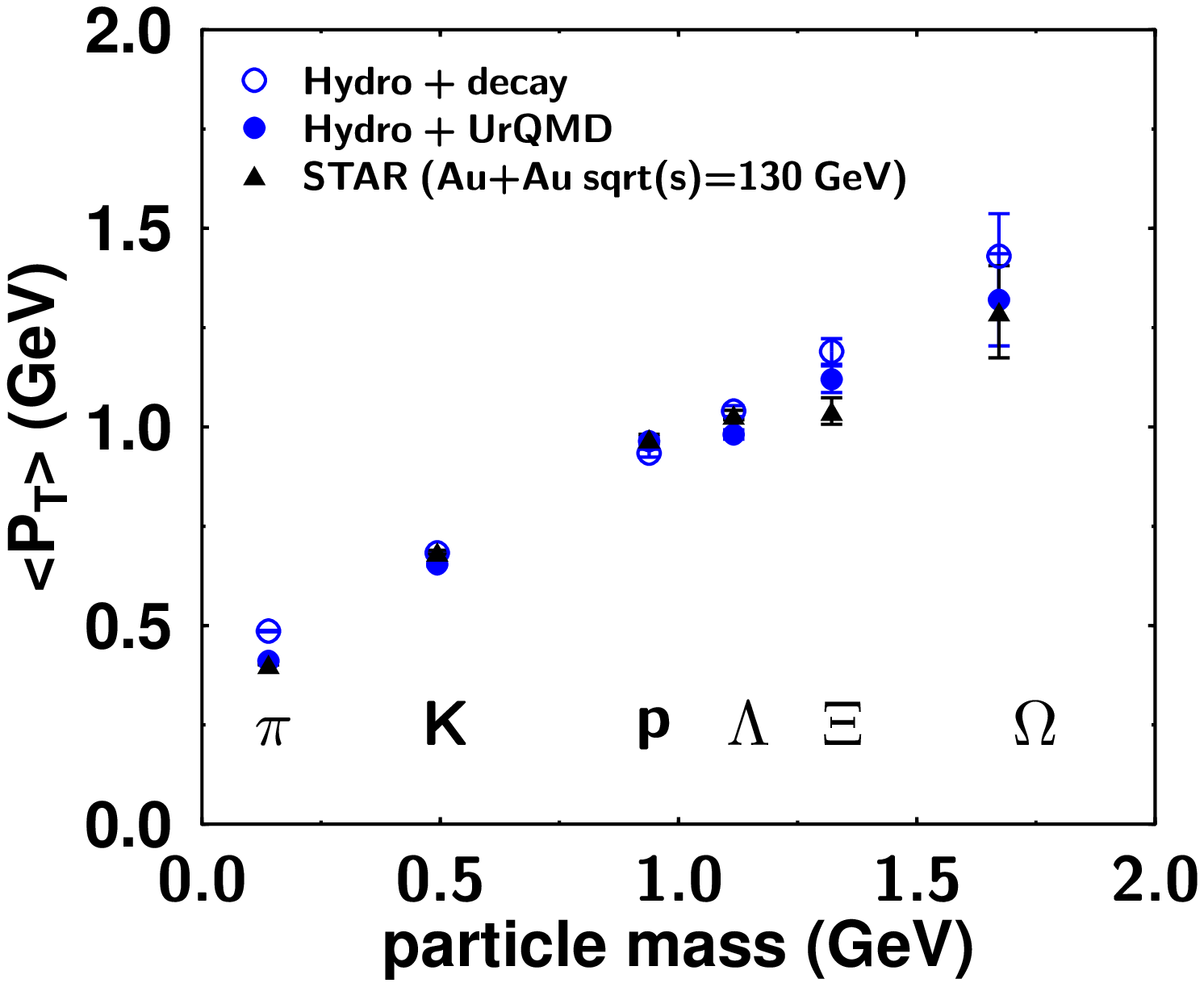, scale=0.5}
\caption{Mean $P_T$ as a function of mass. Figure taken from 
\cite{Nonaka:2006yn}.}
\label{fig_meanpt}
\end{minipage}
\end{figure}
%%%%%%%%%%%%%%%%%%%%%%%%%%%%%%%%%%%%%%%%%%%%%%%%%%%%%%%

%%%%%%%%%%%%%%%%%%%%%%%%%%%%%
\subsection{Hadronization and Quark Recombination}

In hydrodynamic models the hadronization process from the QGP phase to 
the hadron phase is naturally encoded in the equation of state.  
As we have mentioned above hydrodynamic models generally do well
for particles below $P_T \approx 2$ GeV/$c$ at RHIC.
Between that point and the perturbative region above $\approx 6$ GeV/$c$
an intermediate $P_T$ region has been found. It exhibits features
from both domains, without fitting exclusively in any of the two categories.
E.g.\ we find large elliptic flow and baryon over meson ratios that
are reminiscent of hydrodynamics and soft physics, and incompatible with
jet quenching and fragmentation. On the other hand the elliptic flow
is not large enough to follow the ideal hydrodynamic predictions, 
and we find dihadron correlations at intermediate $P_T$ that
exhibit clear jet-like structures. Simple interpolations (hydro+jet models) do
not explain all features, e.g.\ the much smaller elliptic flow of $\phi$
mesons compared to protons. In order to explain hadron production at RHIC,
and in order to understand the breakdown of the pQCD or hydrodynamic
approach it is crucial to understand this kinematic region at intermediate $P_T$. 

Quark recombination or coalescence is the best candidate to explain a large 
amount of experimental data at intermediate $P_T$. Recombination models assume
a universal phase space distributions of quarks at hadronization. Quarks turn
into baryons, $qqq \to B$, and mesons, $q\bar q\to M$ described either by 
using instantaneous projections of quark states onto hadron states  
\cite{Fries:2003vb,Fries:2003kq,Greco:2003xt,Greco:2003mm,
Muller:2005pv}, or a dynamical coalescence process with finite width hadrons
governed by rate equations \cite{Ravagli:2007xx}.
Note that usually only the valence quarks of the hadron are taken into 
account although generalizations have been worked out \cite{Muller:2005pv}.

%%%%%%%%%%%%%%%%%%%%%%%%%%%%%%%%%%%%%%%%%%%%%%%%%%%%%%%
\begin{figure}[tb]
%\begin{center}
\begin{minipage}{8.5cm}
 \epsfig{file=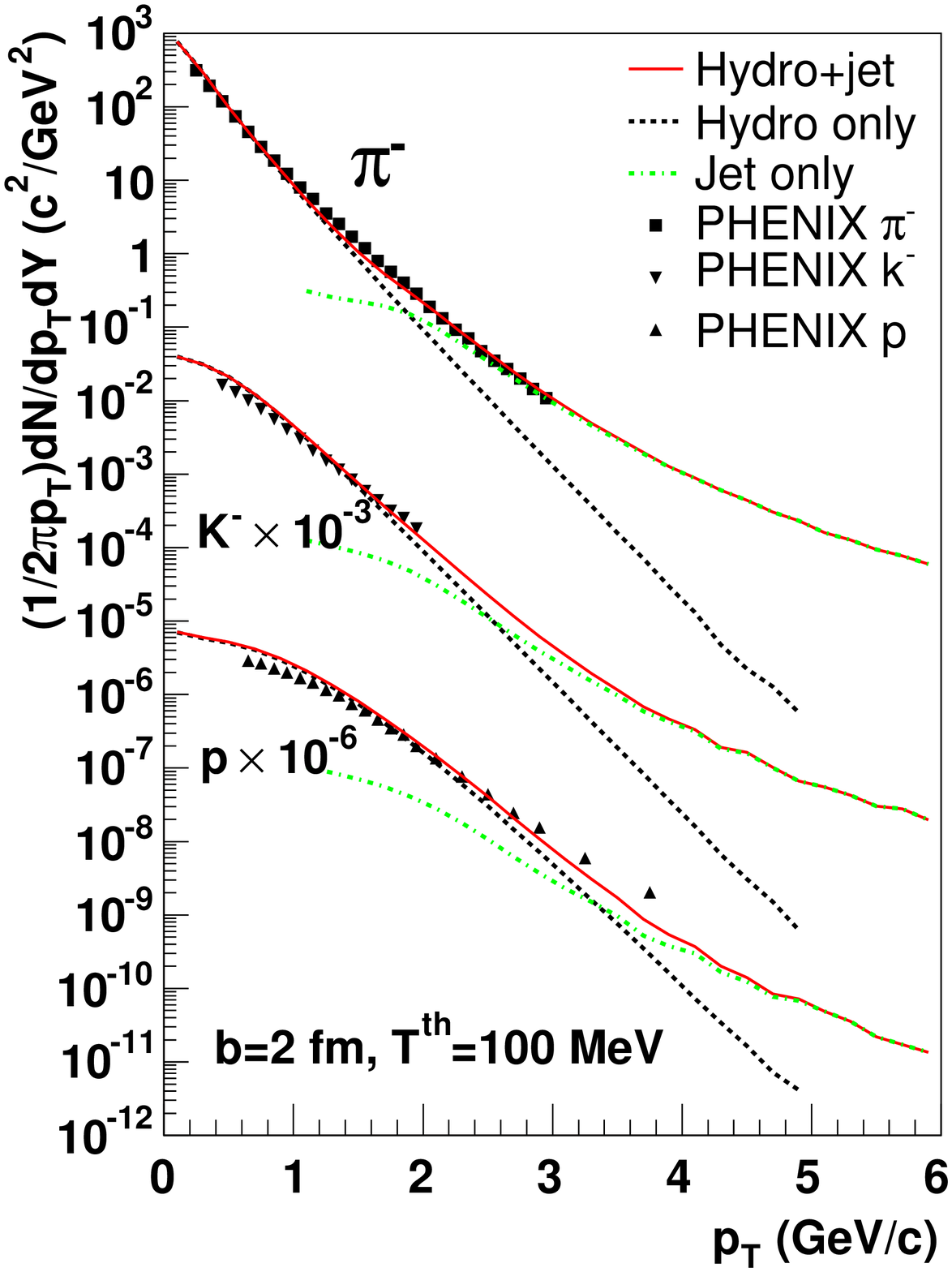, scale=0.45}
       \caption{Spectra from hydrodynamics (dotted), pQCD (dash-dotted),
and their sum (solid line) for $\pi^-$, $K^-$, and $p$ in Au+Au collisions 
at $\sqrt{s_{NN}} =200$ GeV for impact parameter $b=2.0$ fm compared
to data from PHENIX \cite{Adler:2003cb}. Figure reprinted from 
\cite{Hirano:2003pw} with permission from the American Physical Society.}
\label{fig_hydrojet_pt}
\end{minipage}
\hspace{0.5cm}
\begin{minipage}{8.5cm}
\epsfig{file=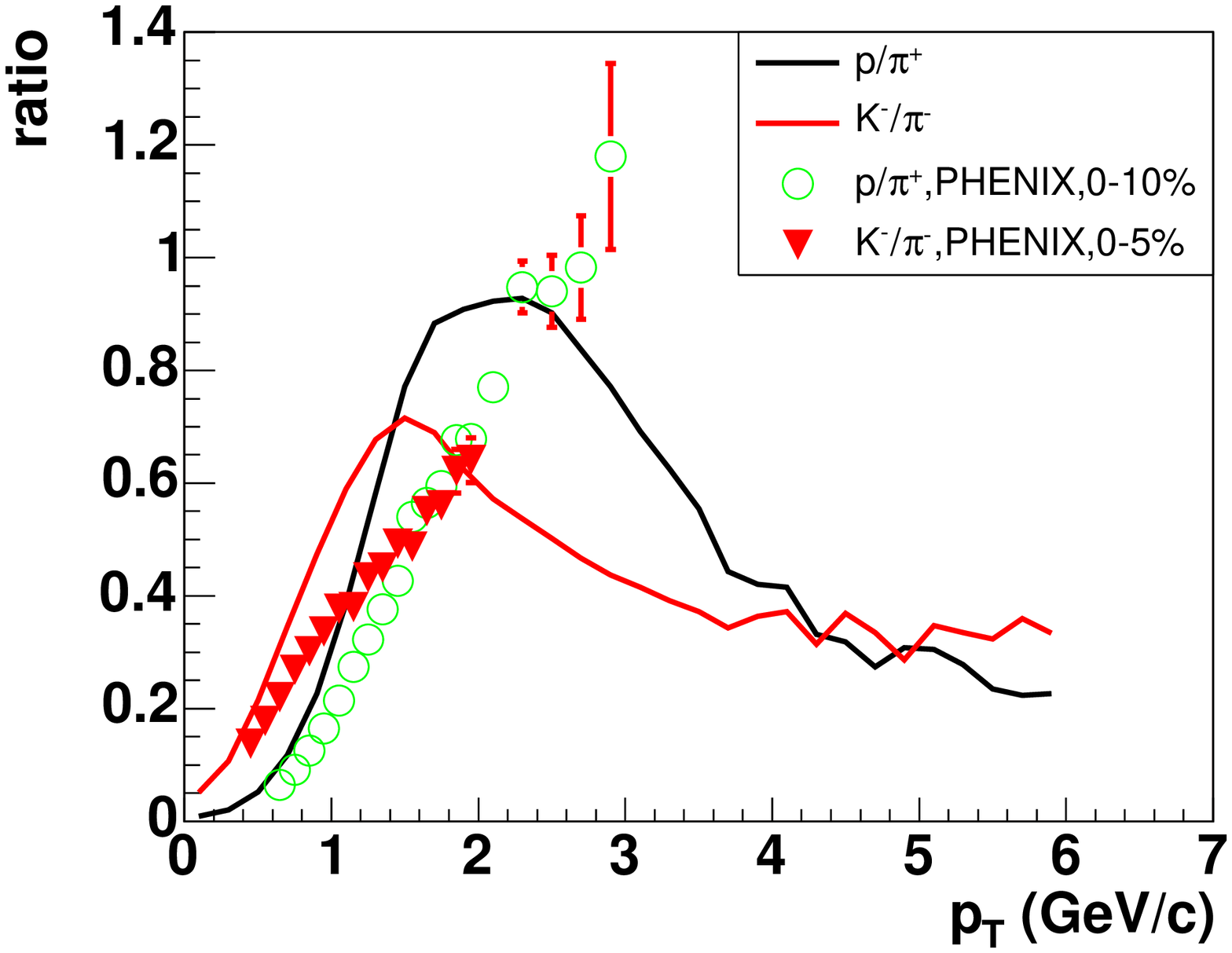,scale=0.45}
\caption{
Ratios $N_p/N_{\pi^-}$ and $N_{K-}/N_{\pi^-}$ as a function of $P_T$ in
Au + Au collisions as in Fig.\ \ref{fig_hydrojet_pt} compared
to data from PHENIX \cite{Adler:2003cb}. Figure reprinted from \cite{Hirano:2003pw}
with permission from the American Physical Society.}
\label{fig_hydrojet_ratio}
\end{minipage}
%\end{center}
\end{figure}
%%%%%%%%%%%%%%%%%%%%%%%%%%%%%%%%%%%%%%%%%%%%%%%%%%%%%%%

The original instantaneous projection models explicitly preserve only three
components of the energy-momentum four-vector in the underlying $2\to1$ and
$3\to 1$ processes. The yield of mesons can be 
expressed through the convolution of the Wigner function $W_{ab}$ for parton 
pairs $a$, $b$ and the Wigner function $\Phi_M$ encoding the meson wave function
\begin{equation}
  \frac{dN_M}{d^3 P} = \int \frac{d^3 R}{(2\pi)^3} \sum_{ab} \int
  \frac{d^3q d^3r}{(2\pi)^3} W_{ab}\left(\mathbf{R}+\frac{\mathbf{r}}{2},
  \frac{\mathbf{P}}{2}+\mathbf{q} ; \mathbf{R}-\frac{\mathbf{r}}{2},
  \frac{\mathbf{P}}{2}-\mathbf{q} \right)
  \Phi_M(\mathbf{r},\mathbf{q}) \, .
\end{equation}
The quark Wigner functions are usually approximated by classical phase space
distributions. Hadron spectra at intermediate $P_T$ are described well by
considering a factorization into thermal quark distributions
\cite{Fries:2003kq}
\begin{equation}
  W_{ab}\left(\mathbf{r}_1,\mathbf{p}_1;\mathbf{r}_2,\mathbf{p}_2\right)
  = f_a\left(\mathbf{r}_1,\mathbf{p}_1\right) f_b\left(\mathbf{r}_2,\mathbf{p}_2\right)
  \, . 
\end{equation}
Correlations between quarks can be introduced to model correlations found 
between hadrons \cite{Fries:2004hd} without interfering with the excellent 
description of spectra and hadron ratios.
 
Dynamical models, like the resonance recombination model 
\cite{Ravagli:2007xx,Ravagli:2008rt,He:2010vw} solve systems of rate or 
Boltzmann equations for the underlying $2\to1$ and $3\to1$ processes. In the
case of resonance recombination mesons and baryons are technically
treated as resonances with widths $\Gamma_H$, and the inverse 
processes (the ``decays'' of mesons and baryons into quarks) are taken into
account to achieve detailed balance. Resonance recombination has
the advantage that is can deliver an equilibrated hadron phase
from an equilibrated quark phase due to energy conservation and detailed
balance. This has opened the possibility to reconcile quark recombination
with hydrodynamics and equilibrium physics at low momenta \cite{He:2010vw}.

At large momenta contact can be made with jet fragmentation by rewriting
fragmentation functions as a process of valence quark coalescence in
a suitably defined jet shower \cite{Hwa:2003ic,Hwa:2004ng,Majumder:2005jy}.
If $S(p)$ is the distribution of quarks in a jet before hadronization and $T(p)$
is the distribution of the underlying event (partially thermalized in heavy
ion collisions) recombination will be applied to the total parton phase space
$S(p)+T(p)$. For mesons this will lead to the following possible combinations:
$SS$ which resembles the fragmentation process within a jet, $TS$ which
is a novel ``soft-hard'' hadronization process, and $TT$ which corresponds
to the usual picture of quark recombination. Shower distributions $S(p)$ can 
be fitted such that $SS$ and $SSS$ recombination fit the known fragmentation
functions for the respective mesons and baryons.

The strength of the quark recombination picture is the predictive power coming
from explaining all measured hadron spectra at intermediate $P_T$ with one
parameterization of the quark phase at hadronization. It has been shown that
at low momenta resonance recombination is compatible with hydrodynamics
and kinetic equilibrium \cite{He:2010vw}, but on the other hand, because a
thermalized state does not retain memories of a previous time evolution, all 
phenomena in the equilibrated region should be explainable by hydrodynamics.
This includes the quark number and kinetic energy scaling observed at RHIC 
at low momenta \cite{He:2010vw}.
The possibility of including quark recombination explicitly into
hydrodynamic model has been studied in \cite{Lee:2008bg}.

%%%%%%%%%%%%%%%%%%%%%%%%%%%%%%%%%%%%%%%%%%%%%%%%%%%%%%%
\begin{figure}
\begin{center}
\begin{minipage}{18cm}
\begin{center}
       \epsfig{file=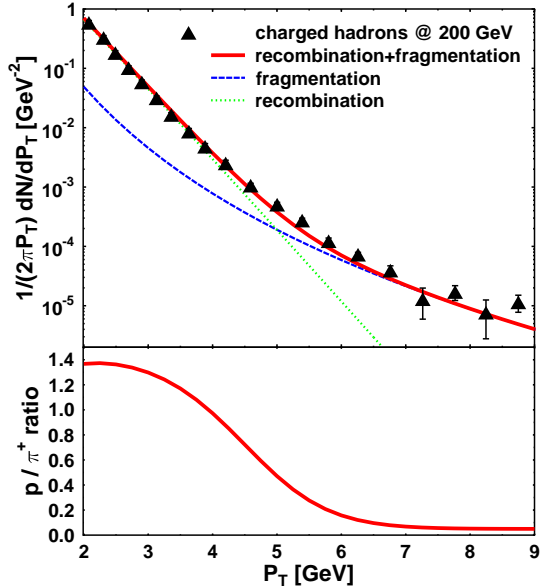, scale=0.5}
       \caption{Top panel: inclusive $P_T$-spectrum of charged hadrons 
in central Au + Au collisions at $\sqrt{s_{NN}}=200$ GeV. Contributions
from a recombination model and perturbative hadron production are shown
compared to experimental data from PHENIX \cite{Jia:2002bka}.
Bottom panel: ratio of protons to positive pions as a function of $P_T$ for
the same Au+Au collisions. 
The region below 4 GeV/$c$ is dominated by recombination, the region above 
6 GeV/$c$ by fragmentation of hadrons from jets. 
Figure taken from \cite{Fries:2003vb}.}
        \label{fig_reco_pt}
        \end{center}
\end{minipage}
\end{center}
\end{figure}
%%%%%%%%%%%%%%%%%%%%%%%%%%%%%%%%%%%%%%%%%%%%%%%%%%%%%%%

The quark number scaling law is a signature feature of quark recombination.
For a quark phase with elliptic flow $v_2^q(P_T)$ at the time of hadronization simple
instantaneous recombination models predict
\begin{equation}
  v_2^H(P_T) = n v_2^q\left(\frac{P_T}{n}\right)
\end{equation}
where $n$ is the number of valence quarks. This scaling law describes a 
key feature of experimental data at intermediate $P_T$ rather accurately.
We make two comments here: (i) the general shape of $v_2$ at intermediate
momenta suggests that in the kinematic range under consideration ($P_T$ between
1.5 and 5 GeV/$c$) hadrons are no longer equilibrated or at least there are
large viscous corrections to hydrodynamics; (ii) at lower momentum the scaling
seems to work rather with kinetic energy instead of $P_T$. This is a rather
accidental feature of $v_2$ for equilibrium hydrodynamics and, as already
brought up above, can not directly be attributed to quark recombination 
\cite{He:2010vw}.

We discuss the situation from the experimental point of view further in the
result section \ref{sec:results}. More profound detours into quark
recombination models are beyond the scope of this review and we refer the 
reader to several review articles on quark recombination for further study 
\cite{Fries:2008hs,Fries:2004ej,Becattini:2009fv}.

%%%%%%%%%%%%%%%%%%%%%%%%%%%%%%%%%%%%%%%%%%%%%%%%%%%%
% Results 
%%%%%%%%%%%%%%%%%%%%%%%%%%%%%%%%%%%%%%%%%%%%%%%%%%%%

\section{Interpretation of Experimental Data from RHIC \label{sec:results}}

We now proceed to discuss some key experimental results from the first
decade of running of the Relativistic Heavy Ion Collider. Many of those
results can be understood, at least qualitatively, within the 
framework of perturbative QCD and hydrodynamics. We will also review 
some attempts to extract quantitative statements about fireball parameters
and properties of quark gluon plasma. We proceed from the simplest
observables, single particle yields and spectra to more intricate ones and
try to take a comprehensive view that includes both bulk and hard probe
particles.

\begin{figure}[tb]
\epsfysize=9.0cm
\begin{center}
\begin{minipage}[t]{10 cm}
\begin{center}
\epsfig{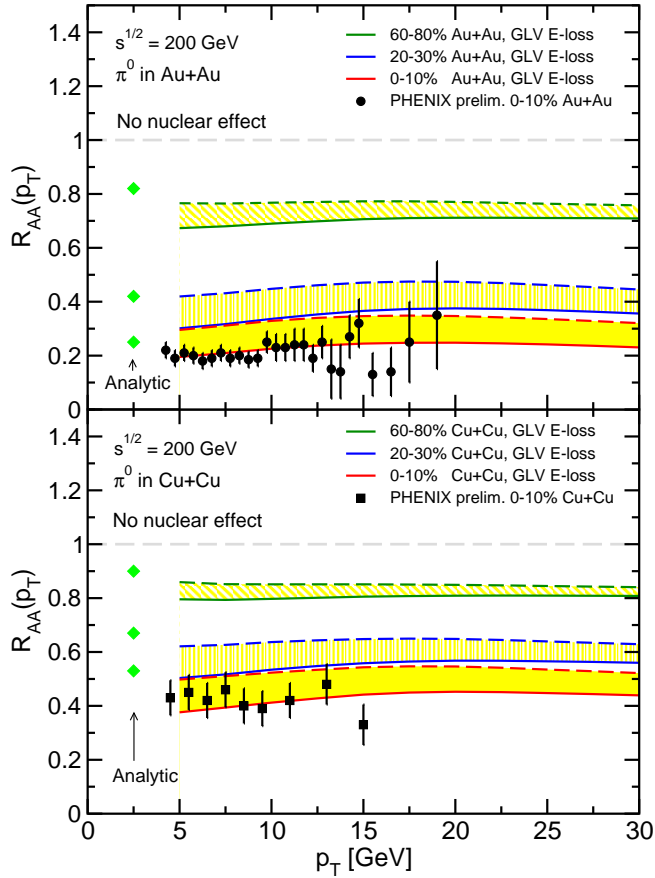}
\end{center}
\end{minipage}
\begin{minipage}[t]{16.5 cm}
\caption{$R_{AA}$ for neutral pions in Au+Au and Cu+Cu collisions at top 
 RHIC energy and for different centralities, calculated in the GLV model. 
 Finalized PHENIX data is now available in \cite{Adare:2008qa}.
 Figure reprinted from \cite{Vitev:2005he} with permission from Elsevier.
 \label{fig:glv}}
\end{minipage}
\end{center}
\end{figure}

\subsection{\it Particle Multiplicities and Single Particle Spectra}

Hydrodynamic models by default deliver the main qualitative features
of bulk spectra measured in heavy ion collisions at RHIC: hadrons at low
transverse momentum $P_T$ are thermalized and exhibit a
relativistic outward collective flow. Nevertheless it is a challenge
to quantitatively describe details, like the subtle differences seen 
between hadron species.
If we want to describe bulk observables with hydrodynamics or hybrid models
based on hydrodynamics the first goal is the determination of realistic
initial conditions. Even if some theoretical modeling of the initial state
is available, there are usually a few parameters that are fitted to single
particle $P_T$-spectra and rapidity distributions in central collisions.

\begin{figure}[tb]
\epsfysize=9.0cm
\begin{center}
\begin{minipage}[t]{18 cm}
\epsfig{file=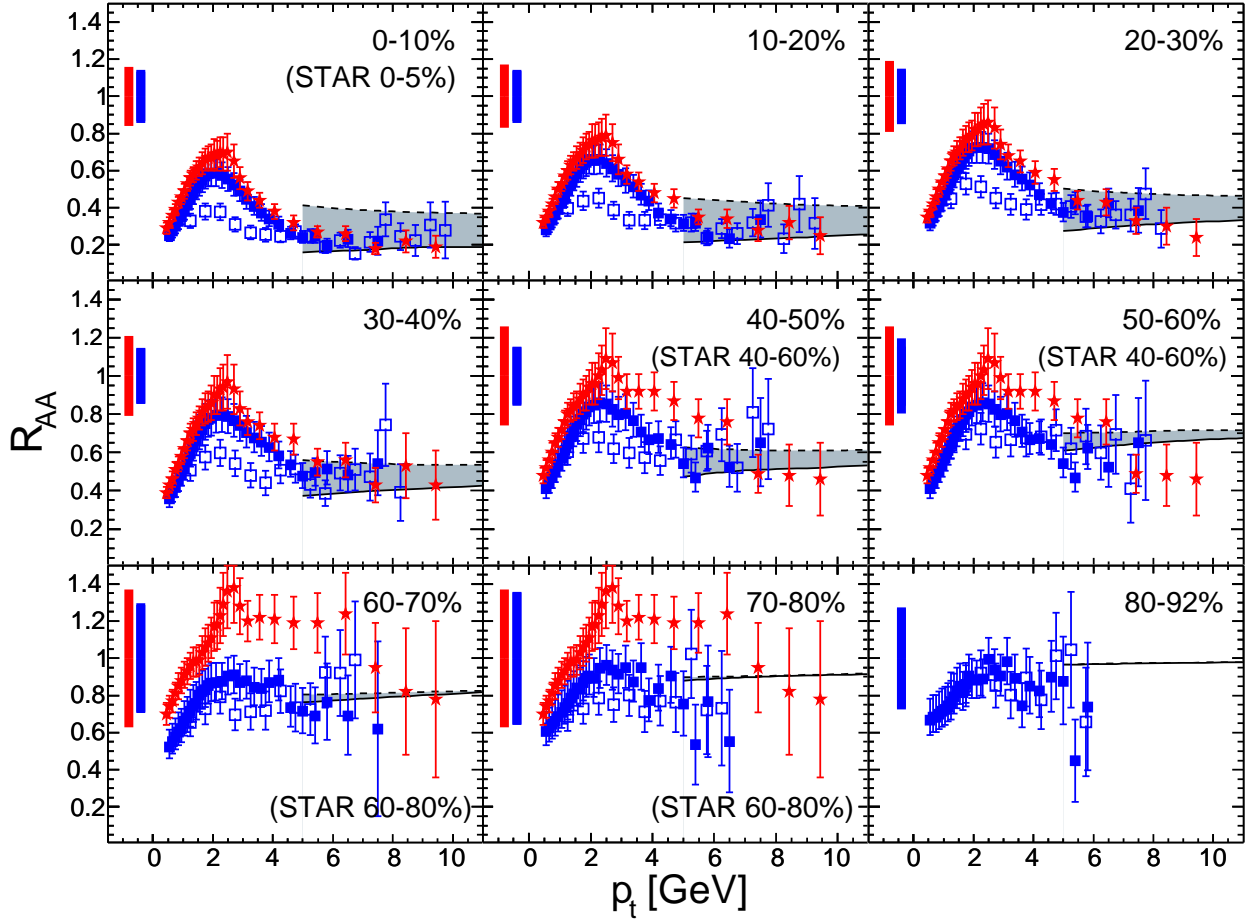,scale=0.9}
\end{minipage}
\begin{minipage}[t]{16.5 cm}
\caption{$R_{AA}$ for several centrality bins for Au+Au collisions at top 
  RHIC energy. Calculations using ASW energy loss with reweighted (dashed)
  and non-reweighted quenching weights (solid lines) are compared to data
  from PHENIX for charged hadrons (closed squares) and neutral pions
  (open squares) \cite{Adler:2003qi,Adler:2003au}, and from STAR for charge
  hadrons \cite{Adams:2003kv}. Figure reprinted from \cite{Dainese:2004te} with 
  kind permission from Springer Science + Business Media.
 \label{fig:daineseraa}}
\end{minipage}
\end{center}
\end{figure}

In Figs.~\ref{fig_rap} and \ref{fig_Pt} we show single particle spectra 
from a hybrid model of (3+1)-dimensional ideal, relativistic hydrodynamics and
UrQMD \cite{Nonaka:2006yn}.
In this model final state interactions at the end of the fireball life
time are taken into account by connecting the hydrodynamic phase to the
hadron-based event generator. 
For the determination of initial conditions, this hybrid model works the 
same way as pure hydrodynamics. Practically, the parameters to be determined 
are the maximum values of energy density and baryon number density which
are fixed from comparison with experimental data of pseudorapidity 
distributions and $P_T$-spectra in central collisions. The
centrality dependence of spectra is then settled and reproduces
the data well as shown in Fig.~\ref{fig_rap}.  

Figure \ref{fig_Pt} compares the $P_T$-spectra for $\pi^+$, $K$ and $p$ 
in central Au + Au collisions. Because the hybrid model correctly treats
chemical and kinetic freeze-out processes, taking into account the different
cross sections of hadrons, the absolute value of the proton spectra shows 
good agreement with the experimental data. This is one of the improvements 
that hybrid models with hadronic transport offer over pure hydrodynamic model 
with only one (thermal) freeze-out process. Such pure hydro calculations
usually fail to explain proton spectra. Clearly, taking into account
the correct freeze-out for each particle species separately is important
and hybrid models with hadronic transport deliver that.

Let us now investigate the detailed effect of resonance decays and hadronic 
rescattering on the shape of the momentum spectra. 
Figure \ref{fig_pt2} compares the $P_T$-spectrum for $\pi^+$ from hydrodynamics
only, i.e.\ at the switching temperature $T_{\rm sw} = 150$ MeV of the
hybrid model (solid line), to the spectrum after resonance decays have 
been taken into account in addition (open symbols). We also show
the result if the full UrQMD transport is run on the hydro result 
(solid circles). The difference between the solid line and open symbols 
directly quantifies the effect of resonance decays on the 
spectrum. They obviously increase the yield of pions, most dominantly in 
the low momentum region $P_T < 1$ GeV/$c$. Furthermore, the difference 
between open and solid symbols quantifies the effect of hadronic rescattering.
Pions with $P_T > 1$ GeV lose momentum resulting in a steeper slope. In other
words hadronic re-interactions cool the spectra. However, they do it
selectively with the appropriate cooling rates for different hadron 
species.

\begin{figure}[tb]
\epsfysize=9.0cm
\begin{center}
\begin{minipage}[t]{19 cm}
\epsfig{file=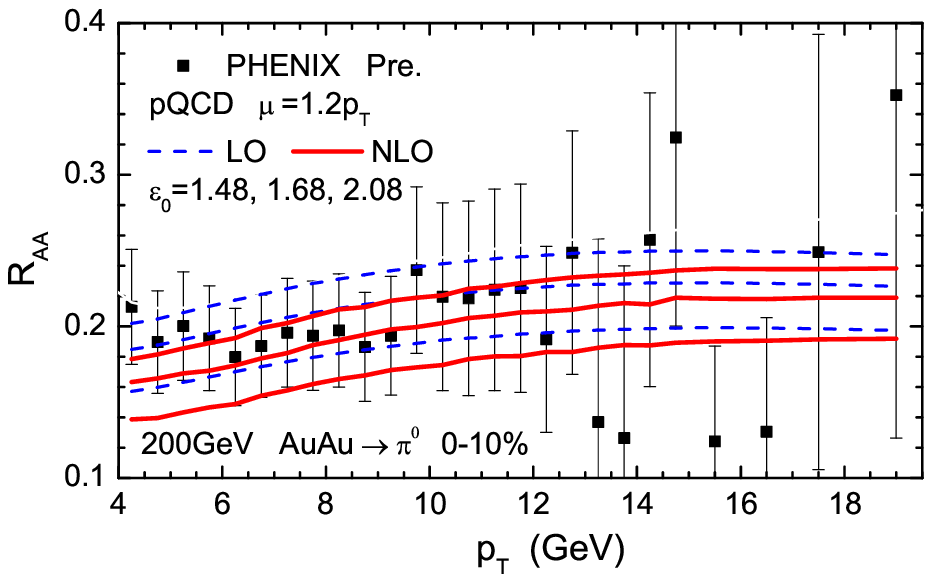,scale=0.9}\hspace{2em}
\epsfig{file=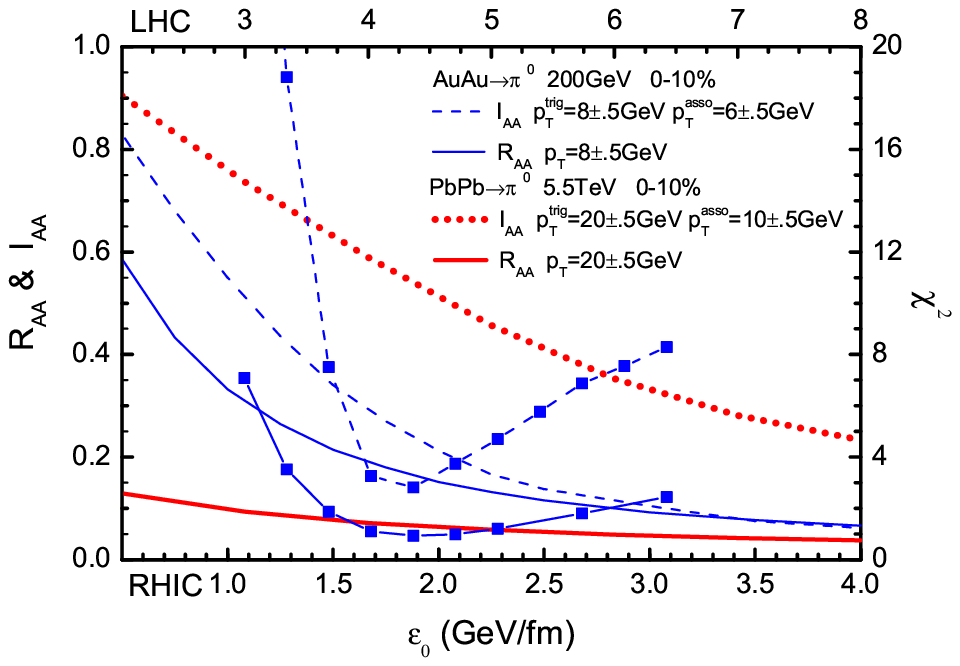,scale=0.9}
\end{minipage}
\begin{minipage}[t]{16.5 cm}
\caption{Left panel: $R_{AA}$ of neutral pions for central Au+Au collisions 
  at top RHIC energies using LO and NLO accuracy hard processes and 
  medium-modified fragmentation functions from the Higher Twist formalism
  compared to data from PHENIX. Finalized PHENIX data available in 
  \cite{Adare:2008qa}. 
  Right panel: $\chi^2$-fit of $\epsilon_0$ to RHIC data on $R_{AA}$ and 
  $I_{AA}$ yielding a consistent value for both observables.
  Figures reprinted from \cite{Zhang:2007ja} with permission from the American 
  Physical Society.
 \label{fig:owens}}
\end{minipage}
\end{center}
\end{figure}

To strengthen this point we show the mean transverse momentum 
$\langle P_T \rangle$ as a function of hadron mass from \cite{Nonaka:2006yn}
in Fig.~\ref{fig_meanpt}. As before we compare results at the switching 
temperature $T_{\rm sw} = 150$ MeV, corrected for hadronic decays (open
symbols) to the results from hydro plus full UrQMD (solid symbols).
In the former case $\langle P_T \rangle$ follows a straight line, as expected
from a hydrodynamic expansion. It is a natural consequence of collective
flow. If hadronic rescattering is taken into account $\langle P_T \rangle $ 
does no longer follow the straight line. The average momentum of pions 
is actually reduced by hadronic rescattering (they act as a heat bath 
in the collective expansion), whereas protons pick up additional 
transverse momentum in the hadronic phase. Data by the STAR Collaboration 
is shown as well (solid triangles). The proper treatment of hadronic final 
state interactions significantly improves the agreement of the
calculation with the data.

Generally, $P_T$-spectra from hydrodynamic calculations show good agreement 
with experimental data up to $P_T \sim 2$ GeV/$c$. For larger $P_T$ 
discrepancies between hydrodynamic results and experimental data appear, 
which suggest the approximate limits of applicability for hydrodynamics.
We can see these discrepancies between hydrodynamic results and experimental 
data in spectra in Fig.\ \ref{fig_hydrojet_pt} and in elliptic flow in
Fig.\ \ref{fig_v2_hydro_jets} (which we discuss in more detail later).
The general features of extended spectra can usually be described in
hydro plus jet models, as e.g.\ shown in Ref.\ \cite{Hirano:2003pw}.

Figure \ref{fig_hydrojet_pt} shows $P_T$-spectra for $\pi^-$, $K^-$ and $p$ 
in Au + Au collisions at $\sqrt{s_{NN}}=200$ GeV. The dotted lines indicate 
ideal hydrodynamics. These spectra exhibit a Boltzmann-like shape 
($\sim \exp(-E/T)$) which is locally boosted by a radial flow velocity $\langle 
u^\mu \rangle$ through $E \to p_\mu u^\mu$. This is the gross feature
of all $P_T$-spectra in hydrodynamic models and comparison with data suggests
that thermalization is reached. The figure also shows the contributions from 
hard (pQCD) hadron production (dash-dotted lines), whose spectra exhibit 
power-law behavior. Because of the exponential suppression of the bulk 
perturbative hadron production dominates at large enough $P_T$.
The sum of both hydro and perturbative (``jet'') contributions describes
the data quite well. The value of $P_T$ where the transition from the soft to
the hard component takes place depends on particle species. In the case of 
pions it happens between $P_T\sim 1$ and $P_T \sim 2$ GeV/$c$ in this
particular calculation. For protons the transition happens gradually
in the range $ 2 \leq P_T \leq 5$ GeV/$c$, but certainly at higher 
momentum than for pions. 
The crossing point is determined by two facts: one is radial 
flow which pushes the soft components toward high $P_T$. The other is the jet 
quenching mechanism which suppresses contributions from jets. However,
the simple hydro+jet model does not explain some striking features of
the systematics of different particle species.

\begin{figure}[tb]
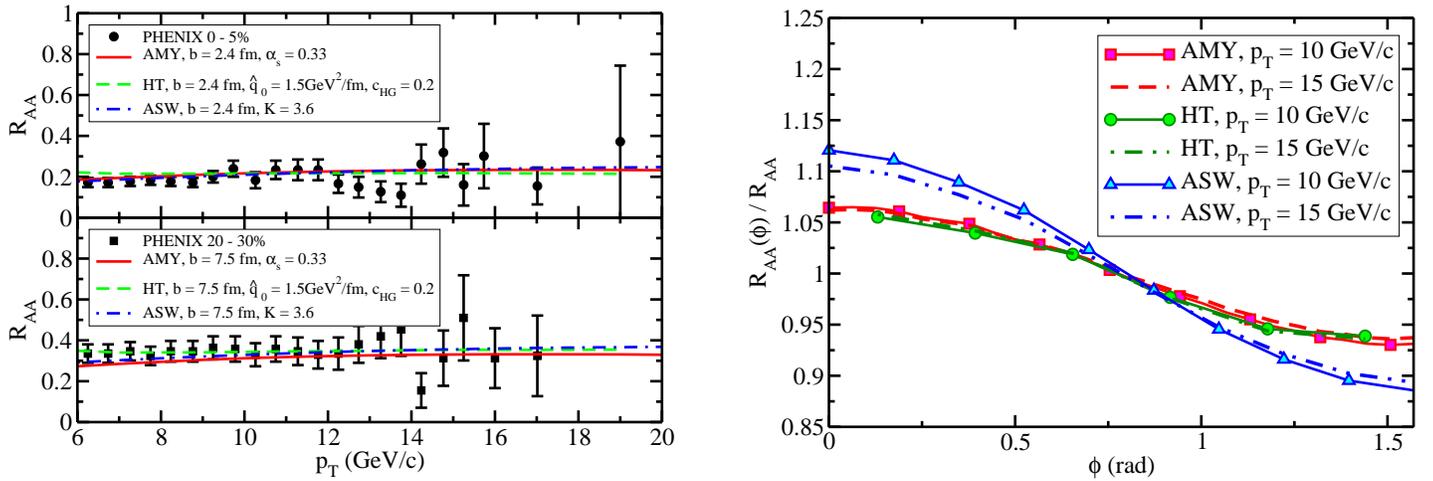

\epsfysize=9.0cm
\begin{center}
\begin{minipage}[t]{19 cm}
\epsfig{file=./plots/bass_RAA_centrality.eps,scale=0.36}\hspace{2em}
\epsfig{file=./plots/bass_RAA_reactionplane_normalized,scale=0.36}
\end{minipage}
\begin{minipage}[t]{16.5 cm}
\caption{Left panels: $R_{AA}$ as function of $P_T$ for central (top) and
mid-central (bottom) collisions calculated from the ASW, Higher
Twist and AMY energy loss models. The single parameter in each model has
been fitted to describe the data by PHENIX (finalized data available in 
\cite{Adare:2008qa}). Right panel:
$R_{AA}$ as function of azimuthal angle $\phi$ normalized by the 
$\phi$-integrated value for two different values of $P_T$. Again all 
three energy loss models are shown. Figures taken from \cite{Bass:2008rv}.
 \label{fig:RAAcomparison}}
\end{minipage}
\end{center}
\end{figure}

The differences in the transition points for different particles come from the
different (mass dependent) effects of radial flow, and from the different
yields of these particles in the fragmentation process (which disfavors
heavier mesons and baryons). The effects of different crossing points 
are most clearly seen in hadron ratios as a function of $P_T$. 
In Fig.\ \ref{fig_hydrojet_ratio} we show the hadron ratios $p/\pi^-$ and 
$K^-/\pi^-$ as a function of $P_T$ from the hydro+ jet model advocated in
\cite{Hirano:2003pw}. A crucial point in this figure is the fact that 
the ratio of $p/\pi^-$ is as large as unity at $P_T \approx 3$ GeV/$c$, which 
can not be understood from pQCD. This anomalously large yield of protons 
at intermediate $P_T$ was the first indication of the so-called baryon 
puzzle.

It was the baryon puzzle that gave birth to quark recombination models
in the RHIC era. Most of the experimental evidence for recombination comes 
from elliptic flow measurements, but the large baryon/meson ratios were 
essential to highlight the necessity for a mechanism that is able to push
the region of soft physics farther out for baryons than for mesons. 
One crucial event was the advent of first data on the $\phi$ meson. In a 
hydro+jet model $\phi$s essentially behave like protons, because of their mass, 
while in data they exhibit the universal behavior of other mesons 
(pions, kaons) at intermediate $P_T$ 
\cite{Fries:2003kq,Abelev:2008fd,Nonaka:2003hx}.

Figure \ref{fig_reco_pt} shows the $P_T$-spectrum of charged particles and 
the $p/\pi^+$ ratio as a function of $P_T$ from the model advocated in
\cite{Fries:2003vb} which includes quark recombination and hadrons from
pQCD (``fragmentation''). The proper hydrodynamic region below
2 GeV/$c$ has been omitted. From comparison with experimental data from
STAR one can conclude that the region below 4 GeV/$c$ is actually dominated by 
quark recombination while the pure pQCD region free of soft or bulk 
hadron production only starts beyond 6 GeV/$c$. Similar conclusions
have been reached in other recombination + jet models.
We will find even stronger evidence when we discuss elliptic flow in the
next subsection, where recombination shows clearly visible and 
experimentally tested differences compared with both hydrodynamics 
and perturbative hadron production.

%%%%%%%%%%%%%%%%%%%%%%%%%%%%%%%%%%%%%%%%%%%%%%%%%%%%%%%
\begin{figure}[tb]
%\begin{center}
\begin{minipage}{8.5cm}
\epsfig{file=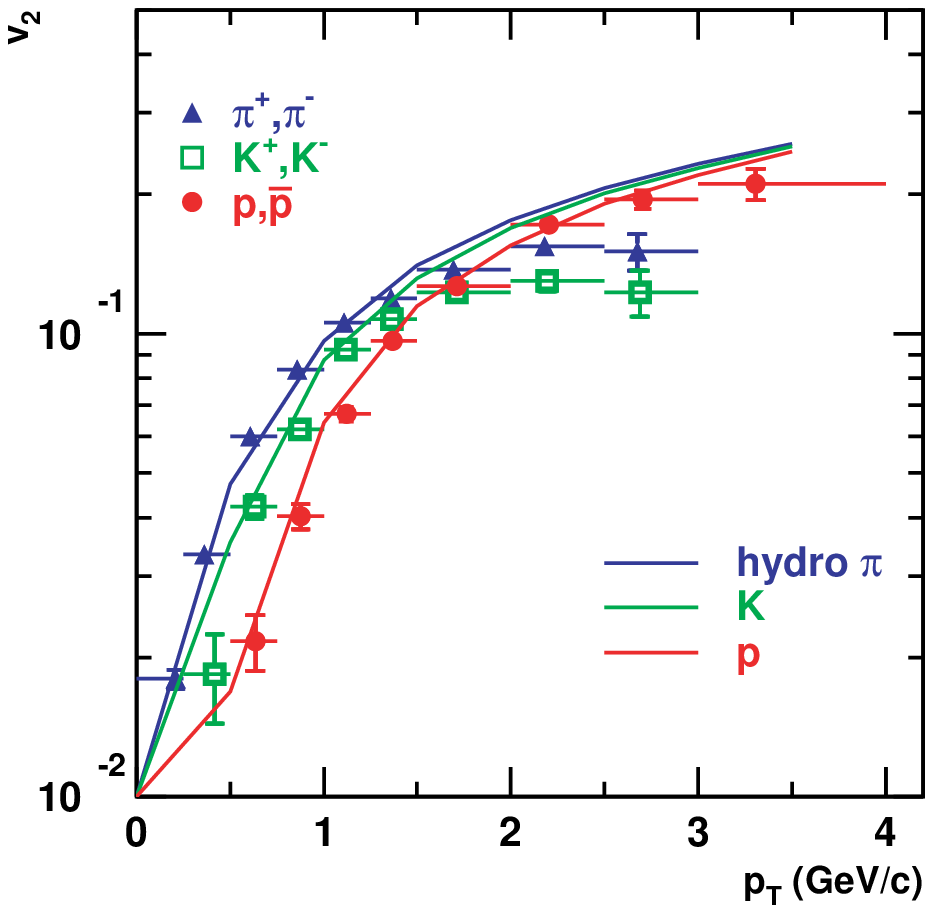,scale=0.8}
\caption{$v_2(P_T)$ for pions, kaons and protons produced in minimum-bias
  collisions at RHIC from an ideal hydro calculation by Huovinen et al.\      
  \cite{Huovinen:2001cy} compared to data from PHENIX \cite{Adcox:2004mh}.
  Figure reprinted from \cite{Adcox:2004mh} with permission from Elsevier.}
\label{fig_v2_pt}
\end{minipage}
\hspace{0.5cm}
\begin{minipage}{8.5cm}
\epsfig{file=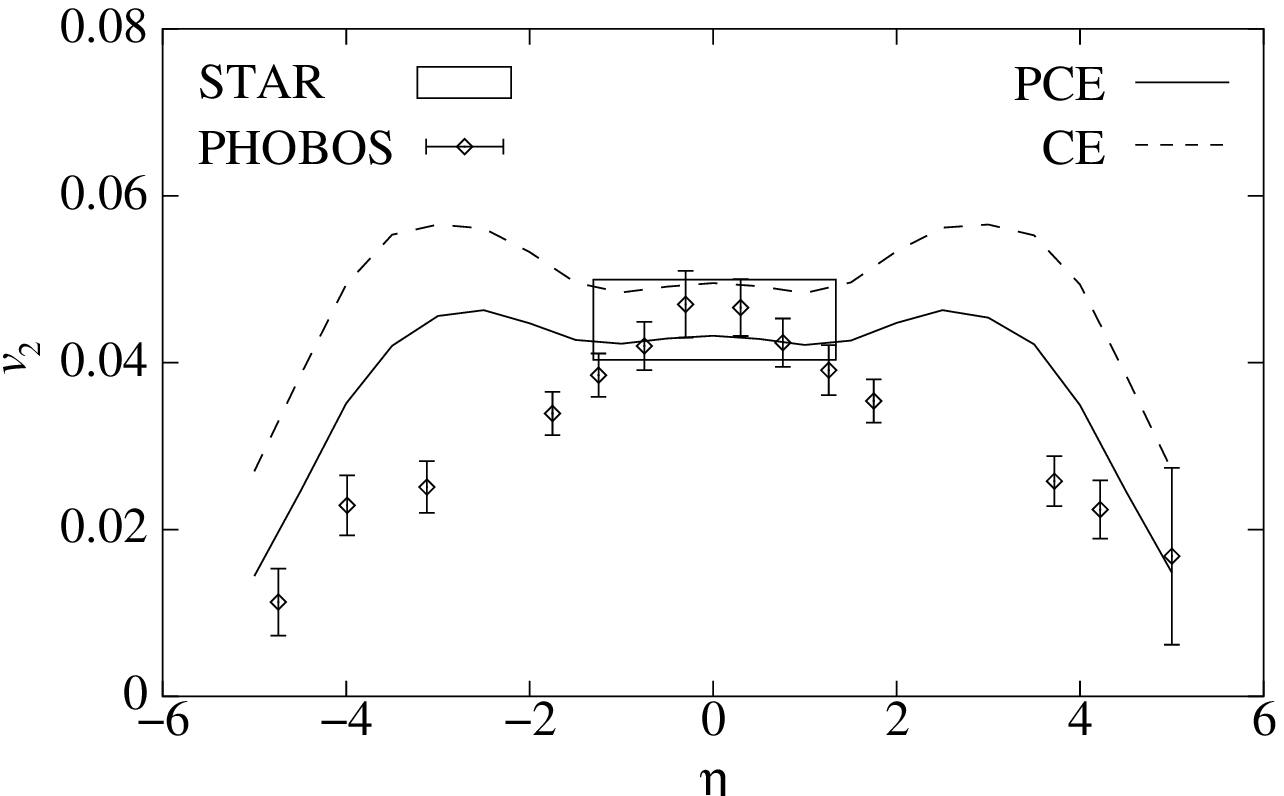,scale=0.65}
\caption{$v_2$ as a function of pseudorapidity $\eta$ for charged hadrons 
 in Au+Au collisions at $\sqrt{s_{NN}}= 130$ GeV. 
 The solid and dashed lines correspond to the hydrodynamic calculations 
 with partial chemical equilibrium (PCE) and full chemical equilibrium (CE) 
 respectively. Data is taken from STAR \cite{Ackermann:2000tr}
 and PHOBOS \cite{Back:2002gz}.
 Figure reprinted from \cite{Hirano:2002ds} with permission from the American 
 Physical Society.
\label{fig_PCE_v2}
}
\end{minipage}
%\end{center}
\end{figure}
%%%%%%%%%%%%%%%%%%%%%%%%%%%%%%%%%%%%%%%%%%%%%%%%%%%%%%%

Beyond 6 GeV/$c$ in transverse momentum hadron spectra are dominated by
perturbative production that includes energy loss through final state
interactions with the medium and fragmentation. Since the interesting
observable is the suppression with respect to high-$P_T$ jets and hadrons
in the vacuum the best way to analyze the data is by considering the
ratio
\begin{equation}
  R_{AA} = \frac{dN^{(AA)}/dP_T}{ \langle N_{\mathrm{coll}}\rangle dN^{(pp)}/dP_T}
\end{equation}
of yields in A+A vs $p+p$ collisions. This nuclear modification factor is
similar to what we had defined in Eq.\ (\ref{eq:rdau}) for $d+$A collisions. 
$\langle N_{\mathrm{coll}}\rangle$ is
the average number of binary nucleon-nucleon collisions that we expect
for a given centrality bin. It is usually determined from Glauber-type
model calculations. An incoherent superposition of nucleon-nucleon collisions
would lead to $R_{AA}= 1$ modulo isospin effects. We have already
discussed in Sec.\ \ref{sec:initstate} how initial state nuclear effects
lead to modest deviations from unity.

As had been predicted, first RHIC data at high $P_T$ revealed a huge
suppression of hadrons \cite{Adler:2003qi,Adams:2003kv,Abelev:2006jr,
Adare:2008qa,Abelev:2009wx}. 
This strong jet quenching is one of the pillars on which a qualitative 
argument for the creation of quark gluon plasma at RHIC energies rests. 
Fig.\ \ref{fig:glv} shows calculations by Vitev within the GLV model 
for the $R_{AA}$ of neutral pions in Au+Au and Cu+Cu at RHIC energies of 
$\sqrt{s_{NN}} = 200$ GeV for different centralities \cite{Vitev:2005he}.
The calculation also takes into account nuclear effects in the initial state. 
The quenching is as large as a factor 5 in central Au+Au collisions as
indicated by the data from PHENIX. Vitev connects the quenching strength
$\hat q$ to a local gluon density $dN_g/dy$ in the medium and he finds
$dN_g/dy \approx 800 - 1175$ for central Au+Au collisions.

%%%%%%%%%%%%%%%%%%%%%%%%%%%%%%%%%%%%%%%%%%%%%%%%%%%%%%%
\begin{figure}[tb]
%\begin{center}
\begin{minipage}{8.5cm}
\epsfig{file=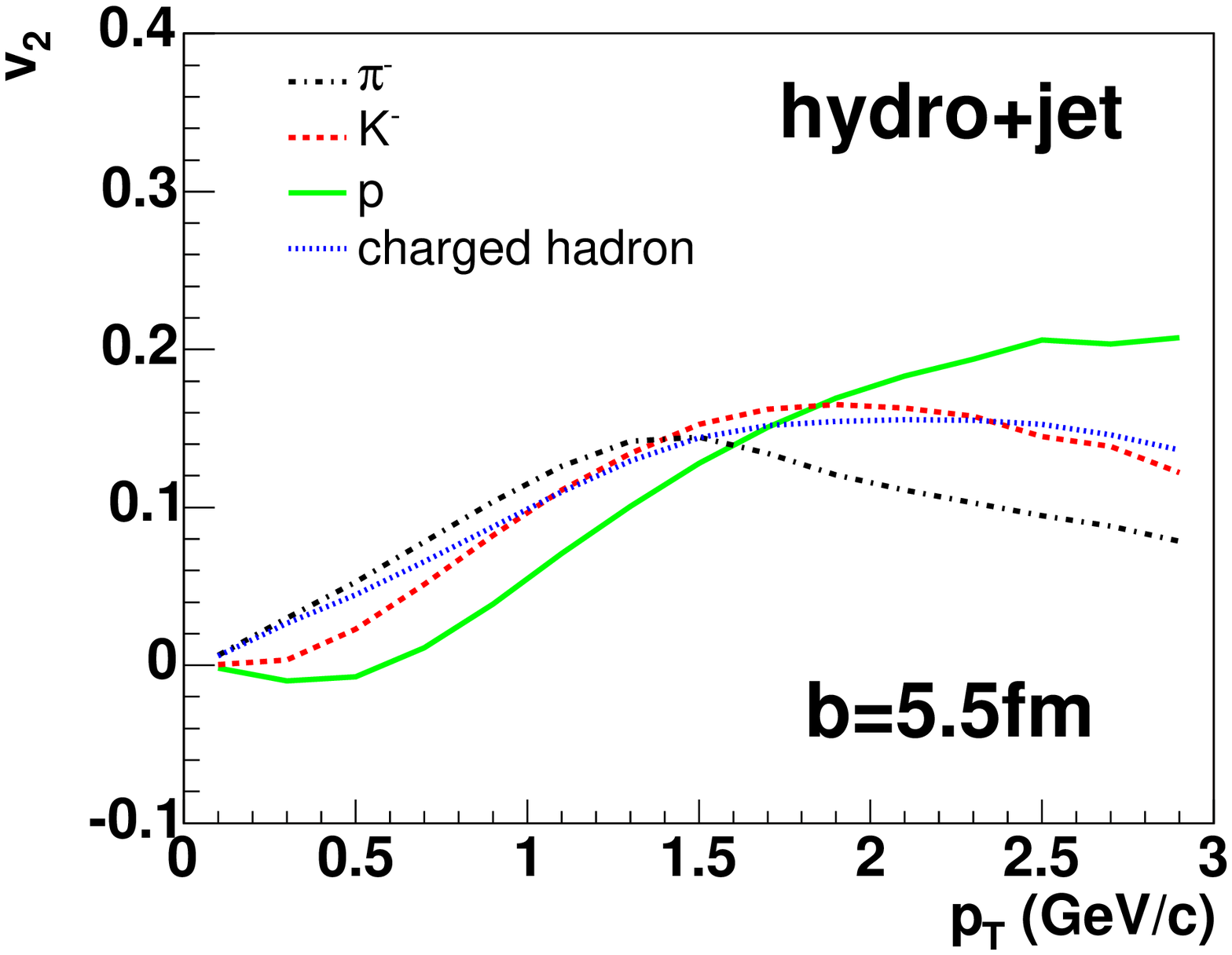, scale=0.45}
\caption{Elliptic flow $v_2(P_T)$ for $\pi^-$, $K^-$, and $p$ in 
Au+Au collisions at $\sqrt{s_{NN}}=200$ GeV at impact parameter $b=5.5$ 
fm from the hydro+jet model \cite{Hirano:2003pw}. Figure reprinted from
\cite{Hirano:2003pw} with permission from the American Physical Society.}
\label{fig_v2_hydro_jets}
\end{minipage}
\hspace{0.5cm}
\begin{minipage}{8.5cm}
\begin{center}
\epsfig{file=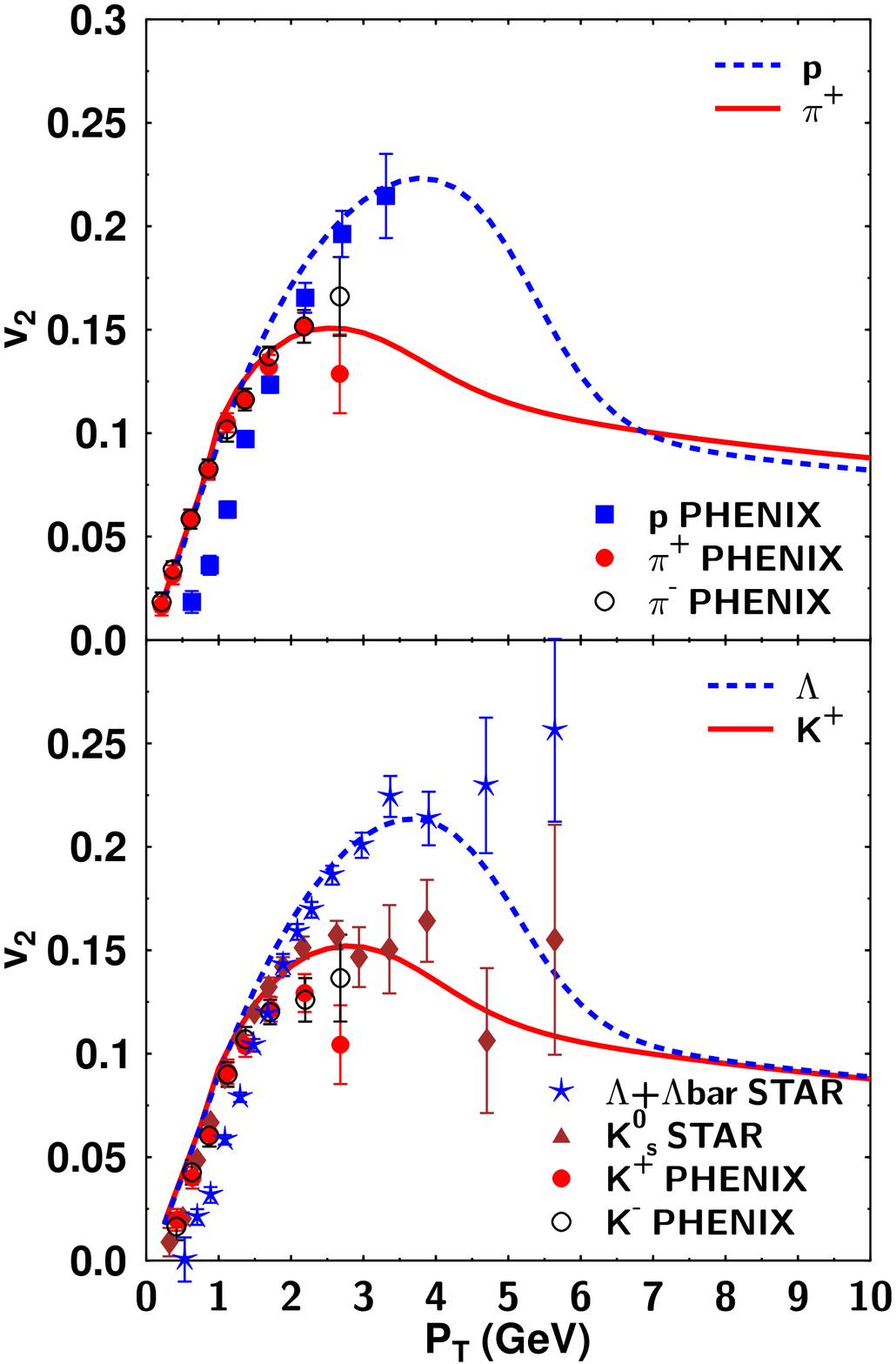, scale=0.42}
\caption{Upper panel: $v_2$ for $p$ and $\pi^+$ from the recombination
plus fragmentation model in \cite{Fries:2003vb}.
Lower panel: $v_2$ for $K^+$ and $\Lambda + \bar{\Lambda}$ 
from the same model. Data shown is from PHENIX \cite{Adler:2003kt} and 
STAR \cite{Snellings:2003mh}. Figure taken from \cite{Fries:2003vb}.}
\label{fig_v2_reco}
\end{center}
\end{minipage}
%\end{center}
\end{figure}
%%%%%%%%%%%%%%%%%%%%%%%%%%%%%%%%%%%%%%%%%%%%%%%%%%%%%%%

Figure \ref{fig:daineseraa} shows a systematic study of the centrality
dependence of the $R_{AA}$ of charged hadrons in Au+Au collisions 
carried out by Dainese, Loizides and Paic using an implementation of
the ASW energy loss model \cite{Dainese:2004te}. They model the
centrality dependence by scaling not only the production of jets but
also the local $\hat q$ by the density of binary nucleon-nucleon collisions,
$\hat q(\mathbf{x}) = k n_{\mathrm{coll}}$. This is different from other
approaches that usually scale soft particle densities in the transverse 
plane by the density of nucleon participants $n_{\mathrm{part}}$. 
The centrality dependence of the data from PHENIX and STAR is
reproduced well by just one adjustable parameter, the normalization $k$.
So far, no conclusive picture has emerged which scaling of $\hat q$ 
is favored by data. 
The fit to data is aided by the large theoretical uncertainty that comes 
from the difference between the reweighted and non-reweighted versions 
of the ASW quenching weights that have been discussed earlier. Those 
two results define the grey bands in Fig.\ \ref{fig:daineseraa}.
The average quenching found by the authors within the ASW formalism
is $\hat q \approx 15$ GeV$^2$/fm. This value, like others obtained with
the ASW model, are rather large compared to those from alternative energy 
loss calculations as we will see below.

Zhang et al.\ have presented one of the first studies of jet quenching
using next-to-leading order hard processes \cite{Zhang:2007ja}. The
left panel of Fig.\ \ref{fig:owens} compares their results for the $R_{AA}$
of $\pi^0$ in central Au+Au collisions at LO and NLO accuracy
with data from PHENIX. They use medium-modified fragmentation functions
inspired by the Higher Twist formalism that are rescaled by the average
energy loss of a parton, and then use these instead of NLO
vacuum fragmentation functions in a next-to-leading order code for hadron
production. They parameterize the energy loss through an integral over
a transverse profile of an expanding fireball along the path of the
parton times a normalization parameter $\epsilon_0$ that characterizes
the stopping power parameter of the medium. The left panel of 
Fig.\ \ref{fig:owens}
shows their result for three different values of $\epsilon$ that 
define an approximate uncertainty band.
Both the LO and NLO calculations can describe the data if $\epsilon_0$ is
a fit parameter. If $\epsilon_0$ is kept fixed the NLO calculation shows
stronger quenching due to the larger ratio of gluon to quark jets which
couple more strongly to the medium.
The right panel of Fig.\ \ref{fig:owens} shows a $\chi^2$-fit of
$\epsilon_0$ to $R_{AA}$ between 4 and 20 GeV/$c$ and to $I_{AA}$ (discussed
below), yielding consistent values in a range $\epsilon_0 = 1.6 \ldots
2.1$ GeV/$c$ \cite{Zhang:2007ja}.

%%%%%%%%%%%%%%%%%%%%%%%%%%%%%%%%%%%%%%%%%%%%%%%%%%%%%%%
\begin{figure}[tb]
\begin{center}
\begin{minipage}{18cm}
\begin{center}
\epsfig{file=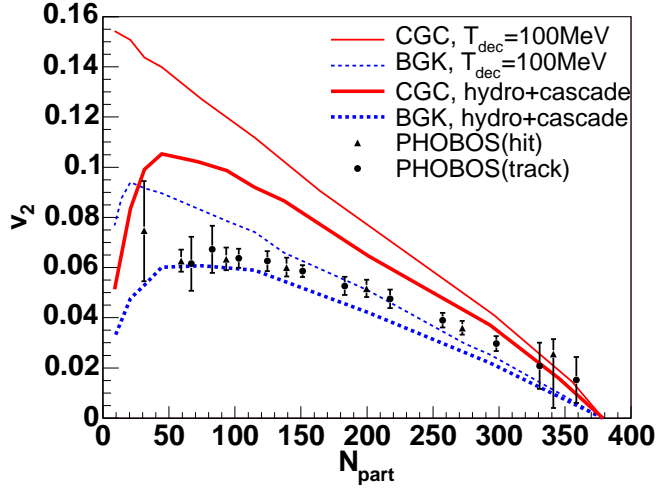, scale=0.45}
\caption{
$P_T$-integrated elliptic flow for charged hadrons at midrapidity
($|\eta | < 1$) from Au + Au collisions at RHIC top energy as a function 
of the number of participating nucleons $N_{\rm part}$. Thin lines show 
the predictions from ideal fluid dynamics only with a freeze-out temperature 
$T_{\rm dec} = 100$ MeV while thick lines indicate the results from the 
hydro+JAM hybrid model. 
For both models two different initial conditions, CGC and BGK, are used.
Figure reprinted from \cite{Hirano:2005xf} with permission from Elsevier.}
 \label{fig_v2_npart}
\end{center}
\end{minipage}
\end{center}
\end{figure}
%%%%%%%%%%%%%%%%%%%%%%%%%%%%%%%%%%%%%%%%%%%%%%%%%%%%%%%

Fig.\ \ref{fig:RAAcomparison} shows results from a comparative study by 
Bass et al. Jets are propagated through a medium described by hydrodynamics,
using three different schemes for energy loss: ASW, Higher Twist, and
AMY \cite{Bass:2008rv}. 
The left panels show $R_{AA}$ as a function of $P_T$
for two different centrality bins. They serve as proof that both
the $P_T$-dependence and the centrality dependence of $R_{AA}$ can 
be described by all three models. Every model has one free
parameter that has been fitted: the strong coupling $\alpha_S$ for AMY,
$\langle \hat q \rangle$ or derived parameters for the HT and ASW formalisms.
In this particular case a normalization factor $K = \hat q/(2\epsilon^{3/4})$ 
as explained in Eq.\ (\ref{eq:bdmpsqhat}) was fitted for ASW, where
$\epsilon$ is the local energy density.

This study confirms the surprisingly large $\hat q$ found in the ASW model
compared to other approaches. For the case that the quenching strength
scales with $\epsilon^{3/4}$ the initial values found for a quark at the 
center of the fireball in central collision are \cite{Bass:2008rv}
\begin{equation}
  \hat q = 18.5\>  \text{GeV}^2/\text{fm for ASW}\, , \qquad
  \hat q = 4.5 \> \text{GeV}^2/\text{fm for HT}
\end{equation}
and for the case that the quenching strength scales like the temperature $T$
it is found that
 \begin{equation}
  \hat q = 10 \> \text{GeV}^2/\text{fm for ASW}\, , \qquad
  \hat q = 2.3 \> \text{GeV}^2/\text{fm for HT}\, , \qquad
  \hat q = 4.1 \> \text{GeV}^2/\text{fm for AMY} \, .
\end{equation}
Recall that the rates in AMY are calculated self-consistently as functions
of the local temperature so there is only one choice to model the
space and time dependence.

%%%%%%%%%%%%%%%%%%%%%%%%%%%%%%%%%%%%%%%%%%%%%%%%%%%%%%%
\begin{figure}[tb]
%\begin{center}
\begin{minipage}{8.5cm}
\epsfig{file=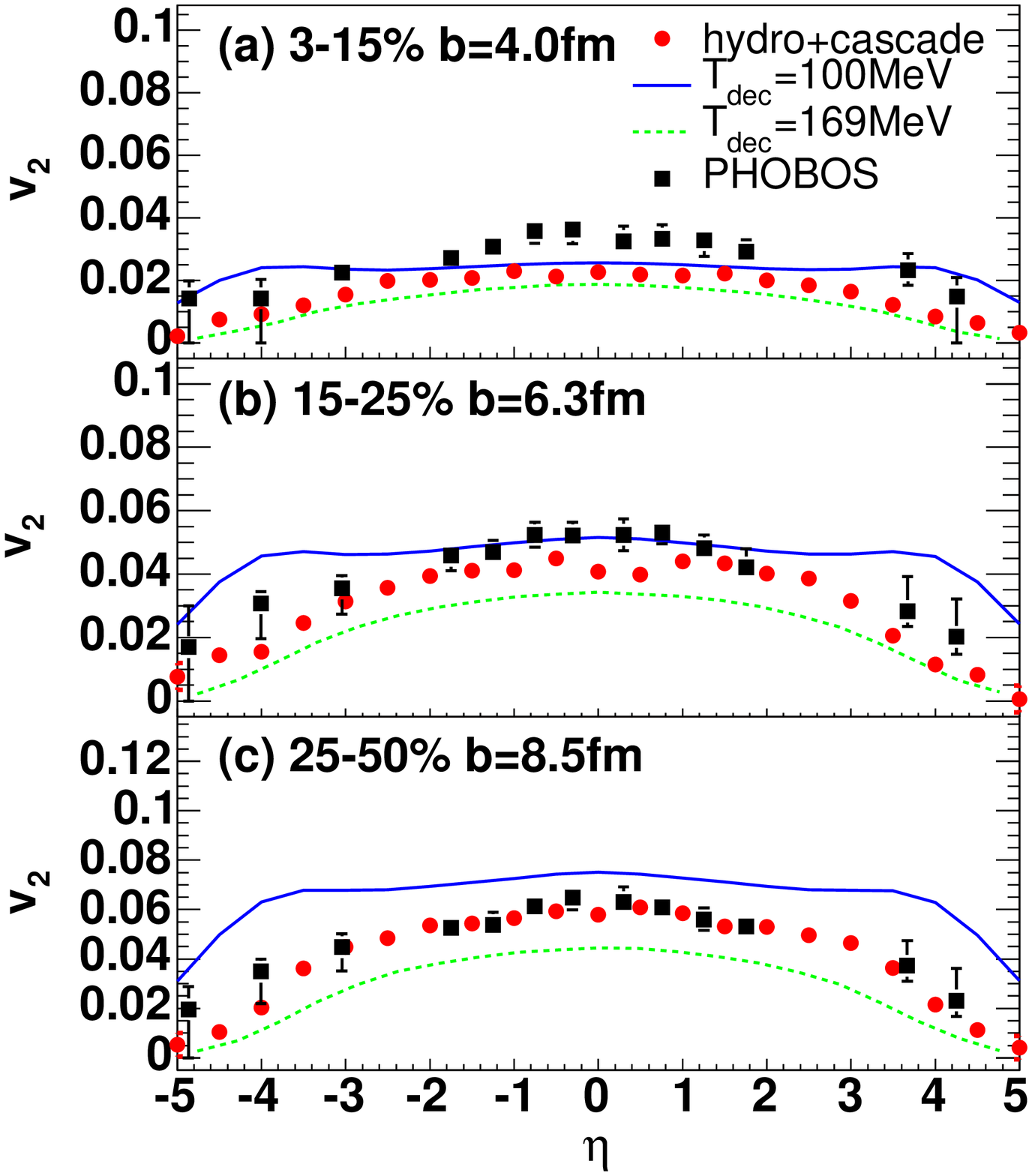,scale=0.42}
\end{minipage}
\hspace{0.5cm}
\begin{minipage}{8.5cm}
\epsfig{file=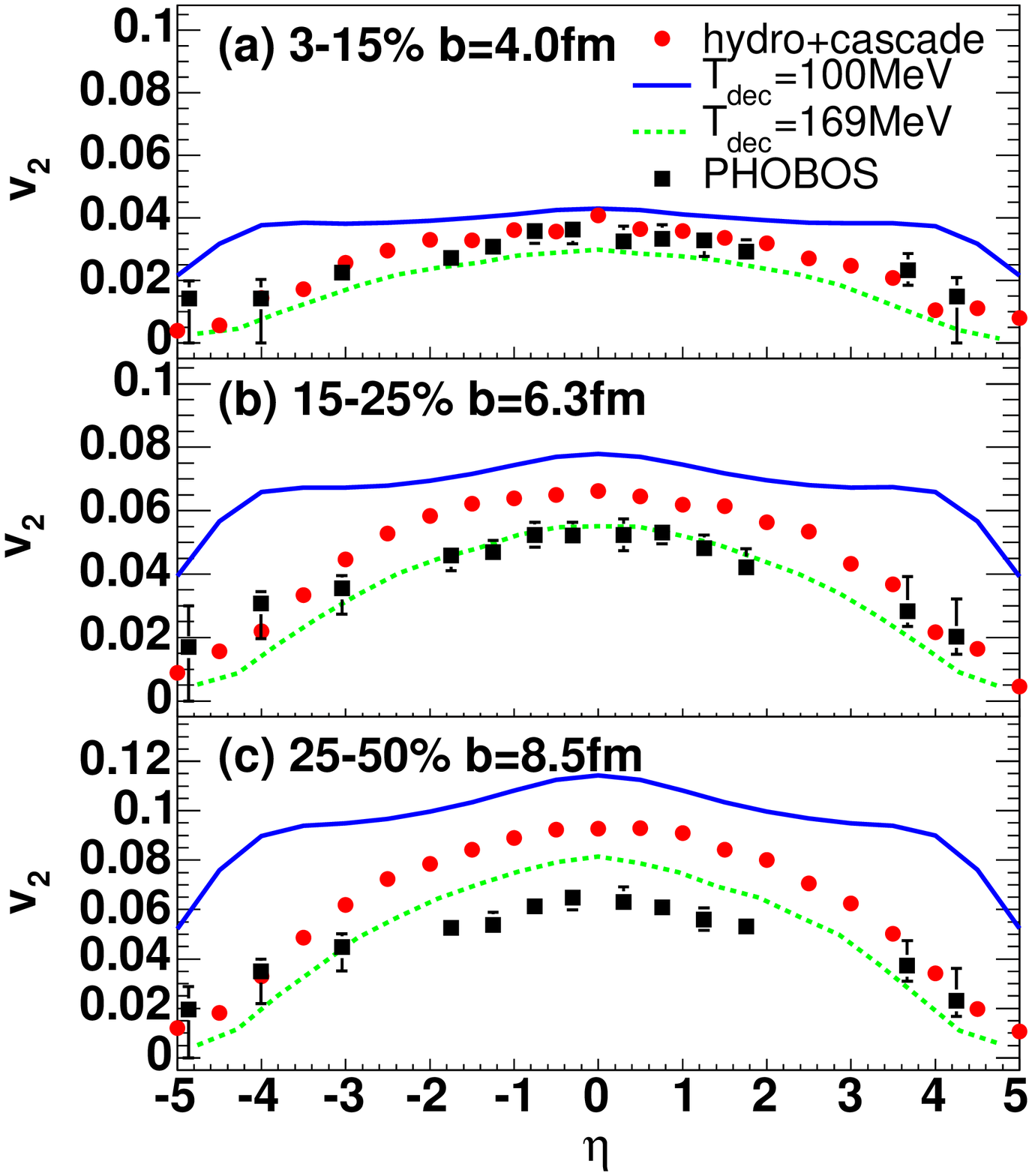, scale=0.42}
\end{minipage}
\caption{Left panel: The pseudorapidity dependence of $v_2$ for charged 
hadrons in (a) central (3-15\%), (b) semi-central (15-25\%), and (c) 
peripheral (25-50\%) Au + Au collisions at $\sqrt{s_{NN}} = 200$ GeV. 
The corresponding impact parameters are $b = 4.0$, 6.3, and 8.5 fm resp. 
The hydrodynamic evolution is initialized with modified BGK initial conditions. 
The lines show the predictions from ideal fluid dynamics only for freeze-out
temperatures of $T_{\rm dec} = 100$ MeV (solid blue) 
and $T_{\rm dec} = 169$ MeV (dashed green). Red circles show the corresponding 
results from the hydro+JAM hybrid model. 
Right panel: Same as left panel, except for using CGC instead of BGK 
initial conditions.
Figures reprinted from \cite{Hirano:2005xf} with permission from Elsevier.}
\label{fig_v2_rap_hc}
%\end{center}
\end{figure}
%%%%%%%%%%%%%%%%%%%%%%%%%%%%%%%%%%%%%%%%%%%%%%%%%%%%%%%

This comparison is unique and very valuable since the same initial
hard cross sections and the same maps for the fireball, from (3+1)-dimensional 
ideal hydrodynamics were used. Any differences in the extracted values of 
$\hat q$ must be due to differences between the calculations themselves, not due to
differences in implementation. One of the conclusions is that our current knowledge
applied to $R_{AA}$ leaves a rather large uncertainty in the determination
of $\hat q$.

The right panel of Fig.\ \ref{fig:RAAcomparison} shows $R_{AA}$ as a
function of the angle $\phi$ with respect to the reaction plane normalized 
by the average $R_{AA}$. Due to the large values of
$\hat q$ the ASW formalism is more strongly dominated by surface emission than 
the other models. This also leads to a stronger angular modulation for
non-spherical fireballs.

%%%%%%%%%%%%%%%%%%%%%%%%%%%%%%%%%%%%%%%%%%%%%%%%%%%%%%%
% Flow, elliptic flow
%%%%%%%%%%%%%%%%%%%%%%%%%%%%%%%%%%%%%%%%%%%%%%%%%%%%%%%

\subsection{\it Azimuthal Anisotropies and Elliptic Flow}

Anisotropies in the azimuthal angle $\phi$, especially elliptic flow, contain 
detailed information on the hot and dense QCD bulk matter created at RHIC.
We will use the notion of elliptic flow for the second harmonic
\begin{equation}
  v_2(P_T) = \frac{\int d\phi \cos 2\phi \, dN/d \phi dP_T}{
        \int d\phi \, dN/d \phi dP_T}
\end{equation}
for any value of $P_T$, even if the asymmetry is not produced through
hydrodynamic flow. 
Large elliptic flow of the bulk matter produced at RHIC had been observed
early on and has led to the claim of perfect fluidity of the quark gluon
plasma just above $T_c$ \cite{Gyulassy:2004zy}. With the advent of 
viscous relativistic hydro codes the interest has shifted toward the 
goal of quantifying the dissipative transport coefficients, in particular 
the shear viscosity $\eta$.

%%%%%%%%%%%%%%%%%%%%%%%%%%%%%%%%%%%%%%%%%%%%%%%%%%%%%%%
\begin{figure}[tb]
\begin{center}
\begin{minipage}{12cm}
\epsfig{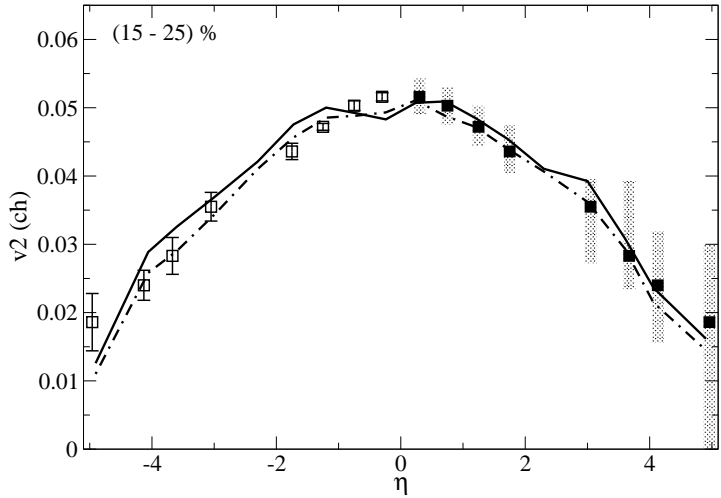}
\caption{
Comparison between true elliptic flow (solid line)
and a suggested method to compute reconstructed elliptic flow
from data $v^{\rm Rec}_2$ (dash-dotted line). Squares represent 
PHOBOS data for mid-central collisions \cite{Back:2002gz,Back:2004mh,
Back:2004zg}. See Ref.~\cite{Andrade:2006yh} for detail. Figure reprinted from  
\cite{Andrade:2006yh} with permission from the American Physical Society.}
\label{fig_v2_fluc_1}
\end{minipage}
\end{center}
\end{figure}
%%%%%%%%%%%%%%%%%%%%%%%%%%%%%%%%%%%%%%%%%%%%%%%%%%%%%%%

In the hydrodynamic regime the asymmetry in the pressure gradient in
and out of the reaction plane for finite impact parameters 
drives a larger acceleration in-plane than out-of-plane.
Figure \ref{fig_v2_pt} shows the elliptic flow $v_2$ as a function of $P_T$ 
for $\pi$, $K$ and $p$ from an early calculation using ideal hydrodynamics
by Huovinen et al.\ \cite{Huovinen:2001cy}.   
The calculations show remarkable agreement with experimental data.
We can observe a clear mass ordering: the $v_2$ for pions is larger 
than that of kaons which in turn is larger that that of protons. The
effect is most pronounced at low $P_T$. This is a natural phenomenon in
hydrodynamics and comes from the interplay of radial flow in and out of the
reaction plane. The effects of flow are more pronounced for more massive
particles due to their smaller thermal velocities at a given temperature,
which translates into smaller asymmetries.
As for spectra the good agreement between hydrodynamic models and 
experimental data is restricted to low $P_T$ as effects of insufficient
thermalization kick in at higher momenta.
Despite the pioneering character of this calculation it is at present no
longer regarded as being very realistic due to the following issues: 
(i) the lack of proper treatment of chemical vs thermal freeze-out, (ii) 
the fact that it assumes boost-invariance and ignores the dynamics 
in longitudinal direction, (iii) the absence of final state interactions, 
(iv) the absence of dissipative corrections.
As we had pointed out earlier hydrodynamic models that do not take into
account the difference of chemical and kinetic freeze-outs can not explain the 
absolute values of proton spectra correctly. However this calculation
gives us the guidelines for the necessary improvements that have been
implemented since then.

Figure \ref{fig_PCE_v2} shows elliptic flow at RHIC as a function of 
pseudorapidity $\eta$. This calculation is carried out with a 
(3+1)-dimensional ideal hydro model with two freeze-out processes 
(labeled PCE = partial chemical equilibrium) \cite{Hirano:2002ds}. These 
features improve the $P_T$-spectra of protons and give reasonable results for 
$v_2(P_T)$ for $\pi$, $K$ and $p$ at midrapidity. However the authors of this
study find that they can not explain the data from RHIC away from midrapidity.
Besides generally overestimating transverse flow, there are humps in forward 
and backward rapidity which do not feature in the experiment data 
\cite{Ackermann:2000tr,Back:2002gz}. It should be noted that the
rapidity spectra are described well after two adjustable parameters in
the initial conditions in longitudinal direction had been fixed. 
This result might indicate that thermalization is reached 
only very close to midrapidity.

%%%%%%%%%%%%%%%%%%%%%%%%%%%%%%%%%%%%%%%%%%%%%%%%%%%%%%%
\begin{figure}[tb]
%\begin{center}
\begin{minipage}{8.5cm}
\epsfig{file=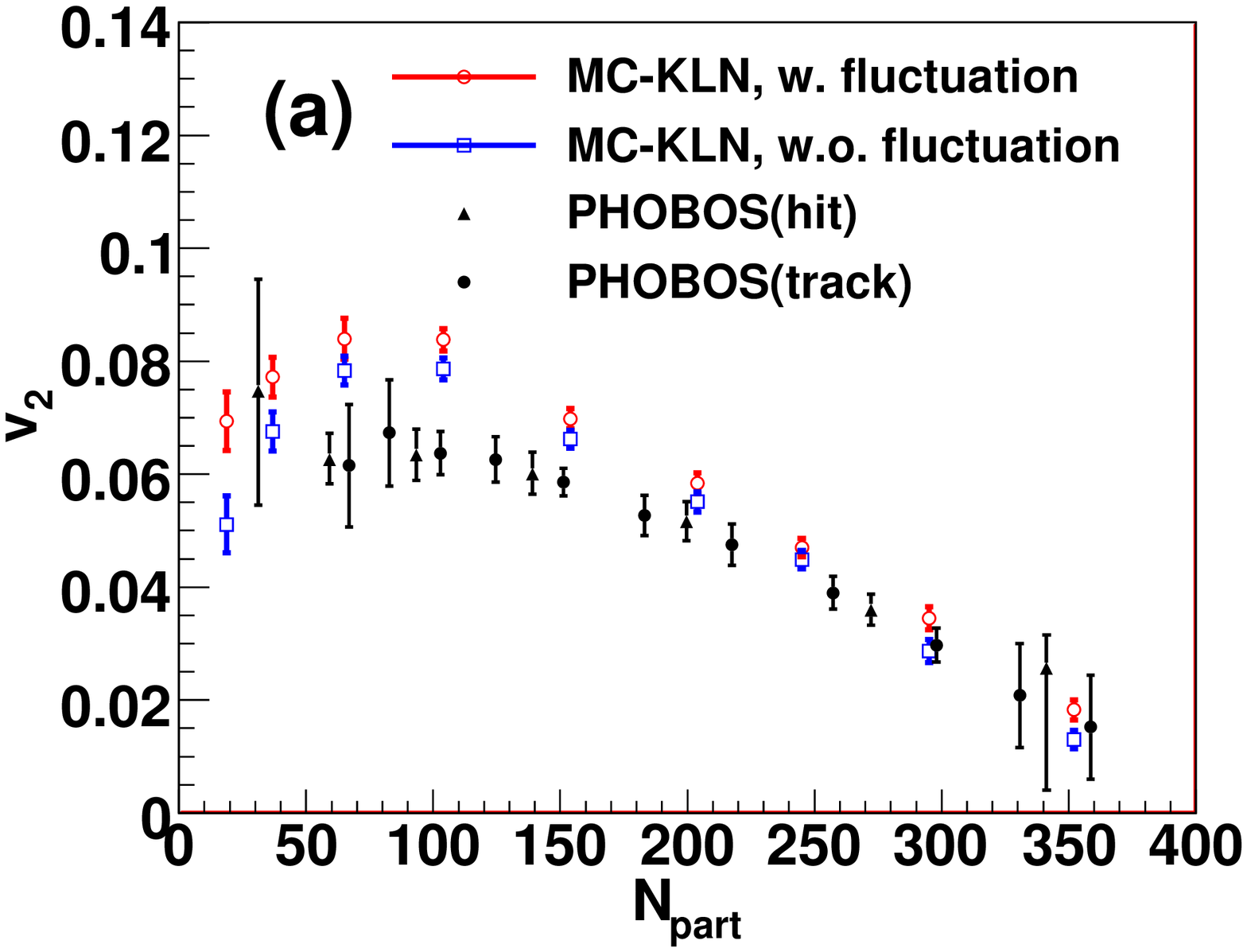,scale=0.4}
\end{minipage}
\hspace{0.5cm}
\begin{minipage}{8.5cm}
\epsfig{file=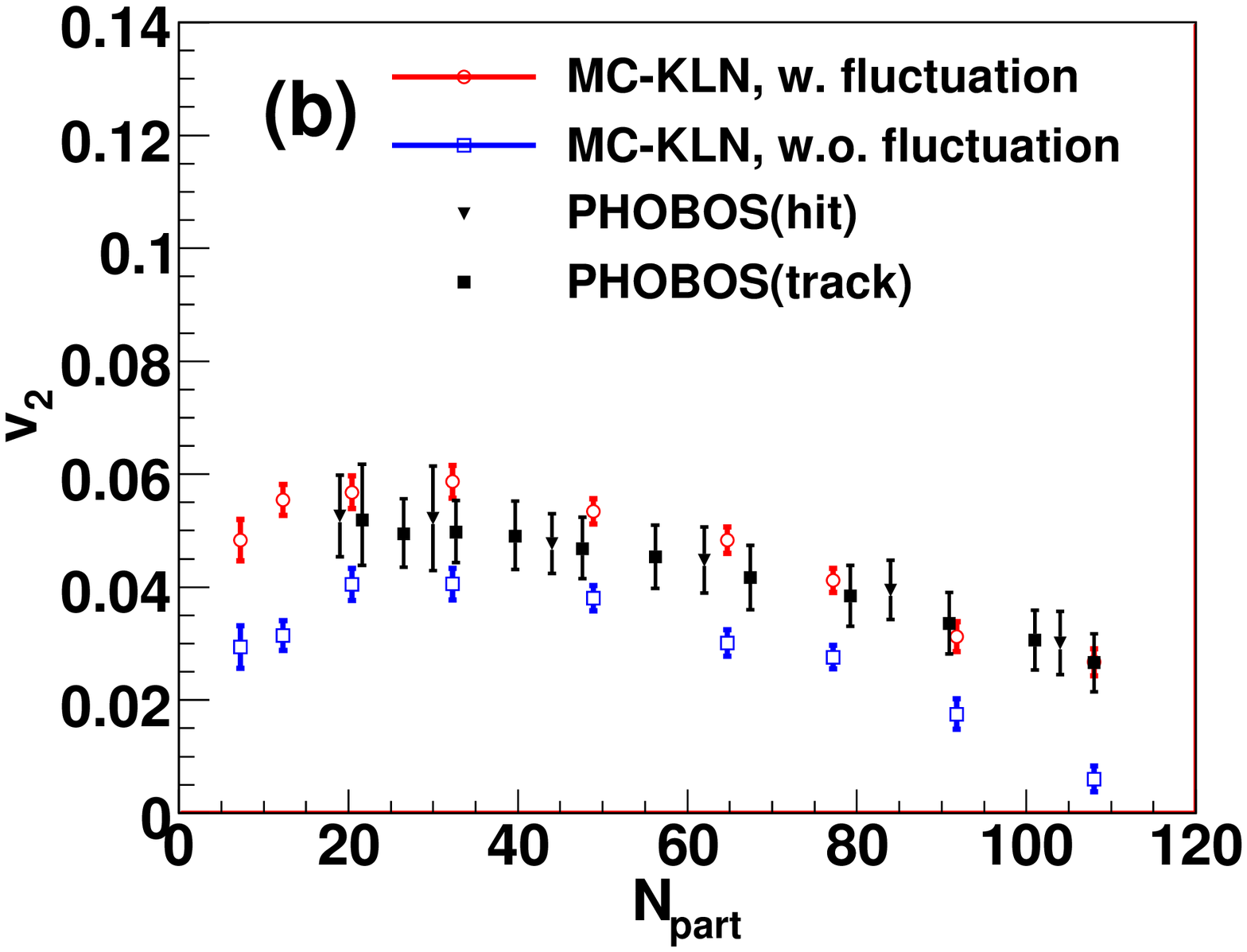, scale=0.4}
\end{minipage}
\caption{Left panel (a): Centrality dependences of $v_2$ for charged 
particles at midrapidity in Au+Au collisions at $\sqrt{s_{NN}} = 200$ 
GeV with CGC initial conditions. Open circles (squares) are results with 
(without) eccentricity fluctuations. 
Right panel (b): The same for Cu+Cu collisions.
Figures reprinted from \cite{Hirano:2009ah} with permission from the American 
Physical Society.}
\label{fig_v2_fluc_2}
%\end{center}
\end{figure}
%%%%%%%%%%%%%%%%%%%%%%%%%%%%%%%%%%%%%%%%%%%%%%%%%%%%%%%

As in the case of spectra one can also infer from comparison of elliptic
flow with data that hydrodynamic models work only at $P_T < 2$ GeV/$c$, see
Fig.\ \ref{fig_v2_pt}. Naturally we expect a transition to a region that
is dominated by perturbative production, with a recombination region in
between.
Note that the mechanism of $v_2$ generation is different for hard probes. 
An azimuthal asymmetry develops simply because of the different amount 
of material jets have to go through in- and out-of-plane, with smaller 
opacity in-plane.
Fig.\ \ref{fig_v2_hydro_jets} shows $v_2$ as a function of $P_T$ for 
$\pi^-$, $K^-$, $p$ and charged hadrons for the hydro + jet model 
in \cite{Hirano:2003pw}. 
In experimental data we observe that the shape of the $P_T$-dependence of $v_2$ 
changes from that of a pure hydro model by bending over at larger
$P_T$, saturating for a while at intermediate $P_T$, and then dipping down 
at even larger values of $P_T$. The transition points, as for spectra, 
depend on particle species.
In addition, the saturation levels at intermediate $P_T$ shows a peculiar
universality with baryons and mesons lining up at two different values
of $v_2$ which scale like 3:2 \cite{Abelev:2008ed,Afanasiev:2009wq}.

Hydro + jet models get some of these basic features, but usually can not
explain the systematics of transition points and the universal baryons 
vs meson saturation levels. This is a strong argument for the presence 
of quark recombination at intermediate $P_T$. Figure \ref{fig_v2_reco} 
shows $v_2$ as a function of $P_T$ for identified particles from the 
recombination plus fragmentation model advocated in Ref.\ \cite{Fries:2003vb}
together with early experimental data. 
Recombination models naturally deliver the universal behavior within
the baryon and meson groups and indeed predict the valence quark
number scaling of $v_2$ as discussed earlier in this review.
The data clearly support this stunning feature and have supported the claim that
traces of collectivity at the parton level can be seen \cite{Fries:2003vb}.
At higher $P_T$ perturbative hadron production takes over. It is generally 
expected that $v_2$ in the perturbative domain does not depend too 
much on hadron species, but reliable data above 6 GeV/$c$ is scarce. 
Large differences between hadrons could be a sign of hadron- instead of 
parton-based jet quenching, or they could indicate rapid changes in 
jet chemistry inside a quark gluon plasma \cite{Liu:2008kj}.
We discuss more theoretical results for $v_2$ at high $P_T$ below.
Figure \ref{fig_v2_reco} shows general agreement of the calculations with
data except for the small $P_T$-region where the mass splitting is not
resolved since no genuine hydrodynamics phase with correct hadron masses
was used.

While recombination models lend a helpful hand to hydrodynamics  
to extend the bulk properties to larger $P_T$, hadronic transport is
able to fix our failing understanding of the pseudorapidity dependence 
discussed earlier, and of the centrality scaling of elliptic flow
\cite{Nonaka:2006yn,Hirano:2005xf}.
In Fig.~\ref{fig_v2_npart} we show the $P_T$-integrated elliptic flow 
for charged hadrons at midrapidity as a function of the number of 
participating nucleons $N_{\rm part}$. The plot compares results
obtained with an ideal (3+1)-dimensional hydrodynamic model with
and without the hadronic cascade model JAM attached as an afterburner
from Hirano et al. \cite{Hirano:2005xf}. In this study the authors also 
compare two
different initial conditions for their hybrid model: a Glauber model (BGK) 
and Color Glass Condensate-based model (CGC). We refer the reader to
Ref.\ \cite{Hirano:2005xf} for more details. We observe that the
hybrid hydro+JAM model with Glauber initial conditions gives the
best description of the centrality dependence of elliptic flow.
The hydro+JAM model with CGC initial conditions which is generally considered 
more realistic is consistent with experimental data only at large 
$N_{\rm part}$. 
   
%%%%%%%%%%%%%%%%%%%%%%%%%%%%%%%%%%%%%%%%%%%%%%%%%%%%%%%
\begin{figure}[tb]
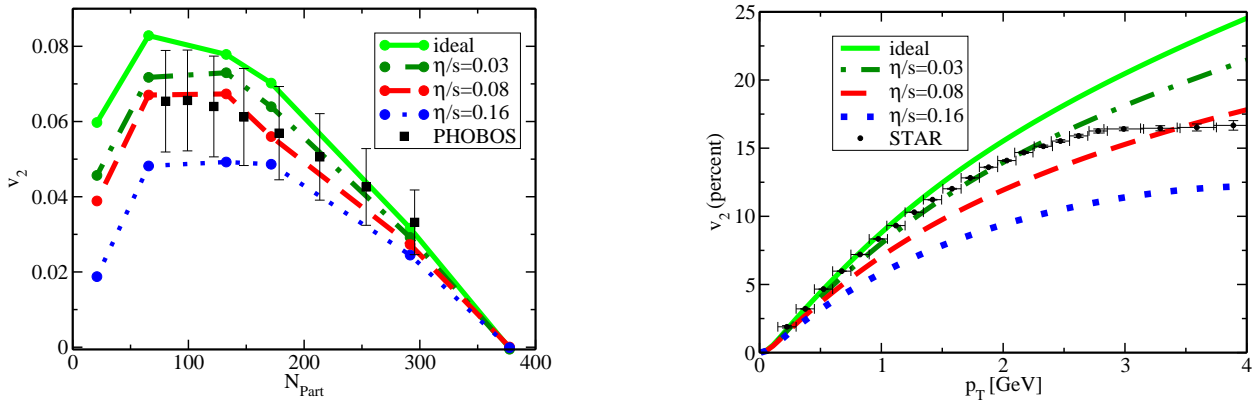

%\begin{center}
\begin{minipage}{8.5cm}
\epsfig{file=./plots/Romatschke_PRL99_figb_v2int.eps, scale=0.3}
\end{minipage}
\hspace{0.5cm}
\begin{minipage}{8,5cm}
\epsfig{file=./plots/Romatschke_PRL99_fig3a_v2again.eps, scale=0.3}
\end{minipage}
\caption{Left panel: Integrated $v_2$ as a function of $N_{\rm part}$ 
for charged particles in Au+Au collisions at $\sqrt{s_{NN}}=200$ GeV,	
compared to viscous hydrodynamics with shear viscosity for various 
viscosity-to-entropy ratios $\eta/s$ and compared to data from PHOBOS
\cite{Alver:2007qw}. 
Right panel: $v_2$ as a function of $P_T$ for the same system
compared to data from STAR \cite{Adams:2003zg}.
Figures reprinted from \cite{Romatschke:2007mq} with permission from the American Physical Society.}
\label{fig_v2_vis}
%\end{center}
\end{figure}
%%%%%%%%%%%%%%%%%%%%%%%%%%%%%%%%%%%%%%%%%%%%%%%%%%%%%%%

The pseudorapidity dependence of $v_2$ for charged hadrons in central, 
semi-central and peripheral Au+Au collisions at $\sqrt{s_{NN}} = 200$ GeV is 
shown in Fig.~\ref{fig_v2_rap_hc}. By comparing results from
pure ideal hydrodynamics at decoupling temperatures of $T_{\rm dec}=100$ MeV 
and $T_{\rm dec}=169$ MeV respectively, we see that the bumps at forward 
and backward rapidities that were already observed in the previous study
\cite{Hirano:2002ds} are larger if the hadronic matter is allowed to 
evolve without dissipation.
On the other hand, $v_2(\eta)$ from the hybrid hydro plus hadronic cascade
model does not allow these structures to build up in $v_2(\eta)$.
We can conclude from all of these hybrid studies with hadronic cascades 
that effects of shear viscosity and dissipation in the hadronic phase,
and proper final state interactions are not negligible.  
We also note that once more the initial conditions based on the Glauber model 
have the upper hand over CGC initial conditions in comparison with data.
Color glass initial conditions would require additional dissipation
during the QGP phase. However, even this much improved investigation did
not take into account the effect of event-by-event fluctuations of the
geometric shape of the density of the initial condition which affects 
the elliptic flow.

The importance of including event-by-event fluctuations was discussed 
systematically by Andrade et al.\ using the hydrodynamic code 
NEXSPHERIO \cite{Andrade:2006yh} which includes a Monte-Carlo generator for
initial conditions. They show that the assumption of symmetry 
of the particle distribution in relation to the reaction plane leads 
to disagreement between the true and reconstructed elliptic flows and  
emphasize that it is important to have a precise experimental determination 
of elliptic flow. Their calculated $v_2$ as a function of pseudorapidity 
shows very nice agreement with experimental data as shown in
Fig.~\ref{fig_v2_fluc_1}. However, they did not connect their hydrodynamic 
model to a transport code for the hadronic phase.

The effect of eccentricity fluctuations on the elliptic flow at midrapidity 
in Au +Au and Cu + Cu collisions was recently also investigated in 
Ref.~\cite{Hirano:2009ah}. Those authors include the effect of initial 
eccentricity fluctuations originating from the nucleon position inside 
the colliding nuclei both for the Glauber model and CGC initial conditions. 
The effect of eccentricity fluctuations is not very large in semi-central 
Au + Au collisions and it does not shift the values of $v_2$ closer to 
experimental data in that region. On the other hand, it enhances $v_2$ in 
Cu + Cu collisions where fluctuations are more important because of the 
smaller system size. As a result $v_2(\eta)$ from CGC initial conditions 
with fluctuations can describe the experimental data quite well.

%%%%%%%%%%%%%%%%%%%%%%%%%%%%%%%%%%%%%%%%%%%%%%%%%%%%%%%
\begin{figure}[tb]
\begin{center}
\begin{minipage}{12cm}
\epsfig{file=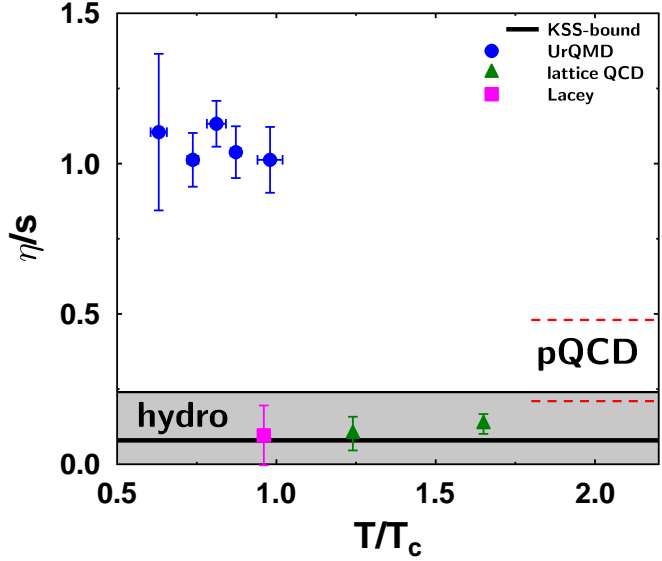, scale=0.6}
\caption{
Estimated band of $\eta/s$ form current viscous hydrodynamics models in 
Tab.\ \ref{table-viscous-hydro} and from several theoretical estimates
discussed in the text. 
}
\label{fig_eta-s}
\end{minipage}
\end{center}
\end{figure}
%%%%%%%%%%%%%%%%%%%%%%%%%%%%%%%%%%%%%%%%%%%%%%%%%%%%%%%

Finally we show results for elliptic flow from one of the recently developed
(2+1)-dimensional relativistic viscous hydrodynamic codes. 
Since both the formulation and the implementation of relativistic, second 
order viscous hydrodynamics is non-trivial, qualitative comparison with
experimental data is just starting.
Figure \ref{fig_v2_vis} shows results for $v_2$ of charged particles
in Au+Au collisions obtained by Romatschke and Romatschke 
\cite{Romatschke:2007mq} together with experimental data from STAR and PHOBOS.  
This study finds that both integrated $v_2$ as a function of 
centrality, and differential $v_2$ as a function of $P_T$ need very small 
values of $\eta/s$, $0.03 \ldots 0.08$, partially violating the 
conjectured KSS bound.
It is too early to reach conclusive results in light of the many small
improvements that ideal hydrodynamics needed to implement over a decade
to become a reliable tool.
E.g.\ the study mentioned here considers only one type of initial
condition (Glauber) and it does not implement final state interactions
\cite{Romatschke:2007mq}.\footnote{
Luzum and Romatschke investigate the initial and freeze-out temperature 
dependence of elliptic flow, using a Glauber-based model and CGC, see
Sec.\ \ref{sec:init}.}   
Nevertheless, the first efforts have jump started the era
of qualitative studies using dissipative hydrodynamics 
\cite{Teaney:2009qa,Song:2007ux}.

In Fig. \ref{fig_eta-s} we show a band for favored values of the shear
viscosity over entropy ratio $\eta/s$ which is extracted from 
the recent viscous hydrodynamic models listed in 
Tab.  \ref{table-viscous-hydro}. We also show results  from lattice QCD 
\cite{Meyer:2007ic}, UrQMD \cite{Demir:2008tr}, pQCD 
(Fig.\ 10 in Ref.\ \cite{Teaney:2009qa}) 
and phenomenological analyses \cite{Lacey:2006bc} for reference.
This $\eta/s$ band from viscous hydrodynamics should not be considered as  
a conclusive result, but it highlights an interesting preliminary finding: 
The $\eta/s$ extracted from comparison of viscous hydrodynamics with RHIC data
is generally in the vicinity of the KSS bound.  
From the comparison with UrQMD and lattice QCD\footnote{The calculation is 
performed with SU(3) pure gauge theory.}, we would conclude that those small values 
of $\eta/s$ must come from the QGP phase.  
These facts support the hypothesis of a sQGP at RHIC.  
In these viscous hydrodynamic calculations the temperature dependence of viscosities 
is usually not taken into account, as shown in Fig. \ref{fig_eta-s}. 
The temperature dependence of transport coefficients may be not very significant 
in the QGP phase at RHIC.  
However the shear viscosity of the hadron phase seems to be much larger than that
of the QGP phase which suggests that it is necessary to take this difference
into account when discussing phenomena related to the QCD phase transition 
\cite{Hirano:2005wx}.

%%%%%%%%%%%%%%%%%%%%%%%%%%%%%%%%%%%%%%%%%%%%%%%%%%%%%%%
\begin{figure}[tb]
\epsfysize=9.0cm
\begin{center}
\begin{minipage}[t]{19 cm}
\epsfig{file=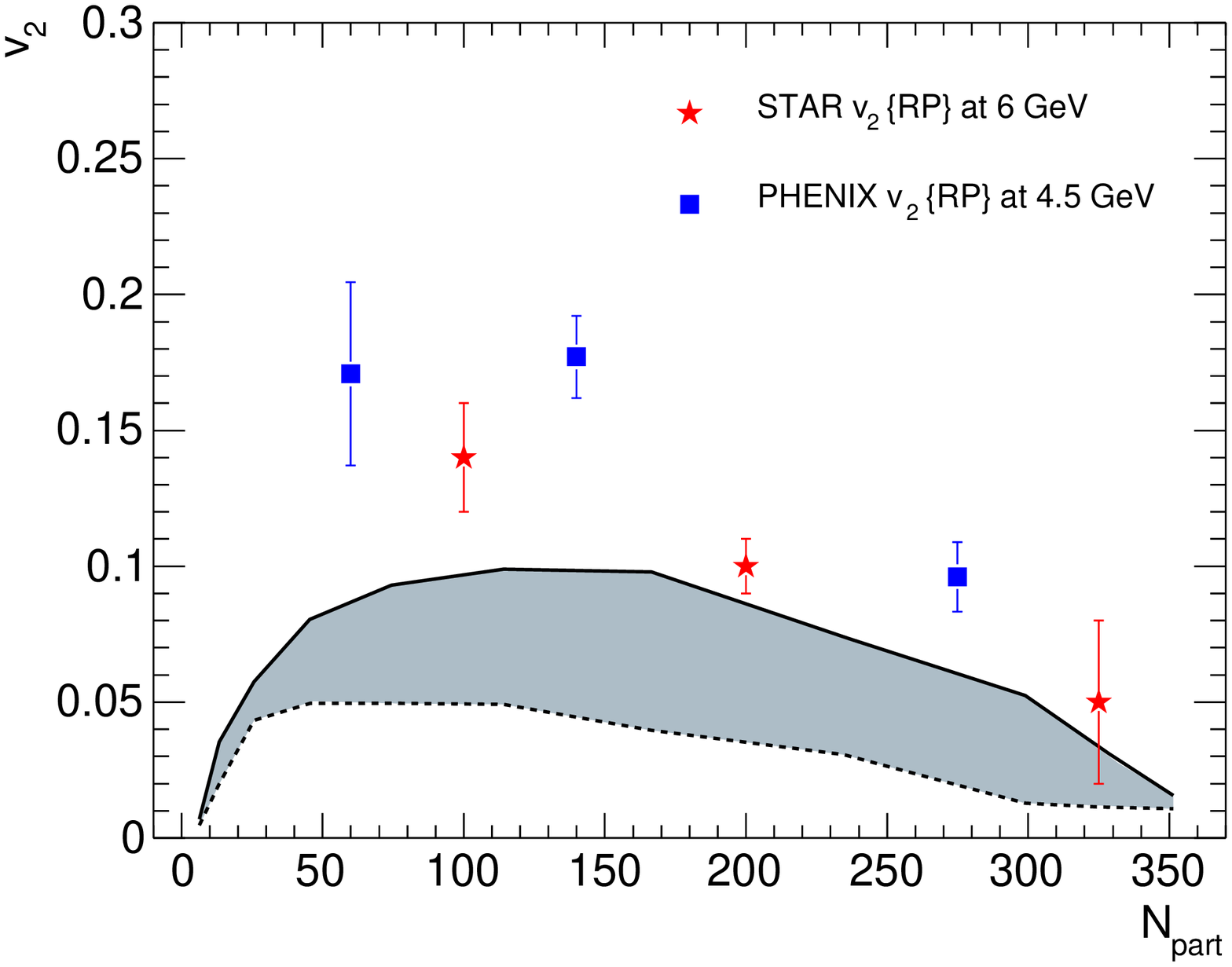,scale=0.42}\hspace{2em}
\epsfig{file=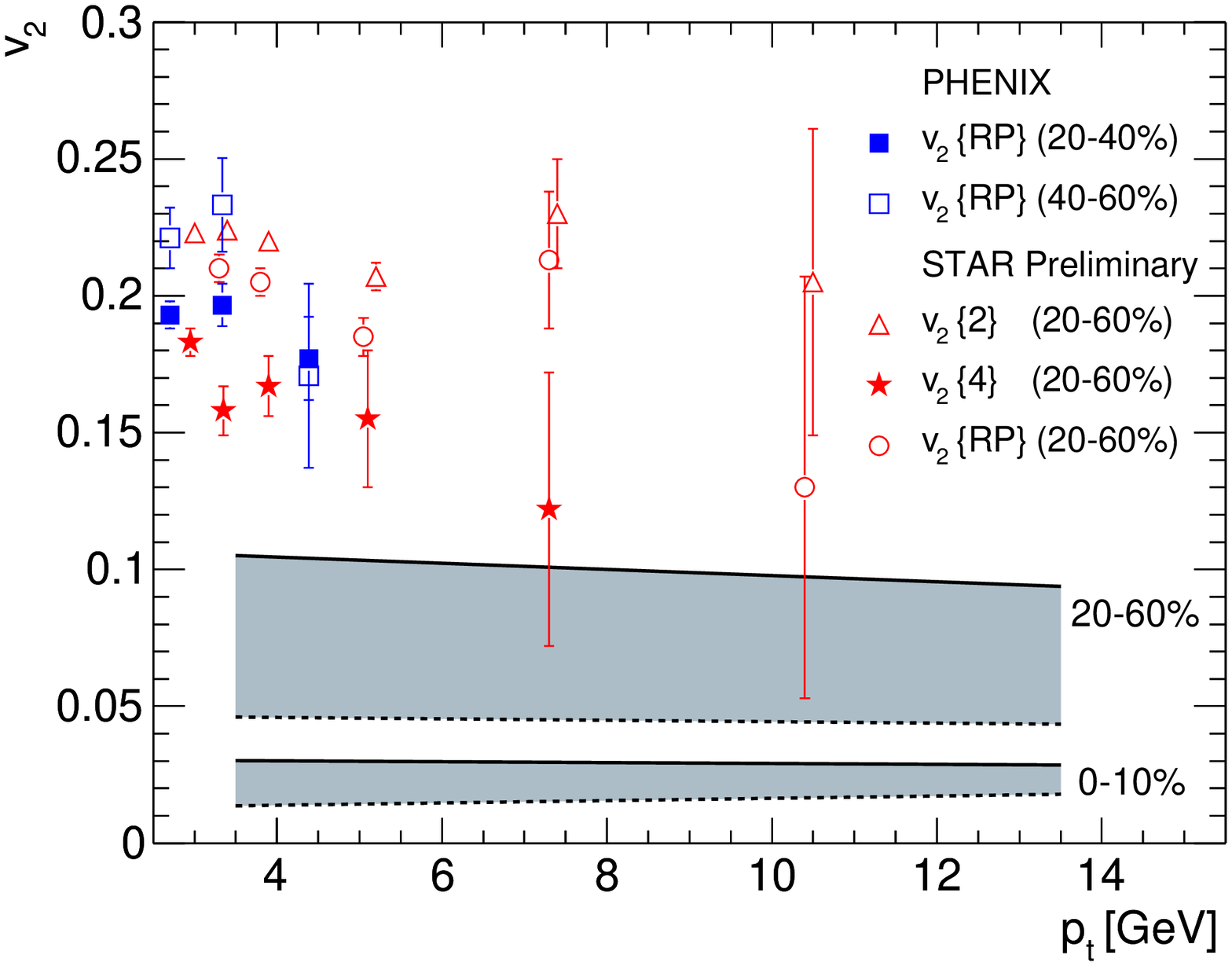,scale=0.42}
\end{minipage}
\begin{minipage}[t]{16.5 cm}
\caption{Left panel: $v_2$ as a function of $N_{\mathrm{part}}$. Right panel:
 $v_2$ as a function of $P_T$ for two different centrality bins.
 As in Fig.\ \ref{fig:daineseraa} dashed lines refer to reweighted and 
 solid lines to non-reweighted ASW quenching weights. STAR and PHENIX 
 data from references \cite{Adler:2003kt,Filimonov:2002xk,Tang:2004vc}.
Figures reprinted from \cite{Dainese:2004te} with kind permission from Springer 
Science + Business Media.
\label{fig:dainesev2}
}
\end{minipage}
\end{center}
\end{figure}
%%%%%%%%%%%%%%%%%%%%%%%%%%%%%%%%%%%%%%%%%%%%%%%%%%%%%%%

Let us come back to the problem of azimuthal anisotropy at large $P_T$.
We have already discussed the basic mechanism how perturbative hadron
production with final state interaction in a medium can lead to positive
$v_2$. We want to close this subsection by showing the $v_2$ obtained in
the study by Dainese, Loizides and Paic, using the ASW formalism, that we 
already discussed for their results on $R_{AA}$ in the previous subsection
\cite{Dainese:2004te}. 
Fig.\ \ref{fig:dainesev2} shows $v_2$ as a function of centrality 
(left panel) and as a function of $P_T$ (right panel) with data from
STAR and PHENIX. We notice that the asymmetry from the difference in opacity 
in- and out-of-plane can be sizable, but the calculation still underestimates
most of the data. The large $v_2$ measured by experiments at large $P_T$ 
has long been puzzling, but the experimental data also exhibits 
large error bars. Note that $\hat q$ was fixed in order to describe 
$R_{AA}$ which leaves no free parameter in this study.

We want to remind the reader of the right panel in Fig.\ 
\ref{fig:RAAcomparison}. In that study $R_{AA}$ was investigated as a function
of the azimuthal angle $\phi$ with respect to the reaction plane, not
just integrated over $\phi$. In principle $R_{AA}(\phi,P_T)$ has more
differential information than either $v_2$ or integrated $R_{AA}$.
We recall that the ASW formalism exhibited the strongest angular modulation
and has therefore the largest $v_2$ in that comparative study.

 %%%%%%%%%%%%%%%%%%%%%%%%%%%%%%%%%%%%%%%%%%%%%%%%%%%%%%%
\begin{figure}[tb]
%\begin{center}
\begin{minipage}{9.5cm}
\begin{center}
\epsfig{file=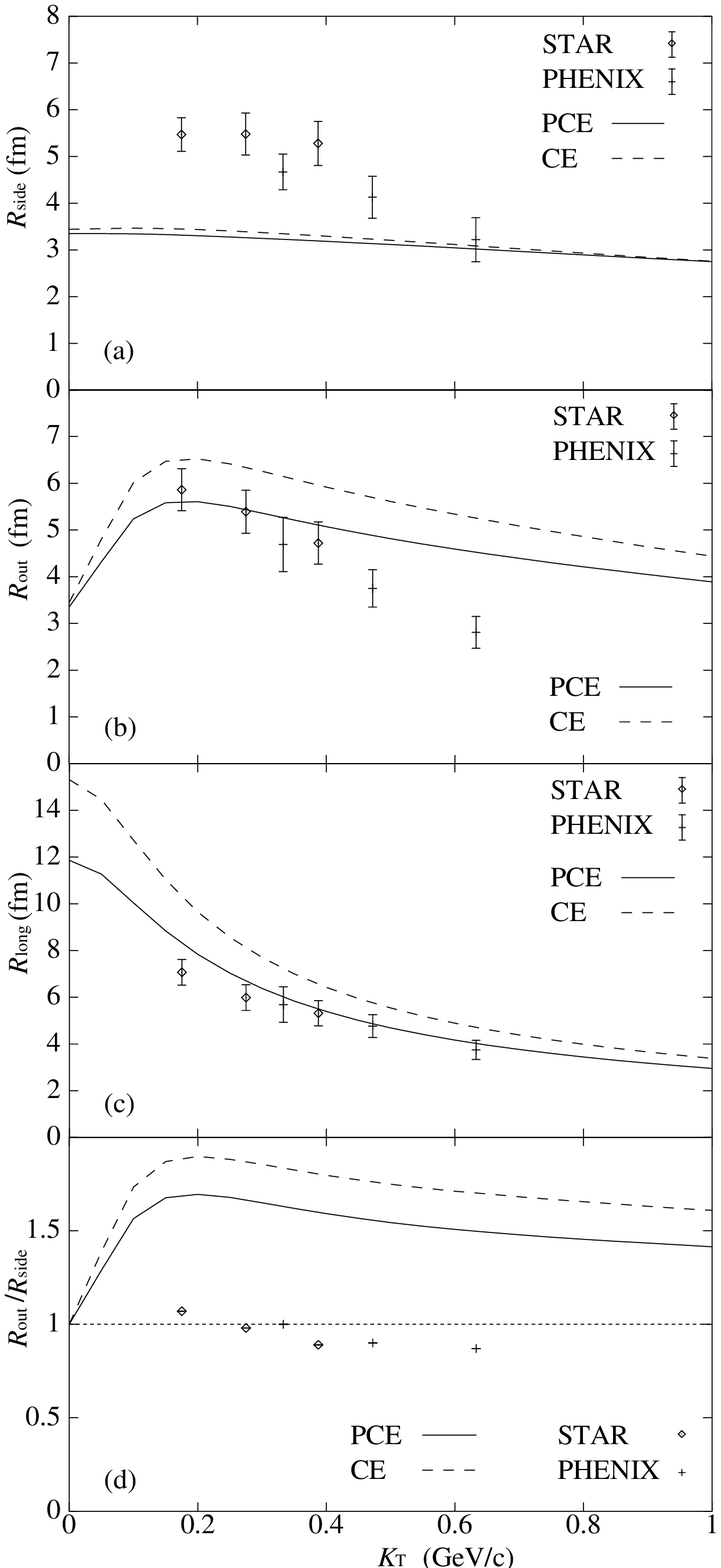,scale=0.44}
\end{center}
\caption{HBT radii for negative pions in (from the top) side, out, and long 
directions and the ratio $R_{\rm out}/R_{\rm side}$ (bottom) as function of 
pair momentum $K_T$. Solid and dashed lines correspond to hydro calculations
using partial chemical equilibrium (PCE) and full chemical equilibrium (CE)
resp. STAR and PHENIX data from Refs.\ \cite{Adler:2001zd,Adcox:2002uc}
Figure reprinted from \cite{Hirano:2002ds} with permission from the American 
Physical Society.}
\label{fig_hbt1}
\end{minipage}
\hspace{0.5cm}
\begin{minipage}{8.0cm}
\vspace{-1cm}
\epsfig{file=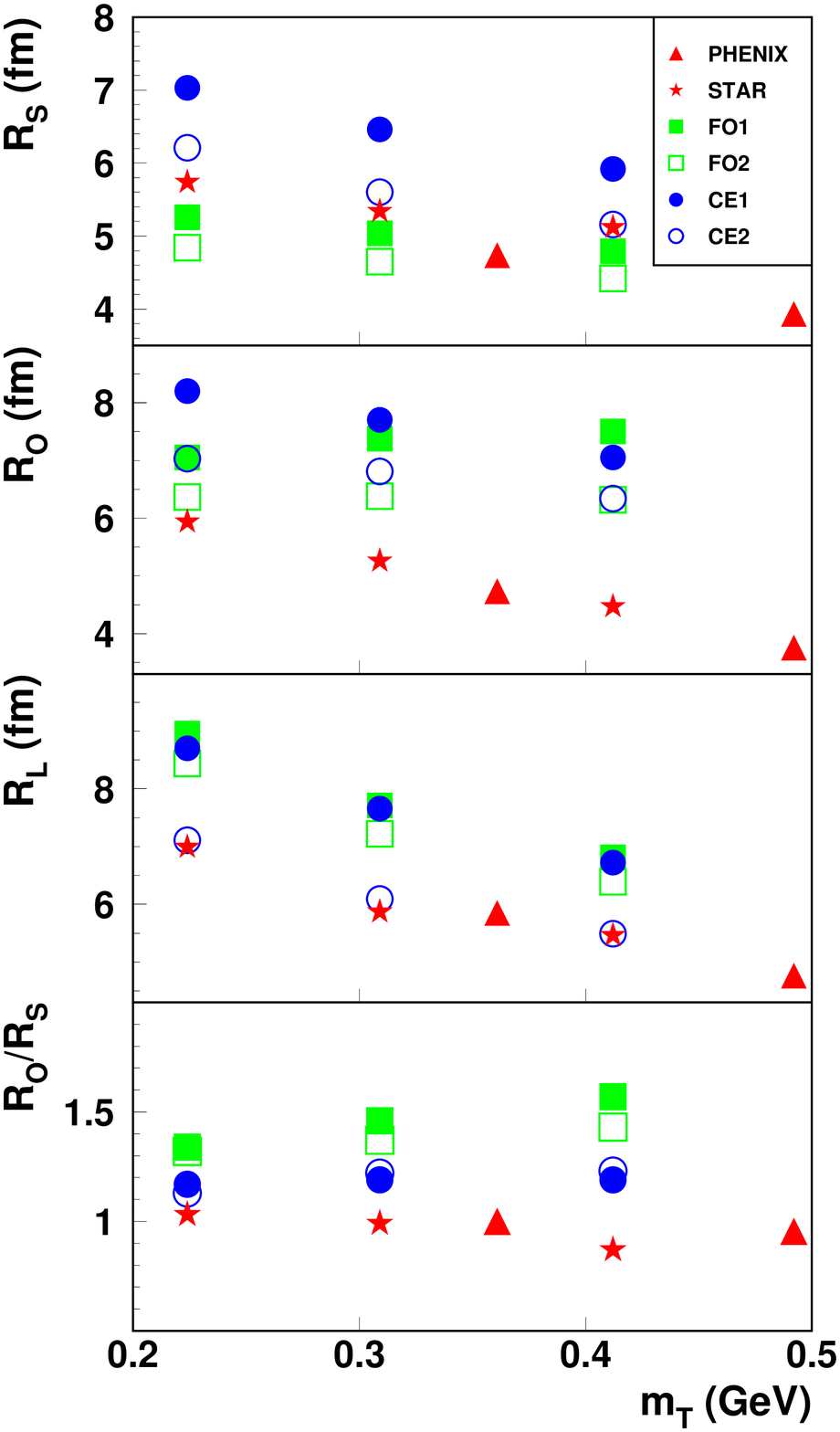, scale=0.4}
\caption{HBT radii and the ratio $R_{\rm out}/R_{\rm side}$ for 
four different scenarios for initial condition and freeze-out, see text
for details. Data for pions from STAR and PHENIX are shown
\cite{Adler:2001zd,Adcox:2002uc}.
Figure reprinted from \cite{Socolowski:2004hw} with permission from the
American Physical Society.}
\label{fig_hbt2}
\end{minipage}
%\end{center}
\end{figure}
%%%%%%%%%%%%%%%%%%%%%%%%%%%%%%%%%%%%%%%%%%%%%%%%%%%%%%%

%%%%%%%%%%%%%%%%%%%%%%%%%%%%%%%%%%%%%%%%%%%%%%%%%%%%%%%
\subsection{\it Two-Particle Correlations}
%%%%%%%%%%%%%%%%%%%%%%%%%%%%%%%%%%%%%%%%%%%%%%%%%%%%%%%

Before RHIC started up two-pion Hanbury Brown-Twiss (HBT) interferometry was 
believed to give us a clear signature of the QCD phase transition.
We expected that an enhancement of the ratio of the inverse width of the 
pion correlation function in out-direction to that in side-direction, 
which results from a prolonged life time of the fire ball with a phase
transition, could be observed at RHIC \cite{Rischke:1996em}.
However what we find in experimental data from RHIC are HBT radii 
that are almost the same as those measured at SPS \cite{Adcox:2004mh}.
Furthermore, most present hydrodynamic models can not describe the 
measured HBT radii correctly although they fit both spectra and 
elliptic flow. 

%%%%%%%%%%%%%%%%%%%%%%%%%%%%%%%%%%%%%%%%%%%%%%%%%%%%%%%
\begin{figure}[tb]
\begin{center}
\begin{minipage}{10cm}
\epsfig{file=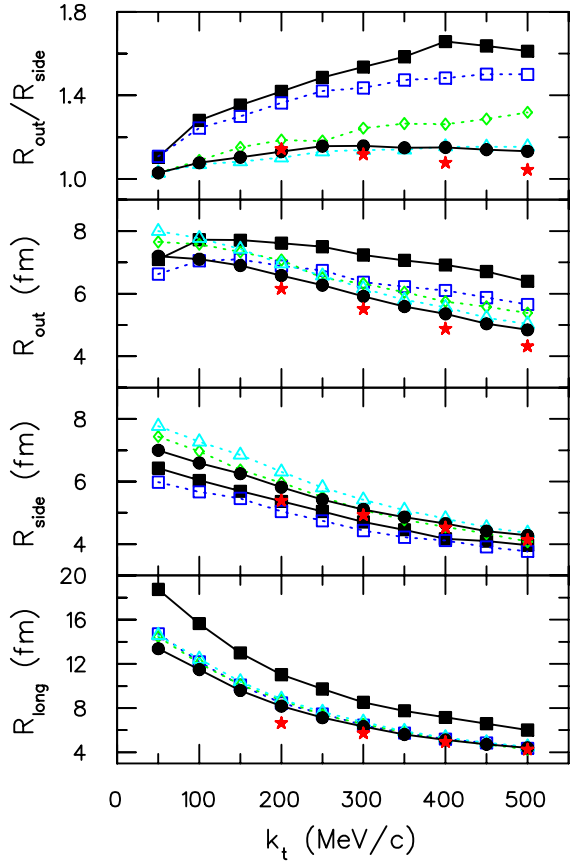,scale=0.5}
\end{minipage}

\hspace{-1cm}
\begin{minipage}{14cm}

\caption{Gaussian HBT radii $R_{\rm out}$, $R_{\rm side}$, and $R_{\rm long}$
from several hydrodynamic calculations explained in the text, together
with data from STAR (red stars).
Figure reprinted from \cite{Pratt:2008qv} with permission from the American 
Physical Society.}
\label{fig_hbt3}
\end{minipage}
\end{center}
%\end{center}
\end{figure}
%%%%%%%%%%%%%%%%%%%%%%%%%%%%%%%%%%%%%%%%%%%%%%%%%%%%%%%

Figure \ref{fig_hbt1} shows the HBT radii $R_{\rm side}$, $R_{\rm out}$,
$R_{\rm long}$ and the ratio $R_{\rm out}/R_{\rm side}$ for negative pions
calculated from (3+1)-dimensional relativistic hydrodynamics for both 
partial chemical equilibrium (PCE) and chemical equilibrium (CE)
in the hadronic phase by Hirano et al.\ \cite{Hirano:2002ds}. 
Except for $R_{\rm long}$ the hydrodynamic calculations fail to reproduce the 
experimental data quantitatively. 
$R_{\rm side}$ ($R_{\rm out}$) from the hydrodynamic model underestimates 
(overestimates) the experimental data. This leads to a large discrepancy 
in the ratio of $R_{\rm out}/R_{\rm side}$ between hydrodynamic models and 
experimental data, which is called the HBT puzzle. The same tendency can be
seen in other hydrodynamic calculations. 
At face value this means that the expansion of the fireball happens more 
rapidly in a shorter time than in hydrodynamic models.  
Partial chemical equilibrium pushes the hydrodynamic calculations closer to 
experimental data, however it is not enough to solve the HBT puzzle. 

Two improvements are taken into account in 
Ref.~\cite{Socolowski:2004hw}. For one, event-by-event fluctuations in 
the initial conditions, and secondly continuous emission instead of the
sudden freeze-out process which is usually used in hydrodynamic models.  
The results are shown in Fig.~\ref{fig_hbt2}. The same radii and ratios
as before are shown in four cases: sudden freeze-out with averaged 
initial condition (FO1), sudden freeze-out with fluctuating initial 
condition (FO2), continuous emission with averaged initial conditions
(CE1) and  continuous emission with fluctuating  initial conditions (CE2). 
The realistic treatments of initial conditions and freeze-out bring a 
significant improvement in $R_{\rm out}$, which also turns into a better 
agreement of $R_{\rm out}/R_{\rm side}$. However, small discrepancies between 
hydrodynamic calculations and the experimental data on $R_{\rm out}$ remain.

A solution to the HBT puzzle was finally proposed by Pratt \cite{Pratt:2008qv}. 
He suggests that the discrepancies are not from a single shortcoming of 
hydrodynamic calculations, but from a combination of several effects: 
mainly prethermalized acceleration, equations of state with inadequate
stiffness, and lacking viscosity. Figure \ref{fig_hbt3} shows results from
his calculations for the HBT radii together with data from STAR (red stars).
The results of many calculations are shown with the extremes being
from a hydro calculation with a first-order phase transition without
pre-thermal flow and without viscosity (black squares and line),
and the gradual improvements culminate in a calculation with
stiffer equation of state and pre-thermal flow and viscosities included
(black circles and line).

\begin{figure}[tb]
\epsfysize=9.0cm
\begin{center}
\begin{minipage}[t]{10 cm}
\epsfig{file=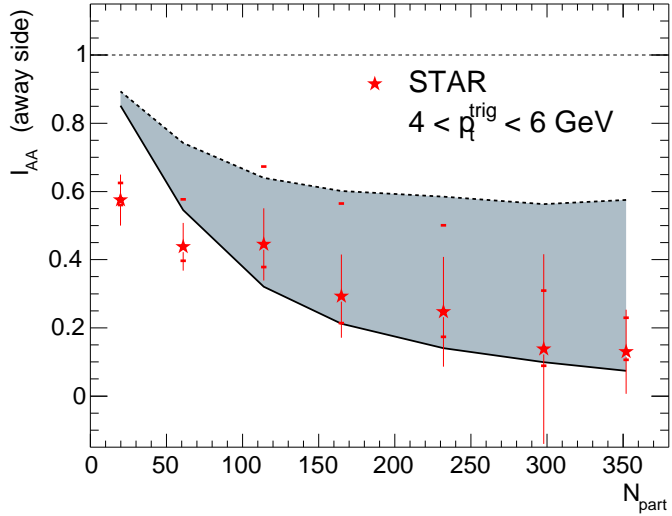,scale=0.45}
\end{minipage}
\begin{minipage}[t]{16.5 cm}
\caption{$I_{AA}$ for charged hadrons as a function of $N_{\mathrm{part}}$
 with trigger particle having $4 < P_T < 6$ GeV/$c$ compared
 to data from STAR \cite{Adler:2002ct}.
 Figure reprinted from \cite{Dainese:2004te} with kind permission from Springer
 Science and Business Media.
 \label{fig:daineseiaa}}
\end{minipage}
\end{center}
\end{figure}

At high $P_T$ 2-particle correlations are measured in heavy ion collisions
not for their information on quantum interference but in order to 
establish kinematic correlations. 
Di-hadron measurements at RHIC have made it possible
to study some properties of jets even though true jet reconstruction is
remarkably difficult. Correlation measurements can be recorded 
by using pairs $T-A$ where $T$ is a trigger hadron and $A$ the associated
particle. Most studies at intermediate and high $P_T$ have been carried
out using such triggered correlations, looking at the per-trigger yield
of associated particles
\begin{equation}
  Y(P_{T},P_{A},\Delta\phi) = \frac{dN/dP_T dP_A d(\Delta\phi)}{dN/dP_T}
\end{equation}
as a function of the relative azimuthal angle $\Delta \phi$ between
trigger and associate. Here $P_T$ and $P_A$ are, in deviation from
our usual notation, the transverse momenta of the trigger and associated
hadrons, resp.
One can then proceed and define a nuclear modification factor for the
associated yield
\begin{equation}
  I_{AA} =
  \frac{Y^{AA}(P_{T},P_{A},\Delta\phi)}{Y^{pp}(P_{T},P_{A},\Delta\phi)}
  \, .
\end{equation}
Like $R_{AA}$ we expect $I_{AA}$ to be unity for a loose superposition of
proton-proton collisions.
Another potentially useful observable are triggered fragmentation
functions. They can be derived from per-trigger yields by 
rewriting the dependence as one on a ``momentum fraction''
$z = P_A/P_T$ and integrating over a narrow bin around $\Delta\phi = \pi$.

The amount of data collected on this topic would merit its own review 
article. We will focus on a few selected examples, mainly to achieve our 
goal to better constrain the transport coefficient $\hat q$. We start
by showing the result from Ref.\ \cite{Dainese:2004te} which is now
well-known from our discussions on $R_{AA}$ and $v_2$. Fig.\ 
\ref{fig:daineseiaa} shows their result for $I_{AA}$ as a function 
of centrality in the ASW model of energy loss compared to data from STAR.
The data fall well into the large error band given by the uncertainty
from the reweighted vs the non-reweighted quenching weights.

\begin{figure}[tb]
\epsfysize=9.0cm
\begin{center}
\begin{minipage}[t]{18 cm}
\epsfig{file=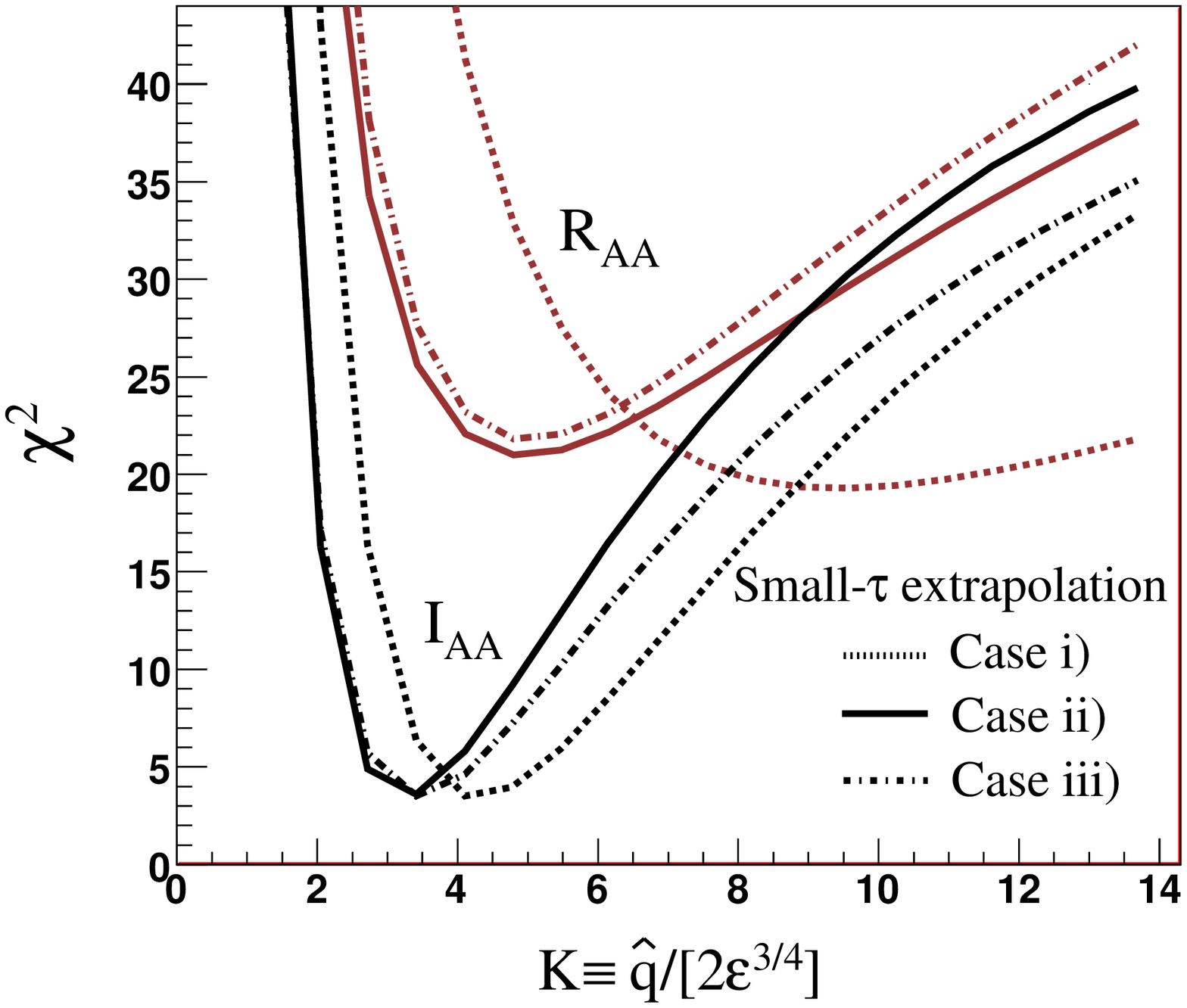,scale=0.4}\hspace{2em}
\epsfig{file=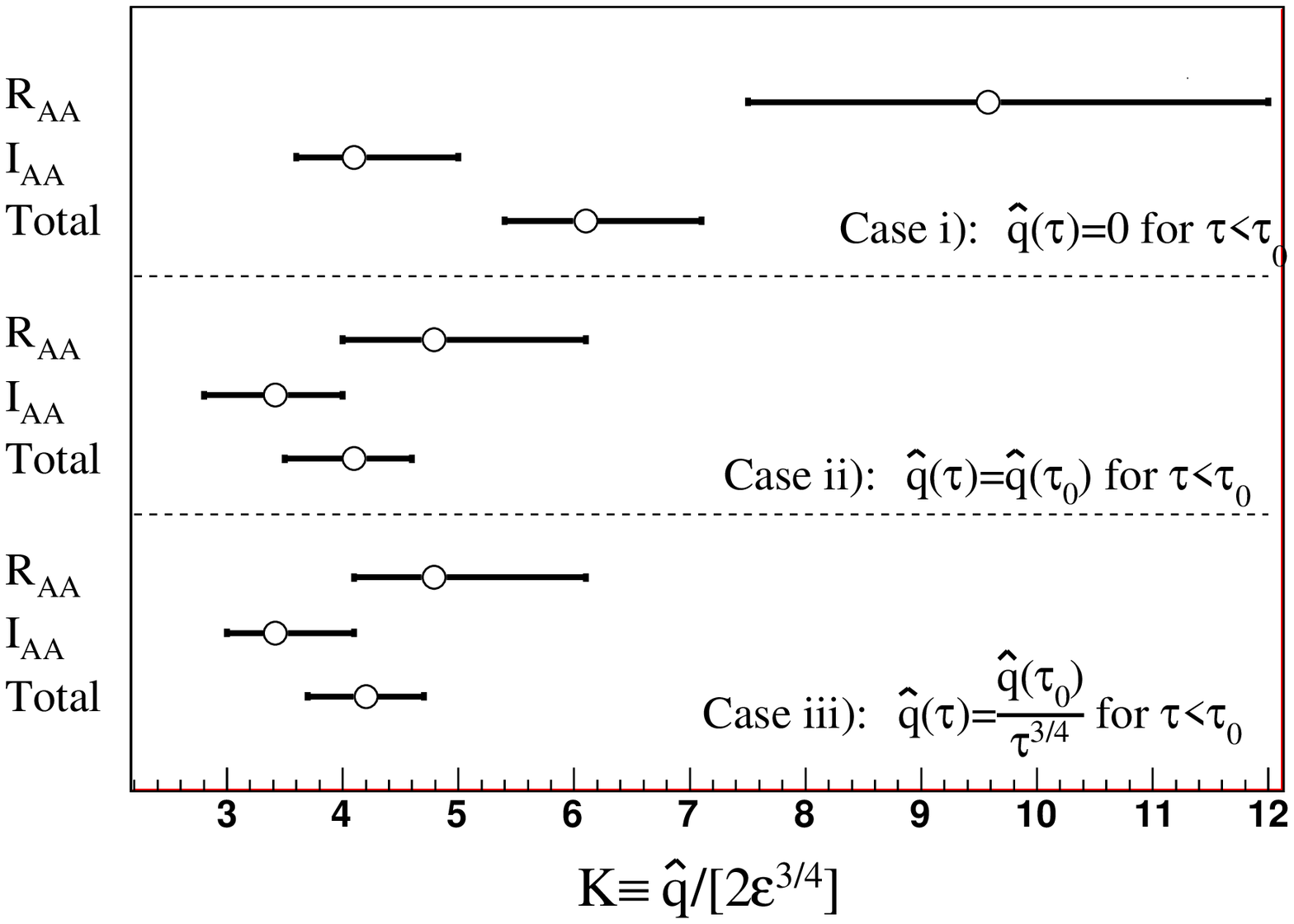,scale=0.45}
\end{minipage}
\begin{minipage}[t]{16.5 cm}
\caption{$\chi^2$-fits of the parameter $K$ in the ASW model of energy loss
  from $R_{AA}$ and $I_{AA}$ for three different extrapolations of jet 
  quenching into the pre-equilibrium phase.
  Figures reprinted from \cite{Armesto:2009zi} with permission from IOP.
 \label{fig:armesto}}
\end{minipage}
\end{center}
\end{figure}

Let us recall the right panel of Fig.\ \ref{fig:owens}. In that study
the Higher Twist formalism was used together with a calculation
of hard processes at NLO accuracy to extract the stopping power 
$\epsilon_0$. As shown in the figure, fitting both $I_{AA}$ and $R_{AA}$
leads to consistent values for $\epsilon_0$. This is somewhat different
for the recent study by Armesto et al.\ in the ASW model where the
consistency of $\hat q$ extracted from $R_{AA}$ and $I_{AA}$ data 
is very sensitive to details of the modeling \cite{Armesto:2009zi}.
Fig.\ \ref{fig:armesto} shows the result for $K = \hat q/(2\epsilon^{3/4})$
extracted from $R_{AA}$ and $I_{AA}$ with varying treatments of the quenching
during the pre-equilibrium phase. The three cases shown are (i) no
quenching before the QGP formation time $\tau_0$, (ii) constant $\hat q$
for $\tau < \tau_0$ and (iii) $\hat q$ increasing as $\tau^{-3/4}$ toward
$\tau =0$. The results is a reminder that the determination of $\hat q$ from
data has larger uncertainties.

\subsection{\it Photons}

Electromagnetic probes are important means to obtain information from 
the quark gluon plasma. Photons and lepton pairs carry the information 
of the whole time-evolution of heavy ion collisions: initial hard collisions,
thermalization, expansion, hadronization, and freeze-out. 
We focus here on photons which encode detailed information of every
process which occurs in heavy ion collisions. A schematic account of
photon sources has already been given, but we want to list the sources
of direct photons once more in one place:
(i) prompt hard photons from initial collisions, (ii) vacuum bremsstrahlung
(fragmentation) photons, (iii) photons from jet conversions and medium-induced
bremsstrahlung (iv) thermal photons from quark gluon plasma (and the hot
hadronic phase). This list also orders photons according to which momentum
regime they are most important for (from the largest $P_T$ to the smallest).

We can deduce the temperature of the fireball from thermal radiation (and
check our understanding of the fireball evolution), and we can infer
information about the medium density and $\hat q$ from jet conversions
and induced photon bremsstrahlung. Photons have also great importance
as triggers in photon-hadron correlation studies at large $P_T$.
Triggered fragmentation functions with photon triggers come close
to real fragmentation functions since the photon, if the source is 
dominated by initial hard photons, carries the same transverse momentum
as its initial partner parton \cite{Wang:1996yh}. This opens a way
to measure real medium-modified fragmentation functions, however one
has to be cautious because of the many other sources that weaken this
kinematic link between the photon and the jet on the other side.

%%%%%%%%%%%%%%%%%%%%%%%%%%%%%%%%%%%%%%%%%%%%%%%%%%%%%%%
\begin{figure}[t]
%\begin{center}
\begin{minipage}{7.5cm}
\epsfig{file=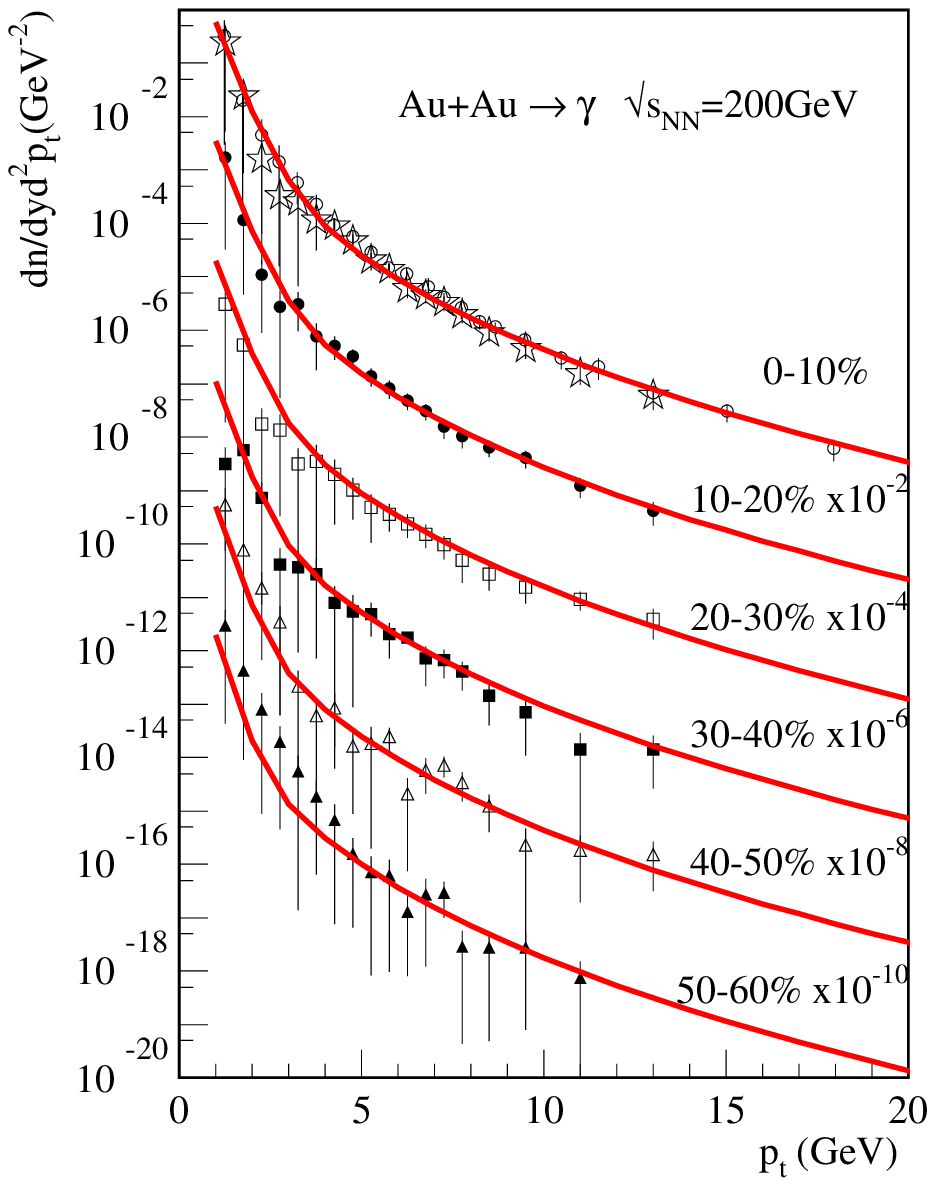,scale=0.7}
\caption{Direct photon production in Au+Au collisions at 
$\sqrt{s_{NN}} = 200$ GeV for five different centrality bins compared to 
data from PHENIX \cite{Adler:2005ig}. 
Figure reprinted from \cite{Liu:2008eh} with permission from the American 
  Physical Society.
\label{fig_photon_liu1}}
\end{minipage}
\hspace{0.5cm}
\begin{minipage}{10cm} \hspace{-4em}
\epsfig{file=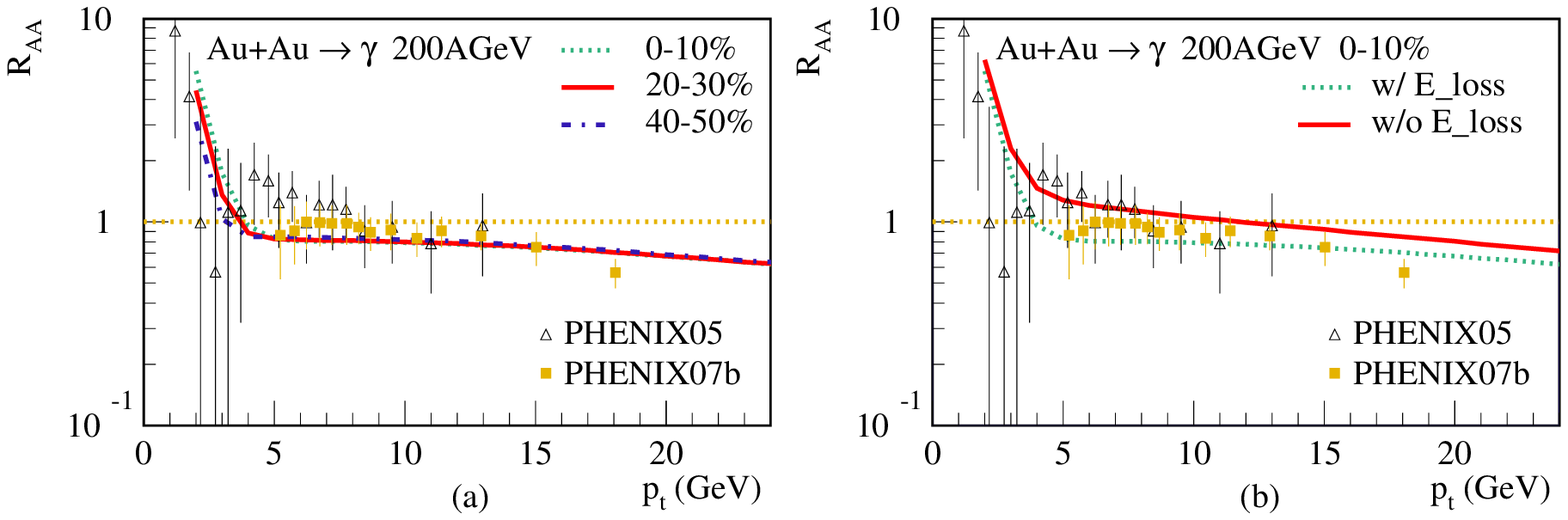,scale=0.64}
\caption{Nuclear modification factor $R_{AA}$ of direct photons as in
  Fig.\ \ref{fig_photon_liu1}. Left panel: three different centrality bins 
  are shown. Right panel: calculations with and without energy loss
  of partons.
  Figures reprinted from \cite{Liu:2008eh} with permission from the American 
  Physical Society.
\label{fig_photon_liu2}}
\end{minipage}
%\end{center}
\end{figure}

\begin{figure}[tb]
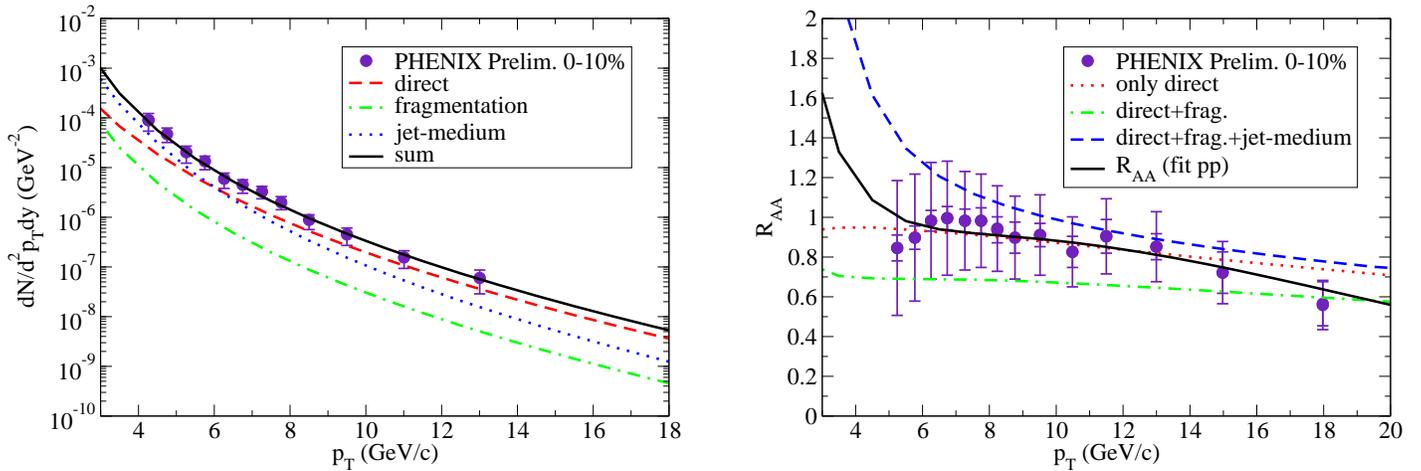

\epsfysize=9.0cm
\begin{center}
\begin{minipage}[t]{19 cm}
\epsfig{file=./plots/qin_photon_yield_0_10,scale=0.35}\hspace{2em}
\epsfig{file=./plots/qin_photon_raa_0_10,scale=0.35}
\end{minipage}
\begin{minipage}[t]{16.5 cm}
\caption{Left panel: direct photon spectra for central Au+Au collisions
at RHIC using contributions from hard collisions, bremsstrahlung and
jet conversions together with data from PHENIX \cite{Isobe:2007ku}.
Right panel: $R_{AA}$ for direct photons from the same sources.
Figures reprinted from \cite{Qin:2009bk} with permission from the American 
Physical Society.
 \label{fig:amyphot}}
\end{minipage}
\end{center}
\end{figure}

\begin{figure}[tb]
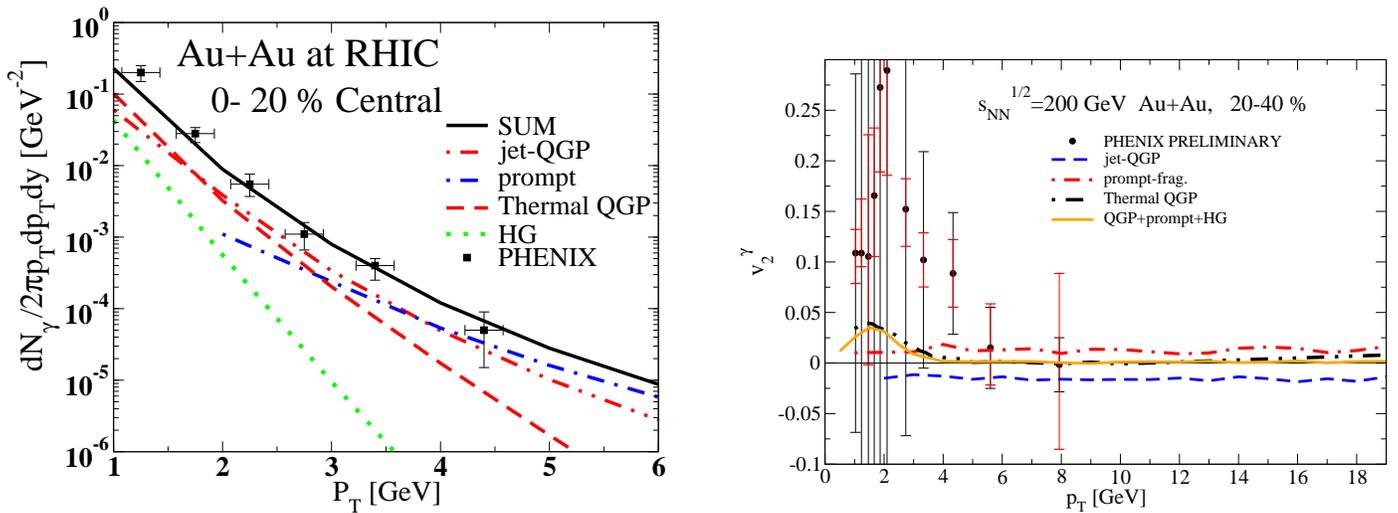

\epsfysize=9.0cm
\begin{center}
\begin{minipage}[t]{19 cm}
\epsfig{file=./plots/turbide_photon_0_20,scale=0.32}\hspace{2em}
\epsfig{file=./plots/turbide_v2_photon,scale=0.32}
\end{minipage}
\begin{minipage}[t]{16.5 cm}
\caption{Left panel: Direct photon spectra from all known sources for 
central Au+Au collisions using the AMY formalism consistently. 
Right panel: direct photon $v_2$ as function of $P_T$.
Figures reprinted from \cite{Turbide:2007mi} with permission from the American 
Physical Society.
 \label{fig:photv2}}
\end{minipage}
\end{center}
\end{figure}

\begin{figure}[t]
\epsfysize=9.0cm
\begin{center}
\begin{minipage}[t]{10 cm}
\epsfig{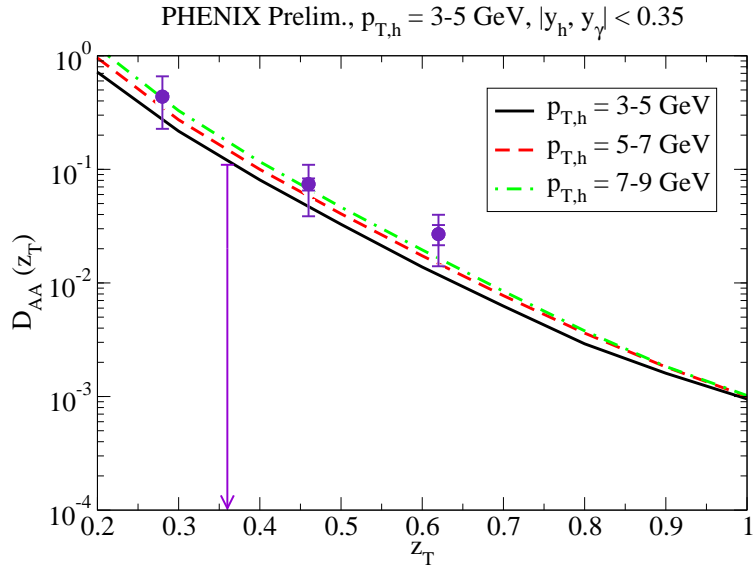}
\end{minipage}
\begin{minipage}[t]{16.5 cm}
\caption{Photon-triggered fragmentation functions $D_{AA}$ for three
 different trigger windows
in the same model as in Fig.\ \ref{fig:amyphot} compare to data from PHENIX
\cite{Adare:2009vd}. Figure reprinted from \cite{Qin:2009bk} with permission
from the American Physical Society.
 \label{fig:amycorr}}
\end{minipage}
\end{center}
\end{figure}

Figures \ref{fig_photon_liu1}  and \ref{fig_photon_liu2} show the 
centrality dependence and the nuclear modification factor of direct 
photons in Au+Au collisions at $\sqrt{s_{NN}} = 200$ GeV, respectively,
in the study by Liu et al. \cite{Liu:2008eh}. They include all
four sources mentioned above and compute thermal radiation 
by using (3+1)-dimensional hydrodynamic fireball evolution. 
Their calculation describes the direct photon data from PHENIX very well.
The same can be said about the calculation of the McGill group 
\cite{Qin:2009bk}.
They use all four sources of photons as well, with energy loss of
jets and induced photon bremsstrahlung taken care of by the AMY
formalism. A successful fit of hadron and photon data together is
hence a crucial consistency test for the AMY formalism.
Fig.\ \ref{fig:amyphot} shows their result for both single inclusive
photon spectra and $R_{AA}$ for large and intermediate $P_T$ (thermal
radiation is therefore omitted in this case).

Fig.\ \ref{fig:photv2} shows a calculation of the McGill group including
thermal photons according to Arnold, Moore, and Yaffe \cite{Turbide:2007mi}. 
In that study the authors also calculated the azimuthal asymmetry $v_2$ 
of direct photons. The $v_2$ of thermal photons and of fragmentation 
photons is expected to be positive, reminiscent of the $v_2$ of bulk 
hadrons and high-$P_T$ hadrons respectively. However, photons from jet 
conversions and induced bremsstrahlung add a negative contribution to the
direct photon $v_2$ as they are produced more abundantly in the direction
where the medium is thicker \cite{Turbide:2005bz}. The left panel
of Fig. \ref{fig:photv2} shows different contributions to $v_2$ with
the contribution from jet conversions indeed being negative. Experimental
data from PHENIX has been inconclusive so far due to large error bars.
Measurements of negative values of $v_2$ would
be direct evidence for jet conversions.

Fig.\ \ref{fig:amycorr} shows photon-triggered fragmentation functions
$D_{AA}$ obtained from photon-hadron correlations calculated in Ref.\
\cite{Qin:2009bk} with the AMY formalism together with data from PHENIX.
This is one of the emerging examples of jet tomography with photon
triggers as originally envisioned in \cite{Wang:1996yh}. In this study
the data is described reasonably well with the parameter in the AMY
energy loss (the coupling constant $\alpha_s$) set to one consistent
value for all observables.

%%%%%%%%%%%%%%%%%%%%%%%%%%%%%%%%%%%%%%%%%%%%%%%%%%%%%%%%%%%%%%%%%%%%%%%

\section{Summary and Conclusions \label{sec:sum}}

Let us briefly summarize. We have laid out the foundations of perturbative
QCD and how they can be used to understand data from the Relativistic
Heavy Ion Collider. We explained the concept of collinear factorization 
widely used in collider physics, and modifications when applied to
colliding nuclei. We have then focused on final state interactions,
and in particular on leading particle energy loss.
Along the way have discussed several popular model calculations for 
energy loss. We have also pointed out some of their basic assumptions
in which they differ. All of them have apparent shortcomings, and even
if jet quenching at RHIC is perfectly perturbative, we should not expect
all of these calculations to apply. However, all of them explain jet
quenching well on a qualitative level. This can be seen positive, we
are on the right track, or negative, we are not in a situation where we
can confidently falsify one or all of those calculations, due to a long
list of uncertainties. We are also not quite ready yet to exclude
non-perturbative quenching scenarios from data. In fact, it is very well
possible that a mixture of perturbative and non-perturbative quenching
co-exist in different temperature ranges.

We showed that all models fit the data on single hadron suppression after 
adjusting a single parameter, including the systematics of the $P_T$- and 
centrality dependence of $R_{AA}$.
On the other hand the quenching strengths extracted from the different
models is quite different. 
The maximum values of $\hat q$ in central collisions that have been found
lie in the range $\approx 2 \ldots 15$ GeV$^2$/fm.
Other observables like $v_2$ at high $P_T$ or $I_{AA}$ have note yet 
cut down this range as it has become clear that the results
are quite sensitive, e.g.\ to details of the treatment of the underlying
fireball.

There is a focused effort to attack the open questions within
the framework of the TECHQM collaboration and the newly founded JET
collaboration. In the future we need rigorous comparisons and a vetting 
process of different calculations, a realistic modeling of the fireball 
and realistic assessments of uncertainties in order to start to falsify
certain assumptions and to reduce the error bar on $\hat q$. We also
have to include more realistic options, like combining perturbative 
quenching with final state effects in the hadronic phase.

We have also reviewed the successful story of hydrodynamics at RHIC,
and the novel developments of the past few years that have brought
the first relativistic viscous hydro codes and hybrid models. We have 
discussed a long list of observables that require the presence
of a thermalized and collectively moving quark gluon plasma phase, 
that can be beautifully explained by hydrodynamic calculations and
their extensions. The most convincing single observations are the mass
ordering of flow and elliptic flow, and the large size of elliptic flow at
RHIC, which are suggested by hydrodynamics and observed in the data.
Starting from basic concepts of ideal and viscous
hydrodynamics we have moved along a variety of steps that brought
vast improvements to our understanding of hydrodynamics, including the
arrival of 3+1 dimensional codes, the correct treatment of freeze-out, etc.

Like in the case of pQCD and jet quenching, hydrodynamics has had some 
difficulties entering a period of precision measurements.
Currently, the equation of state from the comparison between hydrodynamic 
analyses and experimental data favor a first order deconfinement transition,
while lattice QCD prefers a crossover transition.
Clearly, more systematic phenomenological studies, are needed that try
to start from more relaxed assumptions, e.g.\ by including initial radial flow
and dissipative corrections. It is not until both sides, phenomenological
analyses of data and theoretical tools like lattice QCD, 
show the same conclusion that we can pin down the QCD phase diagram.
Routine runs of fully 3+1-dimensional, viscous hydrodynamics are around the
corner and they should bring us a step closer towards an understanding
of QCD thermodynamics and transport coefficients. Current estimates of
the shear viscosity range from the 1 to about 4 times the KSS bound of
$s/4\pi$.

One of the big challenges in this field is the huge amount of data from
RHIC that has remained almost untouched by comprehensive theoretical 
calculations. There are many results on dihadron azimuthal correlations 
at high $P_T$ \cite{Franz:2008ri}, three-particle azimuthal correlations, 
the ridge structure \cite{Mohanty:2008tw}, dihadrons with respect
to the reaction plane \cite{Vale:2009zr}, and the emerging 
full-jet reconstruction \cite{Putschke:2009wr, Salur:2009vz}. 
All of these results should fit into a tent anchored on both sides
by pQCD and hydrodynamics. Multi-module modeling, carefully tested and
endowed with realistic estimates of theoretical uncertainties should
eventually be able to explain those features and in the process the data
should narrow down the error bars on transport coefficients significantly.

On the experimental side two new challenges will arrive shortly.
Energy scans will be conducted at RHIC \cite{STAR-BES} and supplemented
by SPS results and the program at the new FAIR and NICA
facilities \cite{SPS-energy}.
These programs will provide new data on the QCD phase diagram at high
net baryon densities. Reliable hydro+cascade hybrid models will be given
a chance to look for the QCD critical point and a change in the nature
of the phase transition. On the other hand the LHC will vastly improve
the reach of hard probes by colliding ions at unprecedented center of mass
energies. Reconstructed jets and abundant high-$P_T$ probes will provide
much stricter tests of our understanding of how jets interact with quark 
gluon plasma.

As the first 10 years of the high energy heavy ion era come to an end
we find that we have gained much knowledge. It prepares us for
the next decade in which the QCD phase diagram and the quark gluon plasma 
phase will be mapped out quantitatively.

\subsection*{Acknowledgments}

We like to thank R.\ Rodriguez for many useful discussions.
This work was supported by CAREER Award PHY-0847538 from the U.~S.~National
Science Foundation, an Invited Fellowship for Research in Japan by the 
Japanese Society for the Promotion of Science (JSPS), 
RIKEN/BNL Research Center, DOE grant DE-AC02-98CH10886, and
the Global COE Program ``Quest for Fundamental Principles in the Universe" 
of Nagoya University (G07), Grant-in-Aid for Young Scientists (B) (22740156) and 
Grant-in-Aid for Scientific Research (S)(22224003) and 
JSPS Institutional Program for Young Researcher 
Overseas Visits.
R.\ J.\ F.\ would like to express his gratitude to Chiho
Nonaka and the Physics Department at Nagoya University for their kind 
hospitality while part of this work was completed.

\end{document}